\documentclass[twocolumn,prd,floats,preprintnumbers,showpacs]{revtex4}  % Final revtex form

\usepackage[english]{babel}
\usepackage[dvips]{graphicx}
\usepackage{amsmath,amssymb}

\begin{document}

% Put (re)newcommands here.
\newcommand{\ednt}[1]{\textbf{[#1]}}
\newcommand{\etal}{et al.\ }
\renewcommand{\topfraction}{0.95}

% Mathematical typesetting
\newcommand{\nadi}{n_{\rm{ad}}}
\newcommand{\nadiI}{n_{\rm{ad1}}}
\newcommand{\nadiII}{n_{\rm{ad2}}}
\newcommand{\niso}{n_{\rm{iso}}}
\newcommand{\ncor}{n_{\rm{cor}}}
\newcommand{\bark}{{\bar{k}}}       % For  k/k_0
\newcommand{\mc}[1]{\mathcal{#1}}
\newcommand{\mr}[1]{\mathrm{#1}}
\newcommand{\abs}[1]{\vert#1\vert}
\newcommand{\nc}{\newcommand}
 \nc{\sign}{{\mbox{sign}}}
 \nc{\bea}{\begin{eqnarray}} \nc{\eea}{\end{eqnarray}}
 \nc{\beq}{\begin{equation}} \nc{\eeq}{\end{equation}}
 \nc{\nn}{\nonumber}
 \nc{\bi}{\begin{itemize}}\nc{\ei}{\end{itemize}}
 \nc{\ben}{\begin{enumerate}}\nc{\een}{\end{enumerate}}
 \nc{\bb}[1]{\bibitem{#1} (#1)}          % Working form for showing cite labels
% \nc{\bb}[1]{\bibitem{#1}}              % Final form

\title{Correlated Primordial Perturbations in Light of CMB and LSS Data}

\author{Hannu Kurki-Suonio}
\email{hannu.kurki-suonio@helsinki.fi}
\affiliation{Department of Physical Sciences, University of Helsinki,
P.O. Box 64, FIN-00014 University of Helsinki, Finland}

\author{Vesa Muhonen}
\email{vesa.muhonen@helsinki.fi}
\affiliation{Helsinki Institute of Physics, University of Helsinki,
P.O. Box 64, FIN-00014 University of Helsinki, Finland}

\author{Jussi V\"{a}liviita}
\email{jussi.valiviita@helsinki.fi}
\affiliation{Helsinki Institute of Physics, University of Helsinki,
  P.O. Box 64, FIN-00014 University of Helsinki, Finland}

\date{16th December 2004}

\begin{abstract}
We use cosmic microwave background (CMB) and large-scale structure data to
constrain cosmological models where the primordial perturbations have both an
adiabatic and a cold dark matter (CDM) isocurvature component.
We allow for a possible correlation between
the adiabatic and isocurvature modes, and for different spectral indices for
the power in each mode and for their correlation.  We do a likelihood analysis
with 11 independent parameters.  We discuss the effect of choosing the pivot
scale for the definition of amplitude parameters.  The upper limit
for the isocurvature fraction is 18\% around a pivot scale
$k = 0.01$Mpc$^{-1}$. For smaller pivot wavenumbers the limit 
stays about the same.  For larger pivot wavenumbers, very large values of
the isocurvature spectral index are favored, which makes the analysis
problematic, but larger isocurvature fractions seem to be allowed.
For large isocurvature spectral indices $\niso > 2$ a positive correlation
between the adiabatic and isocurvature mode is
favored, and for $\niso < 2$ a negative correlation is favored. The upper
limit to the nonadiabatic contribution to the CMB temperature variance is
7.5\%. Of the standard cosmological parameters, determination
of the CDM density $\omega_c$ and the sound horizon angle $\theta$
(or the Hubble constant $H_0$) are affected most by a possible
presence of a correlated isocurvature contribution. The
baryon density $\omega_b$ nearly retains its ``adiabatic value''.
\end{abstract}

\pacs{PACS numbers: 98.70.Vc, 98.80.Cq}
%\keywords{}
\preprint{HIP-2004-67/TH}
%\preprint{astro-ph/0412XXX}

\maketitle

%
%  -------------------------------------------
%  Article
%  -------------------------------------------
%

\section{Introduction}
\label{sec:intro}

The major part of the present cosmological data, including the cosmic microwave
background (CMB) anisotropy and the large scale distribution of galaxies (large
scale structure, LSS) is fit reasonably well by a simple cosmological model.
This model has a spatially flat ($\Omega = 1$) background geometry.  It has
five energy density components, ``baryons'', photons, massless neutrinos, cold
dark matter (CDM), and a constant vacuum energy (cosmological constant). The
primordial scalar perturbations are gaussian, adiabatic, and scale-free ($\nadi
= \mbox{const.}$).

We call this model the ``adiabatic model'' in this paper.  It has 5 parameters
to be determined from the data, the Hubble constant, $H_0 \equiv h 100\mbox{
km/s/Mpc}$, two density parameters, $\omega_b \equiv \Omega_b h^2$ and
$\omega_c \equiv \Omega_c h^2$ (for baryons and CDM), and the amplitude $A$ and
spectral index $\nadi$ of the primordial scalar perturbations. There is no
evidence in the cosmological data for the presence of additional features or
ingredients beyond this model, like tensor perturbations or neutrino masses,
indicating that they are probably so small as not to show up in the data.
(Actually there is also no evidence for a deviation from scale-invariance,
$\nadi = 1$.) The concordance values of the parameters are $h \sim 0.7$,
$\omega_b \sim 0.023$, $\omega_c \sim 0.12$, $\nadi \sim 1.0$, and
$A \sim 5\times10^{-5}$.

Besides these $5$ fundamental cosmological parameters, there are $2$ additional
parameters needed when the models are compared to CMB and LSS data: the optical
depth $\tau$ due to reionization, and the bias parameter $b$ relating the
observed galaxy power spectrum to the underlying matter power spectrum.

The origin of the primordial perturbations is not known. The favorite
candidate for their generation is quantum fluctuations during a period of
inflation in the very early universe.  While single-field inflation produces
adiabatic perturbations, inflation with more than one field produces also
entropy perturbations $\mc{S}$ in addition to the usual curvature perturbation
$\mc{R}$.

A general perturbation can be divided into an adiabatic mode + an isocurvature
mode, where the adiabatic mode has no initial entropy perturbation, and the
isocurvature mode has no initial curvature perturbation.  Allowing for the
presence of an isocurvature mode does not improve the fit to the existing data
(to the extent of justifying the additional parameters), and thus there is so
far no evidence for the existence of primordial isocurvature perturbations.
However, it is of interest to find out what limits the data set to these
perturbations, as the nature of primordial perturbations is an important clue
to their origin.  Moreover, the presence of an undetected isocurvature
contribution may affect the determination of the main cosmological parameters.

 In principle, there can be different kinds of entropy
perturbations, and thus several different isocurvature modes.  Four different
isocurvature modes were identified in \cite{Bucher:1999re}, the CDM and baryon
isocurvature modes, and two neutrino isocurvature modes. Allowing for the
simultaneous presence of all four kinds would lead to so many parameters that
it would be difficult to obtain meaningful results \cite{Bucher:2000hy}. The
signature of a baryon isocurvature mode in the data is rather similar to the
CDM isocurvature mode, but weaker due to the smaller baryon density parameter.

Here we consider only the CDM isocurvature mode in addition to the adiabatic
mode. We allow the CDM entropy perturbations to have a different spectral index
from the curvature perturbations, and to be (or not to be) correlated with
them.  In comparison to the adiabatic model this brings in 4 new parameters
related to the amplitudes and spectral indices of the entropy perturbations and
their correlation with the curvature perturbations. Thus we have in total $11$
parameters in our cosmological model. For sampling this 11-dimensional
parameter space we use the Markov Chain Monte Carlo (MCMC) method. This is a
follow-up paper of \cite{Valiviita:2003ty} where a preliminary analysis (around
the best-fit adiabatic model) was presented. Here we include more data in the
analysis and sample the likelihood function more accurately.

Before the  Wilkinson Microwave Anisotropy Probe (WMAP) data
\cite{Bennett:2003bz} became available, limits to the
isocurvature contribution in uncorrelated models had been obtained  for the
case $\nadi = \niso = 1$ in \cite{Stompor:1995py} and with $\nadi$ and $\niso$
as independent parameters in \cite{Enqvist:2000hp}, and in correlated models
for one independent spectral index in \cite{Amendola:2002ni}. Pure CDM
isocurvature models had been ruled out  also in the case of a non-flat
background geometry in \cite{Enqvist:2001fu}. Correlated
models were also studied in \cite{Moroi:2001ct,Moroi:2002rd}.

After WMAP, limits to correlated models were first obtained for the case of two
independent spectral indices \cite{Peiris:2003ff,Crotty:2003rz}.  In our
earlier work \cite{Valiviita:2003ty} we obtained preliminary results for the
case of three independent spectral indices using WMAP data only.  Parkinson
\etal \cite{Parkinson:2004yx} considered a particular inflation model producing
correlated perturbations. Moodley \etal \cite{Moodley:2004nz} considered models
with up to three isocurvature modes (CDM and two neutrino modes) present
simultaneously, but all sharing the same spectral index. Ferrer \etal
\cite{Ferrer:2004nv} studied correlated perturbations resulting from inflaton
and curvaton decay. They had two independent spectral indices.

The most similar to the present study is that of Beltran \etal
\cite{Beltran:2004uv}, who consider one isocurvature mode at a time, and allow
separate spectral indices for the adiabatic and isocurvature modes and their
correlation. We compare their approach to ours at the end of
Sec.~\ref{sec:correlation} and their results to ours in Sec.~\ref{sec:beltran}.

Since the adiabatic and isocurvature
components and their correlation are allowed to have different spectral
indices, their relative amplitudes vary as a function of scale $k$.  We define
the amplitude parameters at some chosen pivot scale $k_0$.

When the isocurvature component or the correlation is negligibly small, the
corresponding spectral indices are not constrained by the data.  Such
conditionally unconstrained parameters cause problems also for determining
other parameters from the data.

The way the isocurvature perturbation and correlation is parameterized (e.g.\
the choice of pivot scale $k_0$) affects the integration measure in the
parameter space. Thus different parameterizations correspond to different
priors. When parameters are weakly constrained by the data this ends up in
different posterior likelihoods: When one parametrization (A) is used to obtain
the likelihood function in the parameter space and the results are then
expressed in another parametrization (B), the likelihood function is different
from the case when parametrization B was used initially.  (This difference can
be ``fixed'' by importance weighting using the Jacobian of the parameter
transformation; but this does not address the question which parametrization is
``correct''.) Such effects are discussed in Sec.~\ref{sec:pivotscale}.

We find that the pivot scale should be chosen to be near the middle of the data
sets used (in terms of $\ln k$).
When the isocurvature spectral index $\niso$ is a free parameter a
wrong choice would spoil the analysis. This comes because the data does not
prefer an isocurvature contribution. Then using $k_0$ that is close to the
small $k$ end of the data, $k_\mr{min}$,  leads to extremely small (negative)
$\niso$, in order to minimize the isocurvature contribution. On the other hand,
if $k_0$ is too close to $k_\mr{max}$, then arbitrarily large isocurvature
spectral indices are favored to minimize an overall isocurvature contribution
in the range $[k_\mr{min},k_\mr{max}]$. Unfortunately, the ``standard pivot
scale'' $k_0 = 0.05$Mpc$^{-1}$ (used e.g. by CAMB
\cite{Lewis:2002ah,Lewis:1999bs}) is quite close to $k_\mr{max}$ and another
common choice (see e.g. \cite{Crotty:2003rz}) to give for $k_0$ a value that
corresponds to the present Hubble radius is nearly equal to setting $k_0 =
k_\mr{min}$.

When we started MCMC runs for our model we took $k_0 = 0.05$Mpc$^{-1}$, but
soon realized that our Markov Chains ran towards artificially large $\niso$.
After fixing this problem, when we were finalizing the analysis of better runs
with $k_0 = 0.01$Mpc$^{-1}$, paper \cite{Beltran:2004uv} with pivot scale $k_0
= 0.05$Mpc$^{-1}$ came out. However, they had an ad hoc constraint $\niso < 3$
that saved their main results from most of the artifacts that arise when the
chains run to very large $\niso$. With our choice of the pivot scale the
likelihood for $\niso$ peaks at $\niso \sim 3$ and drops rapidly around
$\niso \sim 4$. Hence, the prior $\niso < 3$ in \cite{Beltran:2004uv} allows a
comparison to our results.

We obtain tight constraints to the CDM isocurvature contribution and find that,
of the main cosmological parameters, only the determination of $\omega_c$ and
$h$ is significantly affected. Compared to the adiabatic models smaller values
of $\omega_c$ and larger values of $h$ become acceptable when allowing for CDM
isocurvature. Interestingly, although we have two additional degrees of freedom
in spectral indices, dertermination of
the baryon density is much less affected than in the
models where all modes share the same spectral index.

In Sec.~\ref{sec:correlation} we introduce and motivate our parametrization of
correlated curvature and entropy perturbations.  In Sec.~\ref{sec:tech} we
write down some technical details of our MCMC study to determine these
parameters, and in Sec.~\ref{sec:results} we give and discuss our results.  In
Sec.~\ref{sec:nonadicon} we discuss the non-adiabatic contribution to the
\emph{observed} CMB and matter power spectra, and in Sec.~\ref{sec:pivotscale}
the effect of changing the pivot scale.  In Sec.~\ref{sec:beltran} we compare
our results to those of \cite{Beltran:2004uv}.

\section{Correlated Perturbations}
\label{sec:correlation}

The calculation of the CMB angular power spectra $C_l$ and the matter power
spectra $P(k)$ starts from ``initial'' values $\mc{R}_{\mr{rad}}$ and
$\mc{S}_{\mr{rad}}$ specified deep in the radiation dominated era
(rad), when all scales of interest are well ``outside the horizon''
(i.e., the Hubble scale $H^{-1}$).  However, this ``initial'' time is well
after inflation, or whatever generated the perturbations, and refers to a time
during and after which the evolution of the universe is assumed to be known. We
denote the time when the perturbations were generated by the subscript
$\ast$.  For inflation, this corresponds to the time when the scale in question
``exited the horizon'' (thus it is different for different scales $k$). Between
$t_{\ast}(k)$ and $t_{\mr{rad}}$ the perturbation is outside the horizon, i.e.,
$k$ is ``superhorizon''.

In the absence of entropy perturbations, curvature perturbations remain
constant at superhorizon scales.  This is not, in general, true for entropy
perturbations, which may evolve at superhorizon scales.  Entropy perturbations
may also seed curvature perturbations.  This happens, e.g., in two-field
inflation, when the background trajectory in field space is curved
\cite{Langlois:1999dw,Langlois:2000ar,Gordon:2000hv}.

Thus the relation between the ``generated'' and ``initial'' values for $\mc{R}$
and $\mc{S}$ can be represented as \cite{Amendola:2002ni}
\begin{equation}
  \label{eq:cor_matrix}
  \begin{pmatrix}
    \mc{R}_{\mr{rad}} \\
    \mc{S}_{\mr{rad}}
  \end{pmatrix}
  =
  \begin{pmatrix}
    1 & T_{\mc{RS}} \\
    0 & T_{\mc{SS}}
  \end{pmatrix}
  \begin{pmatrix}
    \mc{R}_{\ast} \\
    \mc{S}_{\ast}
  \end{pmatrix}.
\end{equation}
The transfer functions $T_{xy}(k)$ describe how the perturbations evolve from
the time of inflation to the beginning of the radiation dominated era. The
exact form of these functions is model dependent and that aspect is not studied
in this work.  We approximate them by power laws.

In the literature there are different sign conventions for the perturbations
${\mc{R}}$ and ${\mc{S}}$.  We define them so that an initial positive comoving
curvature perturbation $\mc{R}_{\mr{rad}}$ corresponds to an initial
overdensity $\delta = \delta\rho/\rho > 0$, and an initial positive entropy
perturbation $\mc{S}_{\mr{rad}}$ corresponds to an initial CDM overdensity. In
terms of the Bardeen potentials, $\Phi$ and $\Psi$, defined so that the metric
in the conformal-Newtonian gauge is
 \begin{equation}
  ds^2 = -(1+2\Phi)dt^2 +
    a(t)^2(1-2\Psi)\delta_{ij}dx^idx^j  \,,
  \label{cNmetric}
 \end{equation}
the comoving curvature perturbation reads
 \begin{equation}
  \mc{R} \equiv -\Psi - \frac{2\rho}{3(\rho+p)}\left(\frac{1}{H}\frac{\partial\Psi}
  {\partial t} + \Phi\right)
  \,,
 \end{equation}
 and the entropy perturbation is
 \begin{equation}
  \mc{S} \equiv \delta_c - \frac34\delta_\gamma \,,
 \end{equation}
where $\delta_c$ and $\delta_\gamma$ are the CDM and photon density
perturbations. With these sign conventions, the ordinary Sachs-Wolfe effect is
 \beq
    \frac{\delta T}{T} \approx -\frac15(\mc{R}_\mr{rad}+2f_c\mc{S}_\mr{rad})
 \eeq
(here $f_c \equiv \omega_c/(\omega_b+\omega_c)$), and a positive correlation
between $\mc{R}_{\mr{rad}}$ and $\mc{S}_{\mr{rad}}$ leads to an additional
positive contribution to the large scale CMB anisotropy, and also to a positive
contribution to the matter power spectrum.

We define the correlation $\mc{C}_{xy}(k)$ between two perturbation quantities
(random variables), $x$ and $y$, as
\begin{equation}
  \label{eq:cor_def}
  \bigl\langle x(\vec{k})y^{\ast}(\vec{k}') \bigr\rangle \bigr\vert_{\mr{rad}} =
  \tfrac{2\pi^{2}}{k^{3}} \mc{C}_{xy}(k) \delta^{(3)}(\vec{k} - \vec{k}').
\end{equation}

The transfer function $T_{\mc{RS}}(k)$ leads to a correlation between
$\mc{R}_{\mr{rad}}$ and $\mc{S}_{\mr{rad}}$ from uncorrelated $\mc{R}_{\ast}$ and
$\mc{S}_{\ast}$,
\begin{align}
  \mc{C}_{\mc{RR}}(k) &= \mc{P}_{\mc{R}}(\ast,k) + T_{\mc{RS}}(k)^2
  \mc{P}_{\mc{S}}(\ast,k)\\
  \mc{C}_{\mc{RS}}(k) &= T_{\mc{RS}}(k)T_{\mc{SS}}(k)\mc{P}_{\mc{S}}(\ast,k)\\
  \mc{C}_{\mc{SS}}(k) &= T_{\mc{SS}}(k)^2 \mc{P}_{\mc{S}}(\ast,k) \,,
\end{align}
where $\mc{P}_{\mc{R}}(\ast,k)$ and $\mc{P}_{\mc{S}}(\ast,k)$ are the power
spectra of $\mc{R}_{\ast}$ and $\mc{S}_{\ast}$.

Approximating the power spectra $\mc{P}_{\mc{R}}(\ast,k)$,
$\mc{P}_{\mc{S}}(\ast,k)$ and the transfer functions $T_{\mc{RS}}(k)$,
$T_{\mc{SS}}(k)$ by power laws with spectral indices $m_1$, $m_2$, $m_3$, and
$m_4$, respectively, we get that the autocorrelations (power spectra) have the
form
\begin{equation}
  \label{eq:autocor}
  \begin{split}
    \mc{P}_{\mc{R}}(k) &\equiv \mc{C}_{\mc{RR}}(k) =
    A^{2}_{r}\biggl(\frac{k}{k_{0}} \biggr)^{\nadiI-1} +
    A^{2}_{s}\biggl(\frac{k}{k_{0}} \biggr)^{\nadiII-1} \\
    \mc{P}_{\mc{S}}(k) &\equiv \mc{C}_{\mc{SS}}({k}) =
    B^{2}\biggl(\frac{k}{k_{0}} \biggr)^{\niso-1},
  \end{split}
\end{equation}
where $\nadiI = m_1+1$, $\nadiII = m_2+2m_3+1$, and $\niso
= m_2+2m_4+1$ and the epoch (rad) is implied. The three components
are the usual adiabatic mode, a second adiabatic mode generated by the entropy
perturbation, and the usual isocurvature mode, with amplitudes $A_{r}$, $A_{s}$
and $B$ at the pivot scale $k=k_{0}$, respectively.

The cross-correlation between the adiabatic and the isocurvature
component is now
\begin{equation}
  \label{eq:crosscor}
  \mc{C}_{\mc{RS}}(k) = \mc{C}_{\mc{SR}}(k) =
  A_{s}B\biggl(\frac{k}{k_{0}} \biggr)^{\ncor-1},
\end{equation}
where $\ncor = m_2+m_3+m_4+1 = (\niso + \nadiII)/2$. The correlation is between
the second adiabatic and the isocurvature component as is natural since these
components have the same source.

We have chosen the pivot scale $k_0 = 0.01~\mbox{Mpc}^{-1}$, but we also
consider pivot scales 0.002~Mpc$^{-1}$ and 0.05~Mpc$^{-1}$ in
Sec.~\ref{sec:pivotscale}.
%The choice of the pivot scale is of crucial
%importance in obtaining the results from the data fitting.
We shorten the notation by defining $\bark = k/k_{0}$.

The CMB angular power spectrum is given by
\begin{multline}
  \label{eq:Cl_from_cor}
  C_{l}^{(\mr{T/E/B})(\mr{T/E/B})} = \\
  4\pi
  \sum_{xy}\int\frac{dk}{k} \mc{C}_{xy}(k) g^{(\mr{T/E/B})}_{xl}(k)
  g^{(\mr{T/E/B})}_{yl}(k),
\end{multline}
where the $g_{l}$'s are the transfer functions that describe how an initial
perturbation evolves to a temperature (T) or polarization (E- or B-mode)
anisotropy multipole $l$.

Now, using the equations \eqref{eq:autocor}, \eqref{eq:crosscor} and
\eqref{eq:Cl_from_cor} we obtain for the temperature angular power spectrum
\begin{align}
  \label{eq:ClTT}
  C^{\mr{TT}}_{l} &= 4\pi \int \frac{dk}{k} \Bigl[ A^{2}_{r}
  \bigl(g^{\mr{T}}_{\mc{R}\:l}\bigr)^{2} \bark^{\nadiI-1} +
  A^{2}_{s} \bigl(g^{\mr{T}}_{\mc{R}\:l}\bigr)^{2} \bark^{\nadiII-1} \nonumber\\
  &\quad+ B^{2} \bigl(g^{\mr{T}}_{\mc{S}\:l}\bigr)^{2} \bark^{\niso-1} +
  2A_{s}B g^{\mr{T}}_{\mc{R}\:l}g^{\mr{T}}_{\mc{S}\:l} \bark^{\ncor-1}
  \Bigr]  \nonumber\\
  \equiv & A_r^2\hat{C}^{\mr{TTad1}}_l + A_s^2\hat{C}^{\mr{TTad2}}_l +
  B^2\hat{C}^{\mr{TTiso}}_l
  + A_sB\hat{C}^{\mr{TTcor}}_l ,
\end{align}
and for the TE cross-correlation spectrum
\begin{align}
  \label{eq:ClTE}
%  \begin{split}
    C^{\mr{TE}}_{l} &= 4\pi \int \frac{dk}{k} \Bigl[
    A^{2}_{r} g^{\mr{T}}_{\mc{R}\:l}g^{\mr{E}}_{\mc{R}\:l}\bark^{\nadiI-1} +
    A^{2}_{s} g^{\mr{T}}_{\mc{R}\:l}g^{\mr{E}}_{\mc{R}\:l}\bark^{\nadiII-1} \nn\\
    &\quad+ B^{2} g^{\mr{T}}_{\mc{S}\:l}g^{\mr{E}}_{\mc{S}\:l} \bark^{\niso-1} \nn\\
    &\quad+ A_{s}B \bigl(g^{\mr{T}}_{\mc{R}\:l}g^{\mr{E}}_{\mc{S}\:l} +
    g^{\mr{T}}_{\mc{S}\:l}g^{\mr{E}}_{\mc{R}\:l}\bigr) \bark^{\ncor-1}
    \Bigr] \nn\\
    \equiv & A_r^2\hat{C}^{\mr{TEad1}}_l + A_s^2\hat{C}^{\mr{TEad2}}_l +
  B^2\hat{C}^{\mr{TEiso}}_l
  + A_sB\hat{C}^{\mr{TEcor}}_l \,.
% \end{split}
\end{align}

There are thus $3$ amplitude parameters (three absolute values and one sign,
the relative sign of $A_s$ and $B$.)  Now we need to choose the amplitude
parametrization to be used in the likelihood analysis, i.e., what shall we use
as the three independent parameters with flat prior likelihoods.  One choice
would be just $A_r$, $A_s$, and $B$. However, we would rather express our
results in terms of a total amplitude, a relative isocurvature contribution and
a correlation.

In \cite{Valiviita:2003ty} we followed \cite{Peiris:2003ff} and used
 \beq
    f_\mr{iso} \equiv \sqrt{\frac{B^2}{A_r^2+A_s^2}} \in [0,\infty)
 \eeq
for the isocurvature contribution and
 \beq
    \cos\Delta \equiv \sign(A_sB)\sqrt{\frac{A_s^2}{A_r^2+A_s^2}}
 \eeq
for the correlation (with the sign convention opposite to that of
\cite{Peiris:2003ff}). The data is quadratic in these parameters (see
Eq.~(\ref{eq:ClTT})), meaning that fairly large values of $f_{\mr{iso}}$ and
$\cos\Delta$ are needed for the effect to show up in the data. This exacerbates
the problem that models with a small $A_s$ and $B$ get a lot of weight in the
likelihood function, since the spectral indices $\nadiII$ and
$\niso$ are not constrained.

A flat prior for $\cos\Delta$ leads to a non-flat prior distribution for
$\cos^2\!\!\Delta$. Thus the parametrization by $\cos\Delta$ favors small
multiplier $\cos^2\!\!\Delta$ in front of the second adiabatic component in
\cite{Valiviita:2003ty,Peiris:2003ff}. Moreover, large values of $\sin^2\!\!\Delta
= 1-\cos^2\!\!\Delta$ are then favored, so that even without any data the first
adiabatic component will be favored in the likelihood analysis. Likewise, the
parametrization by $ f_\mr{iso}$ (instead of $f_\mr{iso}^2$) favor small
multiplier in front of the isocurvature component. All in all, there was an
implicit bias towards pure adiabatic models in
\cite{Valiviita:2003ty,Peiris:2003ff}. A similar caveat applies to
\cite{Ferrer:2004nv}.

We would prefer amplitude parameters for which the data has a linear response.
We define a total amplitude parameter $A$ by
\begin{equation}
  A^2 \ \equiv \ A_r^2 + A_s^2 + B^2
  \label{eq:defA}
\end{equation}
and the isocurvature fraction and correlation parameters
\begin{align}
    \label{eq:defalpha}
    \alpha &\equiv \frac{B^2}{A^2} \in [0, 1]\\
    \label{eq:defgamma}
    \gamma &\equiv \sign(A_sB)\frac{A_s^2}{A_r^2+A_s^2} \in [-1, 1]\,.
\end{align}
Now the total angular power spectrum can be written as:
\begin{align}
  \label{eq:totCl}
  C_{l} &= A^{2} \bigl[ (1-\alpha)(1-\abs{\gamma})\hat{C}^{\mr{ad1}}_{l} +
  (1-\alpha)\abs{\gamma}\hat{C}^{\mr{ad2}}_{l} \nonumber \\
  & + \alpha \hat{C}^{\mr{iso}}_{l} +
  \mr{sign}(\gamma)\sqrt{\alpha(1-\alpha)\abs{\gamma}}\hat{C}^{\mr{cor}}_{l}
  \bigr] \nonumber \\
  & \equiv
  C^{\mr{ad1}}_{l} + C^{\mr{ad2}}_{l} + C^{\mr{iso}}_{l} + C^{\mr{cor}}_{l}
  \,.
\end{align}

\begin{figure}[th]
  \centering
  \includegraphics[angle=270,width=0.45\textwidth]{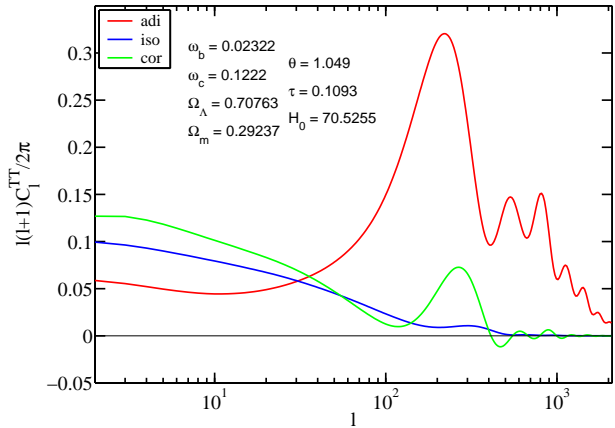}
  \caption{The unit-amplitude component angular power spectra
    $\hat{C}^{\mr{ad}}_{l}$ (red), $\hat{C}^{\mr{iso}}_{l}$ (blue), and
    $\hat{C}^{\mr{cor}}_{l}$ (green) of Eqs.~(\ref{eq:ClTT}) and
    (\ref{eq:totCl}) for the case of spectral indices $\nadi =
    \niso = 1$ and other cosmological parameters representing median
    values of their marginalized likelihoods from our 11-parameter model.
    These curves would represent the relative contributions to the total $C_l$
    for the case $\alpha = 0.5$, $\gamma = 1$, i.e., ``equal'' weights for the
    adiabatic and isocurvature contributions and a maximal positive
    correlation between them.}
  \label{fig:cltt}
\end{figure}

\begin{figure}[th]
  \centering
  \includegraphics[angle=270,width=0.40\textwidth]{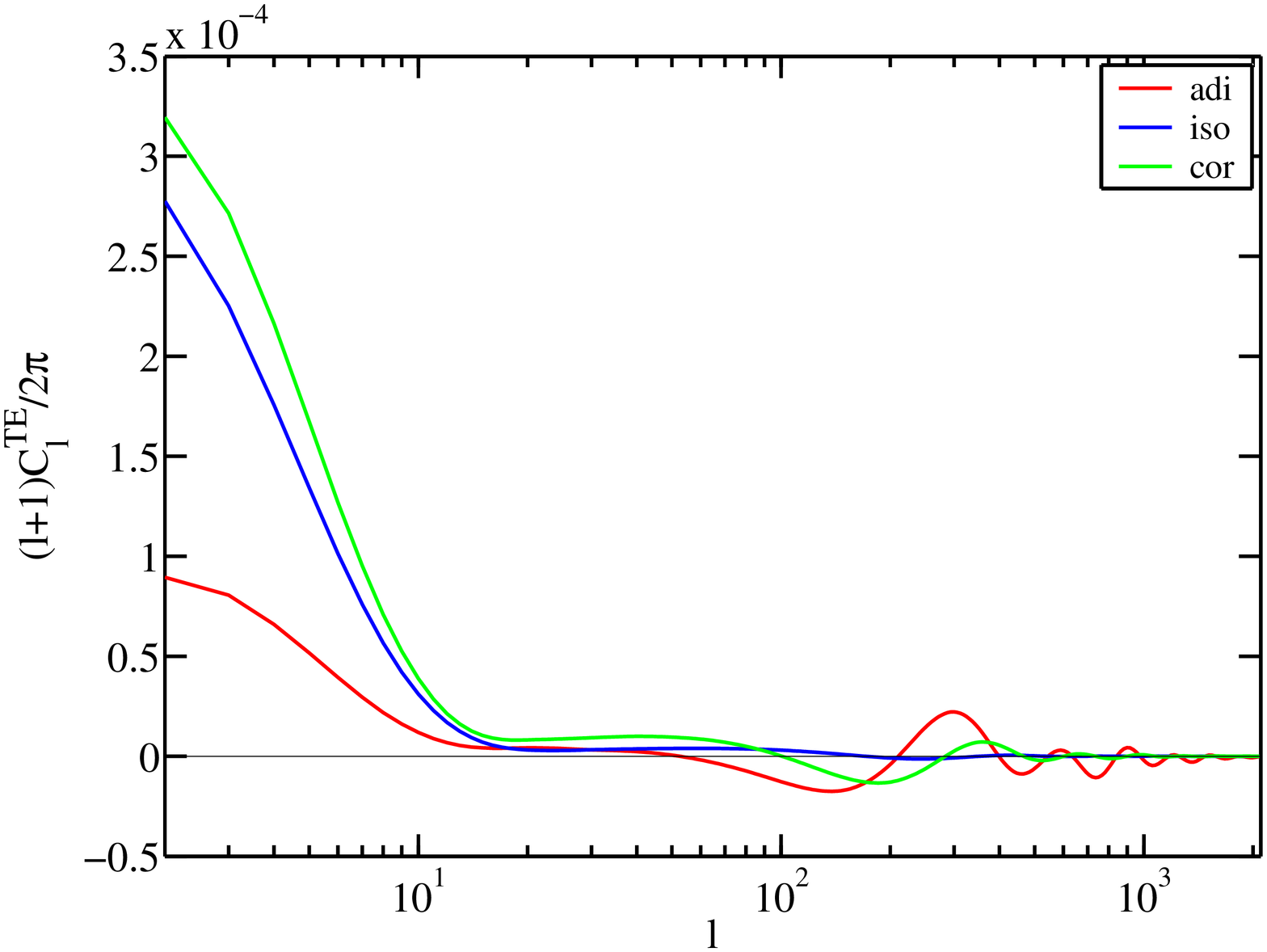}
  \includegraphics[angle=270,width=0.40\textwidth]{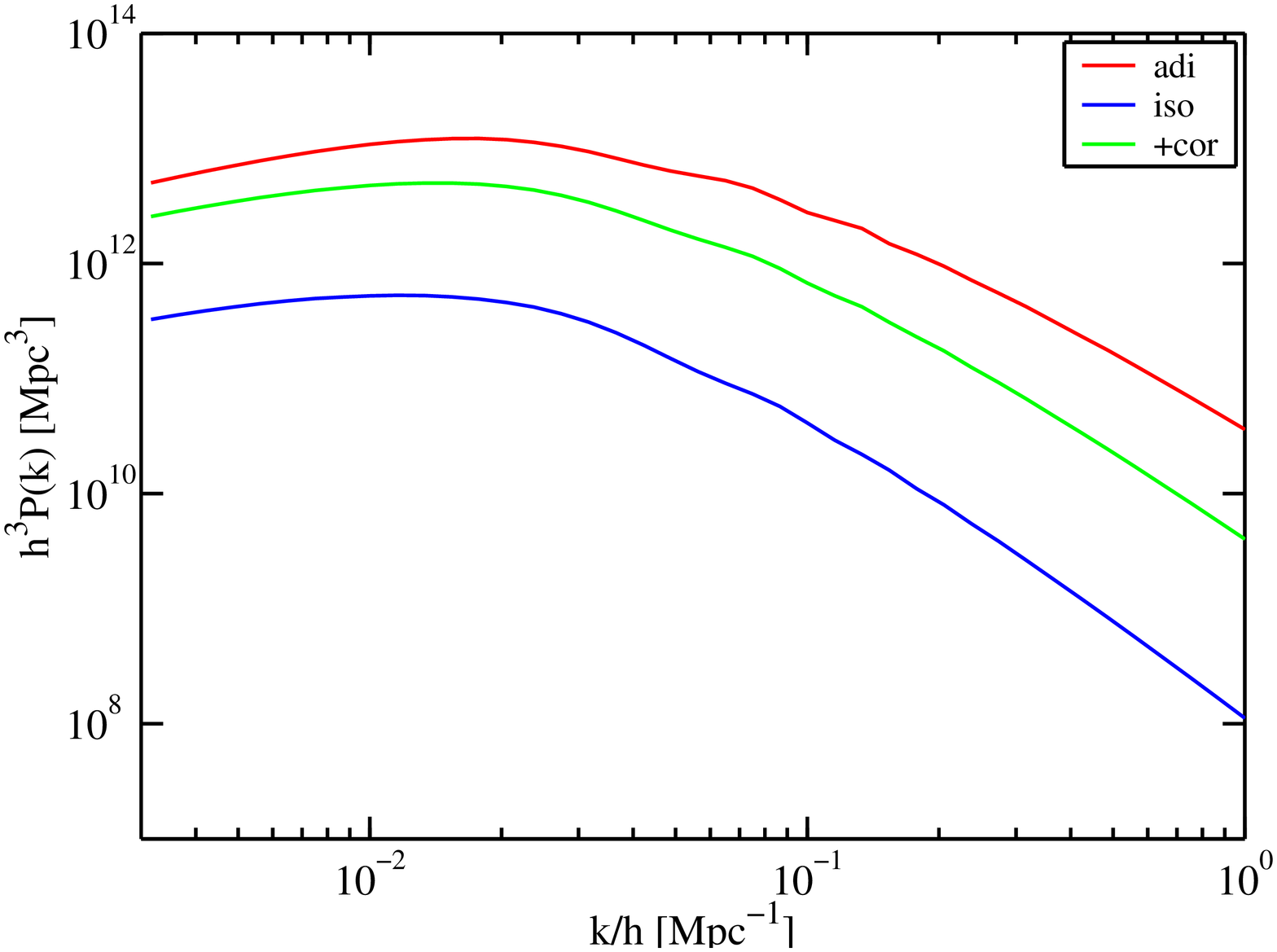}
  \includegraphics[angle=270,width=0.40\textwidth]{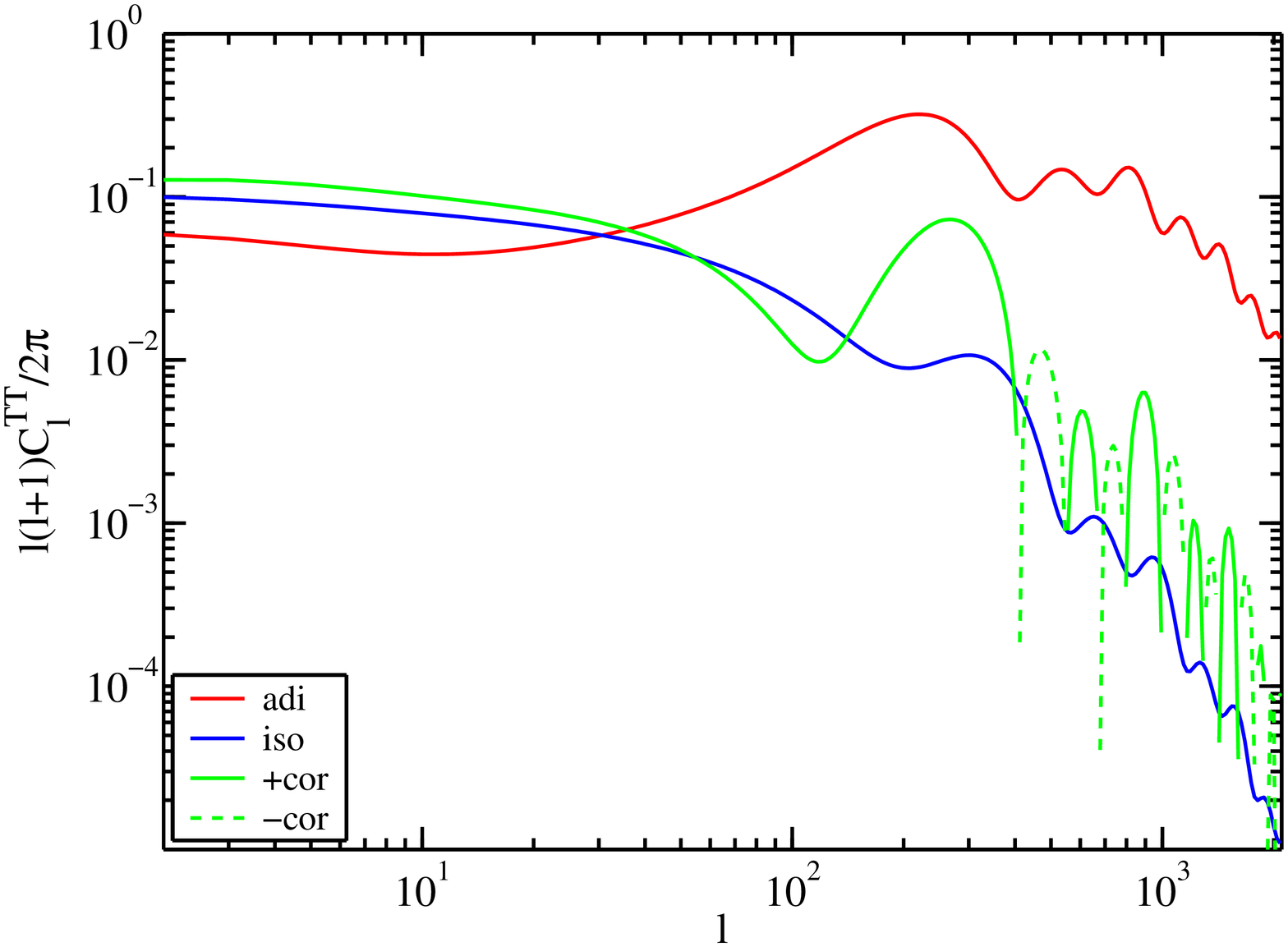}
  \caption{The same as Fig.~(\ref{fig:cltt}), but for (a) $\hat{C}_l^{\mr{TE}}$ and (b)
    the matter power spectrum $\hat{P}(k)$.  We also show (c) the
    $\hat{C}_l^{\mr{TT}}$ of Fig.~\ref{fig:cltt} with a logarithmic scale, so
    that the effect of changing the spectral indices can be readily estimated
    from the figure.  The pivot scale $k_0 = 0.01 \mbox{Mpc}^{-1}$ becomes $k_0/h
    = 0.01418$Mpc$^{-1}$ for the parameter values used ($h = 0.7053$)
    for this plot.}
  \label{fig:clothers}
\end{figure}

Here $\hat{C}^{\mr{ad1}}_{l}$ and $\hat{C}^{\mr{ad2}}_{l}$ represent
adiabatic spectra which would result from a curvature perturbation
$\mc{R}_{\mr{rad}}$ with unit amplitude ($A_r = 1$ or $A_s = 1$) at the pivot
scale $k_0$. (They are otherwise the same, but have spectral indices
$\nadiI$ and $\nadiII$.) Likewise, $\hat{C}^{\mr{iso}}_{l}$
represents an isocurvature spectrum from a CDM entropy perturbation of unit
amplitude ($B=1$), and $\hat{C}^{\mr{cor}}_{l}$ the extra contribution from
correlation for $A_sB = 1$. (See Figs.~\ref{fig:cltt} and \ref{fig:clothers},
which represent the case of scale-invariant perturbations.).  The ``hatless''
$C^{\mr{ad1}}_{l}$, $C^{\mr{ad2}}_{l}$, $C^{\mr{iso}}_{l}$, which are
necessarily non-negative, and $C^{\mr{cor}}_{l}$, which can also be negative,
are the contributions to the total $C_l$.
A relation similar to (\ref{eq:totCl}) holds for the matter power spectrum $P(k)$.

Note that, e.g., $\alpha = 0.5$ does not mean that the adiabatic and
isocurvature contributions would be equal at any particular scale.  Since
$\alpha$ refers to the ratio of primordial perturbations, to which the $C_l$
contributions are related through the transfer functions, the situation is
different for different scales, and depends on the other cosmological
parameters. In particular, if the spectral indices are very different, a very
small isocurvature fraction can still correspond to a large isocurvature
contribution at some scales and vice versa.

We define a shorthand notation
\begin{equation}
  \alpha_{\mr{cor}} \equiv \mr{sign}(\gamma)\sqrt{\alpha(1-\alpha)\abs{\gamma}}
\end{equation}
for the relative ``weight'' of the correlation spectrum $C^{\mr{cor}}_{l}$.

The problem remains that when some multiplier in (\ref{eq:totCl}) is close
to zero, the spectral index of the corresponding component becomes
unconstrained leading to more volume in parameter space upon marginalization.
This may introduce a bias towards ``pure'' models where the isocurvature or
correlation amplitude is zero.

We want the pivot scale to be roughly in the middle of the data set used, and
have chosen $k_0 = 0.01 \mbox{Mpc}^{-1}$ as our pivot wavenumber.  For the
concordance values of the cosmological parameters, $\Omega_\Lambda = h = 0.7$,
this corresponds to ``pivot multipole'' $l_0 \sim 140$.
[The correspondence is $l_0 \sim D_\ast k_0$, where $D_\ast=D_\ast(h,\Omega_\Lambda,\Omega_m)$
is the angular diameter distance to last scattering.
$D_\ast(h,0.7,0.3) \approx h^{-1}$10~000Mpc$^{-1}$ while the
``old day's standard value'' was
$D_\ast(h,0,1) \approx h^{-1}$6~000Mpc$^{-1}$.]

This work is similar to a recently published study by Beltran et
al.~\cite{Beltran:2004uv}.  The main differences are:
 1) Different
parametrization of correlation. When we divide the adiabatic spectrum in a
correlated and an uncorrelated part, they consider the total adiabatic spectrum
$\mc{P}_\mc{R}$ and the correlation spectrum $\mc{C}_\mc{RS}$ as the basic
entities, which they approximate by power laws.  This leads to constraints on
the correlation spectral index $\ncor$, which depend on the correlation
amplitude, and therefore they introduce a related parameter,
``$\delta_\mr{cor}$'', to be the independent parameter, leaving $\ncor$ as a
derived parameter.
 2) They have set an upper limit $\niso \leq 3$, whereas we allow $\niso$ to
vary over a wider range.
 3) They use a pivot scale
$k_0 = 0.05\mbox{ Mpc}^{-1}$ ($l_0 \sim 700$). We use $k_0 = 0.01\mbox{
Mpc}^{-1}$ ($l_0 \sim 140$),
but consider also the effect of changing the pivot scale.
 4) They use a larger data set, including 
    type Ia Supernova (SNIa) data
    \cite{Riess:2004nr}, whereas we use CMB and
    LSS data only.
%%%%%%%%%%%%%%%%%%%%%%%%%%%%%%%
%
%
%%%%%%%%%%%%%%%%%%%%%%%%%%%%%%%%
 5) They include an equation-of-state parameter $w$ for dark energy, wheras we
keep $w = -1$.
 6) They consider neutrino isocurvature modes also.

 Crotty \etal \cite{Crotty:2003rz} and Beltran \etal
\cite{Beltran:2004uv} use the same isocurvature parameter $\alpha$ as we
use, but they use the correlation parameter
\begin{equation}
    \beta \equiv -\cos\Delta \equiv\! -\sign(A_sB)\sqrt{\frac{A_s^2}{A_r^2+A_s^2}}
    \equiv\! -\sign(\gamma)\sqrt{\abs{\gamma}}
\end{equation}
In \cite{Crotty:2003rz} $\beta$ is assumed scale invariant, whereas in
\cite{Beltran:2004uv} it is approximated by a power law with index $\ncor$ so
that our $\ncor \equiv \frac12(\nadiII+\niso)$ corresponds to their
$\ncor+\frac12(\nadi+\niso)$.

\section{Technical Details of the Analysis}
\label{sec:tech}

The model we are studying has 11 parameters.
We have chosen to use the following
independent parameters (primary parameters)
for the likelihood analysis: the baryon
density $\omega_{\mr{b}}$, the CDM density $\omega_{\mr{c}}$, the sound horizon
angle $\theta$, the optical depth due to reionization $\tau$, the bias
parameter $b$, the uncorrelated adiabatic spectral index $\nadiI$, the
correlated adiabatic spectral index $\nadiII$, the isocurvature spectral index
$\niso$, the logarithm of the overall amplitude $\ln(10^{10}A^{2})$, the
isocurvature fraction $\alpha$ and the correlated fraction $\gamma$ of the
adiabatic perturbations.
%We use $\ln(10^{10}A^{2})$ instead of $A^{2}$ since now all the parameters are
%all roughly of the same order.  (Talla lienee muutakin merkitysta, se antaa erilaisen
%priorin ja uskoisin etta flat prior logaritmille on toivottavampi. HKS)

The sound horizon angle (in units of $\frac{1}{100}$ radian)
 \beq
    \theta = \theta(\omega_b,\omega_c,h) \equiv 100 \times \frac{s_\ast}{D_\ast} \,,
 \eeq
where $s_\ast$ is the sound horizon at last scattering and $D_\ast$ is the angular
diameter distance to last scattering \cite{Hu:2000ti}, is used as an independent
parameter instead of $h$ (or $\Omega_\Lambda$), since it is more tightly
constrained by the data.

The bias $b$ is defined by
\begin{equation}
\left. \phantom{\Big(}P_\mr{gal}^\mr{SDSS}(k)\right|_{z_\mr{eff}\simeq0.15}
= \left.\phantom{\Big)}b^2 P(k)\right|_{z=0}\,.
\end{equation}
So we multiply the present-day theoretical matter power
$P(k)$ by $b^2$ before comparing to the galaxy
power spectrum observed by the Sloan Digital Sky Survey (SDSS)
\cite{Tegmark:2003uf}
at effective redshift $z_\mr{eff}$.
In the figures, we actually plot $b^2 P(k)$.

We find the posterior likelihoods for the primary parameters and a number of
derived parameters using the Markov Chain Monte Carlo (MCMC) method. The chains
are generated using our modified version of the publicly available CosmoMC code
\cite{Lewis:2002ah}. The CMB angular power spectra and the matter power spectra
are calculated by the CAMB code
\cite{Lewis:1999bs,Lewis:2002nc} (see also \cite{Gordon:2002gv}).
It needed some modifications for faster treatment of correlation.

CosmoMC/CAMB evaluates the matter power spectrum in linear
perturbation theory. However, very small scales
($k/h \gtrsim 0.15$Mpc$^{-1}$) have already become non-linear.
The publicly available code HALOfit utilizes results from lattice
simulations of clustering \cite{Smith:2002dz}. However,
the applicability of the HALOfit to our model is not granted,
since the lattice simulations have been performed
in adiabatic models with moderate spectral indices.
Hence, following the recipe of \cite{Tegmark:2003uf},
we calculate the matter power spectra in linear theory
and compare them only to the first 17 data points 
($k/h \lesssim 0.15$Mpc$^{-1}$) of the
SDSS galaxy survey  \cite{Tegmark:2003uf}.

For the observational CMB data we take
% compare our CMB spectra to are
the WMAP temperature autocorrelation (TT) and
temperature-polarization cross-correlation (TE) data
\cite{Hinshaw:2003ex, Kogut:2003et, Verde:2003ey}. To extend the
coverage of the data to higher multipoles we use 
the TT data from CBI \cite{Readhead:2004gy} and ACBAR \cite{Kuo:2002ua},
which we later call ``other CMB data''.
% For the
%matter power spectra we use the SDSS galaxy survey \cite{Tegmark:2003uf}.

Details of the data sets are:
\begin{itemize}
\item WMAP TT, {\bf 899} data points, $l = 2$ -- $900$,\\ ($k \sim 1.4\times 10^{-4}$Mpc$^{-1}$ -- $0.064$Mpc$^{-1}$).
\item WMAP TE, {\bf 449} data points $l = 2$ -- $450$,\\ ($k \sim 1.4\times 10^{-4}$Mpc$^{-1}$ -- $0.032$Mpc$^{-1}$).
\item ACBAR TT, {\bf 7} $l$-bands, $l_{\rm eff} = 991$ -- $1831$,\\ ($k \sim 0.071$Mpc$^{-1}$ -- $0.131$Mpc$^{-1}$).
\item CBI TT, {\bf 13}  $l$-bands, $l_{\rm eff} = 369$ -- $1884$,\\ ($k \sim 0.026$Mpc$^{-1}$ -- $0.135$Mpc$^{-1}$).
\item SDSS galaxy power, {\bf 17} $k$-bands, $k_{\rm eff}/h = 0.016$Mpc$^{-1}$ -- $0.15$Mpc$^{-1}$.
\end{itemize}
In parenthesis we indicate what wave numbers the given multipole ranges
correspond in models that have $\Omega_\Lambda=h=0.7$, i.e. $D_\ast \sim 14 000$Mpc$^{-1}$.
The total number of data points (1385) leads to the reduced number
of degrees of freedom $\nu = 1385-11=1374$ for our model and
$\nu=1385-7=1378$ for the adiabatic model.

First we did several 8-chain runs to see what happens in 
a MCMC study of our model. Finally, we chose a suitable parametrization,
described above, and performed an 8-chain initialization run with the option
to update the proposal matrix (jump function) turned on in CosmoMC.
We used this run to obtain a good proposal matrix for our full run.
In our full run
%To obtain a reasonable sampling of a parameter space this large 
we ran the code
on an IBM AIX cluster utilizing 32 processors for 12 days
to produce 32 chains that started from separate randomly picked
points in parameter space.
%Hui:2003hn
%. This produced 32
%chains which contained (
After cutting off the burn-in periods
the total number of accepted steps, i.e., different combinations of
our primary parameters, was 266~651. The total number of different models tried
(step trials) was 8~005~143. The option to update the proposal density while
generating the chains was not used in order to produce pure MCMC chains.
In addition to this main run, another set of 8 chains with 60~254 different models with
continuously updated proposal density is used as additional data when
discussing the effect of the pivot scale in Section \ref{sec:pivotscale}.
For a clear review of steps included in MCMC analysis,
especially the meaning of marginalized likelihoods,
see the Appendix of Tegmark \etal \cite{Tegmark:2003ud}.

The parameters were allowed to vary within the following ranges:
\begin{gather}
  \omega_b \in [0.005, 0.1],\quad \omega_c \in [0.01, 0.99],
  \quad \theta \in [0.3, 10.0], \nonumber \\
  \nadiI \in [-3, 4],\quad \nadiII \in [-3, 4],
  \quad \niso \in [-3, 12],\nonumber \\
  \tau \in [0.01, 0.3],\quad \ln(10^{10}A^{2}) \in [1, 7],
  \quad b \in [0.1, 2.5], \nonumber \\
  \gamma \in [-1, 1] \quad \text{and} \quad  \alpha \in [0, 1] \nonumber
\end{gather}

The MCMC method implicitly assigns flat priors for these independent parameters.
The ranges for $\alpha$ and $\gamma$ follow from
their definitions. For the other parameters, except $\tau$, we have set very
wide ranges, so that the likelihood is negligible at the boundaries. However,
we also imposed a top-hat prior for the Hubble constant: $0.4 \leq h \leq 1.0$,
which cuts off some models that would otherwise be acceptable (at 95 \% C.L.).

We have constrained $\tau$ to be less
than 0.3. We found in our preliminary studies that there are models with $\tau
> 0.3$ that fit well to the data. These models form a separate region in the
parameter space, and have also a high baryon density, of the order of
$\omega_{\mr{b}} \sim 0.03$. This high baryon density is much above the values
obtained from big bang nucleosynthesis (BBN) calculations \cite{Burles:2000zk}
and we decided not to consider such models in this paper. Including the $\tau
> 0.3$ region would be problematic with the MCMC method as it is not well suited for
such bimodal distributions. Moreover, $\tau > 0.3$ leads to a very high
reionization redshift, which is not favored by
astrophysical considerations \cite{Hui:2003hn}.

To cover our parameter space as well as possible, within the limits of
available computational resources, the starting point for each of the 32 chains
was randomly selected from the following Gaussian distributions:
\begin{align}
  \omega_b & = 0.0236\pm0.005,& \omega_c & = 0.124\pm0.035,
  \nonumber \\
  \theta & = 1.047\pm0.038,& \tau & = 0.11\pm0.229, \nonumber \\
  \nadiI & = 0.97\pm0.27,& \nadiII & = 0.97\pm1.60, \nonumber \\
  \niso & = 2.09\pm3.70,&   b & = 0.99\pm0.34,
  \nonumber \\
  \gamma & = 0.01\pm1.3,& \alpha & = 0.035\pm0.24, \nonumber \\
  \ln(10^{10}A^{2}) & = 3.20\pm0.4\,. & &  \nonumber
\end{align}
The width for a given parameter is
four times the width of the posterior distribution of the same parameter from
our preliminary runs.

\section{Results}
\label{sec:results}

In Fig.~\ref{fig:adiparam} we show the marginalized (``1-d") likelihoods for
those 7 of our independent parameters, which correspond to the 7 parameters of
the adiabatic model. In Fig.~\ref{fig:adiparamd} we show likelihoods for some
derived parameters related to them.

In Fig.~\ref{fig:isoparam} we show the marginalized likelihoods for our
remaining 4 independent parameters, and in Fig.~\ref{fig:isoparamd} for some
related derived parameters.

Flat priors for our independent parameters lead to non-flat priors for the
derived parameters, which contribute to some features in the distributions of
the latter.

The best-fit
%\footnote{
% We have excluded an even better-fit model with $\chi^2 = 1459.20$ for reasons
%explained in Sect.~\ref{sec:largeniso}.
% }
(11-parameter) model has $\chi^2 = 1459.29$, just slightly better than the
best-fit (7-parameter) adiabatic model $\chi^2 = 1459.65$. Thus there is
clearly no indication in the data for the presence of an isocurvature
contribution. Our results should be considered in terms of upper limits to
isocurvature perturbations and uncertainties in the determination of
cosmological parameters due to the possibility of an isocurvature
contribution.

We first discuss the effect of allowing a (possibly correlated) isocurvature
contribution, on the determination of the standard cosmological parameters.
% Those seven of our independent parameters, which exist also in the
%adiabatic model, namely
The likelihoods of
$\omega_b$, $\omega_c$, $\theta$, $\nadiI$,
$\ln(10^{10} A^2)$, $\tau$, and $b$, are compared
with the corresponding likelihoods of the adiabatic
model in Fig.~\ref{fig:adiparam}.

The amplitude $A$ has now a different meaning than in the adiabatic model, as
it includes the isocurvature contribution also.  Since the isocurvature
transfer functions lead to less power in most of the data from a given
primordial amplitude than the adiabatic transfer functions (see
Figs.~\ref{fig:cltt} and \ref{fig:clothers}), larger total amplitudes $A$ are
allowed for models with a significant isocurvature contribution.
%The
%possibility of a negative contribution to the data from correlation also acts
%in this direction.  {\em Itse asiassa tama jalkimmainen ilmio ei nayta
%  tarkealta, $\gamma$-prior ei juurikaan vaikuttanut $A$:n leveyteen. Jatetaan
%  edellinen lause pois.}

The distribution for the adiabatic spectral index $\nadiI$ has become
much wider.  The reason for this is that the correlated adiabatic component
(``ad2'') may take the role of the adiabatic perturbation of the adiabatic
model:
If $|\gamma| \sim 1$, but $\alpha$ is small, the model looks like the
adiabatic model; that the adiabatic mode is correlated with the isocurvature
mode does not have much significance, if the isocurvature component itself is
negligible. In this case $\nadiII$ is then constrained to be close to the
spectral index value of the adiabatic model, but $\nadiI$ becomes
unconstrained, as this contribution has negligible amplitude.  We discuss the
question of the adiabatic spectral index further in Sec.~\ref{sec:adiindex}

The uncertainties in the determination of $\omega_b$, $\tau$, and $b$ are
increased somewhat.  We discuss $\omega_c$ and $\theta$ in
Sec.~\ref{sec:largelambda}, and $\omega_b$ in Sec.~\ref{sec:obH}. We
devote Secs.~\ref{sec:isoparam} and \ref{sec:corparam} to the isocurvature
and correlation parameters, respectively. In Sec.~\ref{sec:largeniso}
we discuss a warning example of a model with very large $\niso$
that must be rejected for several reasons.

 %%%%%%%%%%%%%%%%%%%%%%%%%%%%%%%%%%%%%%%%%%%%%%%%%%%%%%%%%%%%%%%%%%%%%
\subsection{Adiabatic spectral index}
 \label{sec:adiindex}

We can define an effective single adiabatic spectral
index by
\begin{multline*}
  \nadi^{\mr{eff}}(\bar k) - 1 =
  \frac{d\ln{\cal P}_{\cal R}(\bar k)}{d\ln\bar k}\\
  = \frac{(\nadiI - 1) (1 - |\gamma|) \bar k^{\nadiI-1} +
    (\nadiII - 1) |\gamma| \bar k^{\nadiII-1}}{(1 - |\gamma|)\bar
    k^{\nadiI-1} + |\gamma| \bar k^{\nadiII-1}}\,,
%  \label{eqn:nadieffk}
\end{multline*}
which is scale dependent. The first derivative
\begin{equation*}
  \frac{d n_{\mr{adi}}^{\mr{eff}}(\bar k)}{d\ln\bar k}
  =
  \frac{(1-|\gamma|)|\gamma|(\nadiI - \nadiII)^2
       \bar k^{\nadiI+ \nadiII}}{[(1-|\gamma|)
       \bar k^{\nadiI} + |\gamma|\bar
    k^{\nadiII}]^2}
%  \label{eqn:nadiderk}
\end{equation*}
is zero only when $\nadiI = \nadiII$ or $\gamma = 0, \pm1$.
Otherwise it is positive \cite{Valiviita:2003ty}.

\begin{figure}[th]
  \centering
  \includegraphics[angle=270,width=0.45\textwidth]{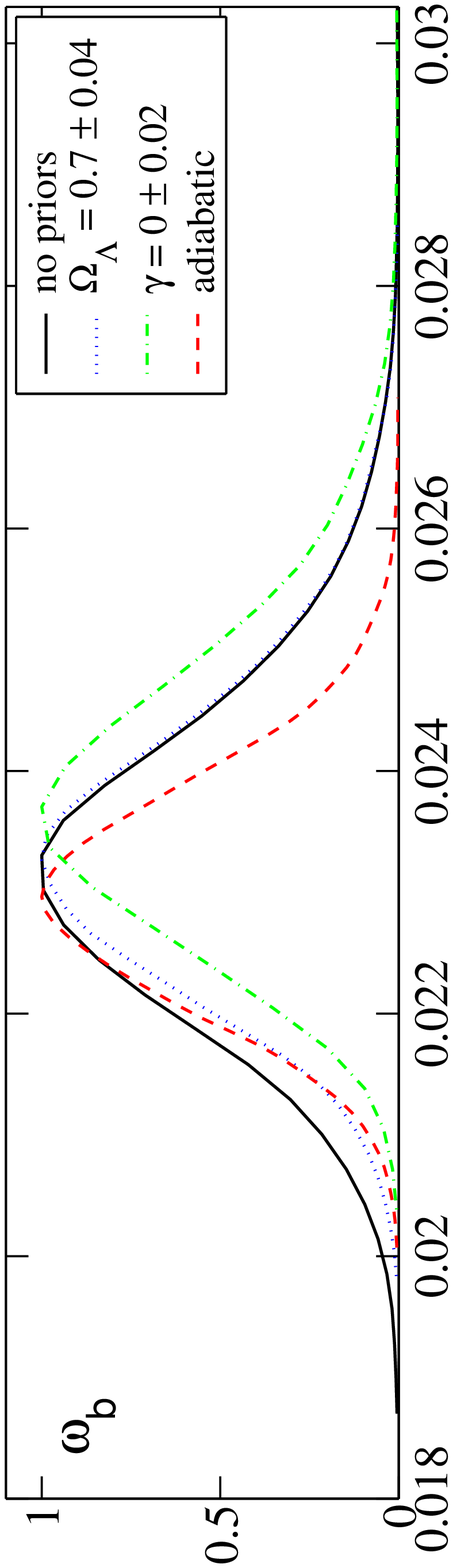}
  \includegraphics[angle=270,width=0.45\textwidth]{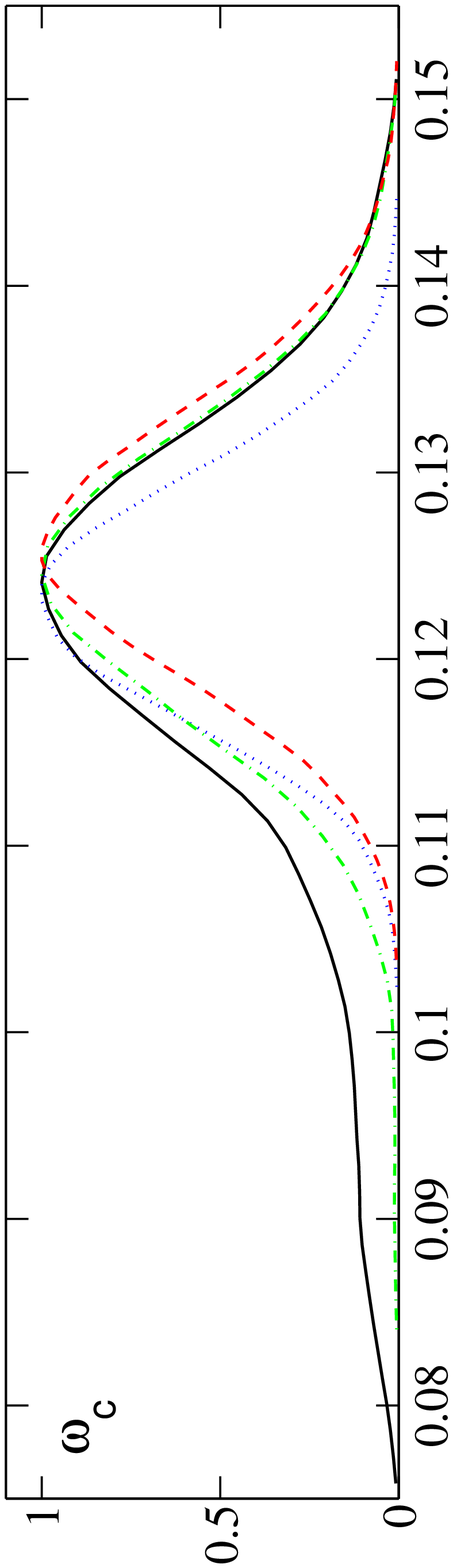}
  \includegraphics[angle=270,width=0.45\textwidth]{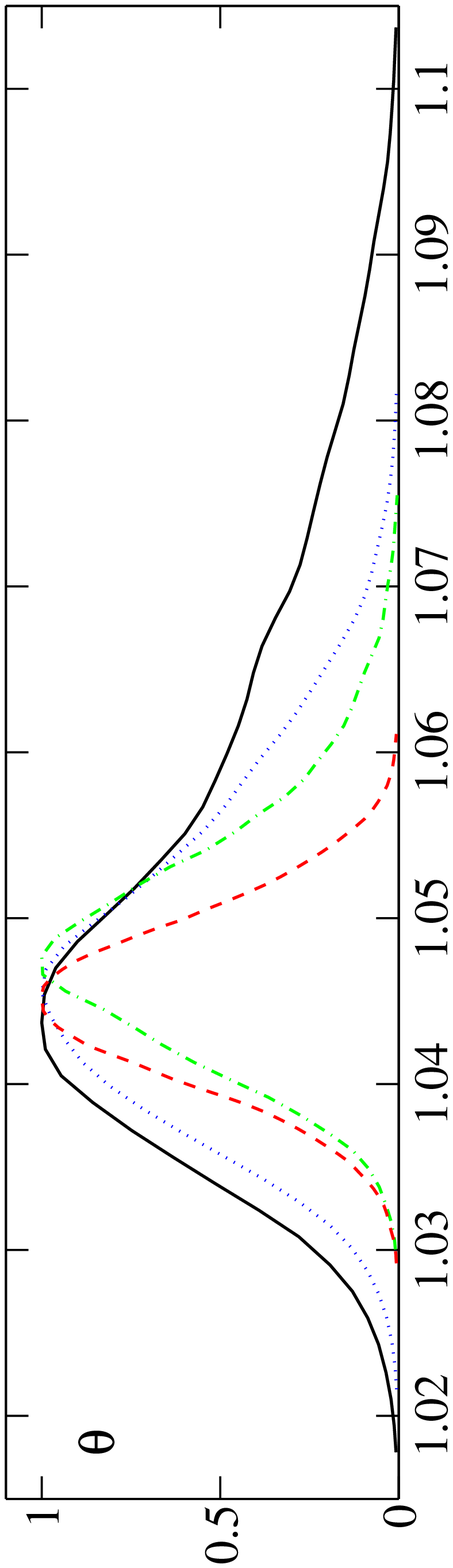}
  \includegraphics[angle=270,width=0.45\textwidth]{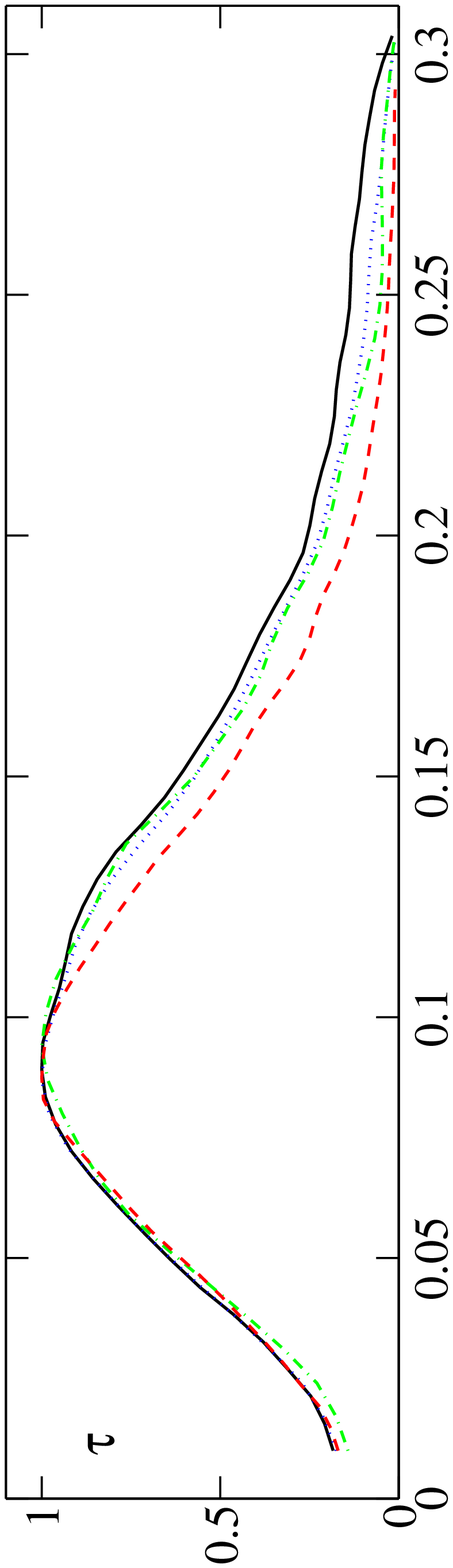}
  \includegraphics[angle=270,width=0.45\textwidth]{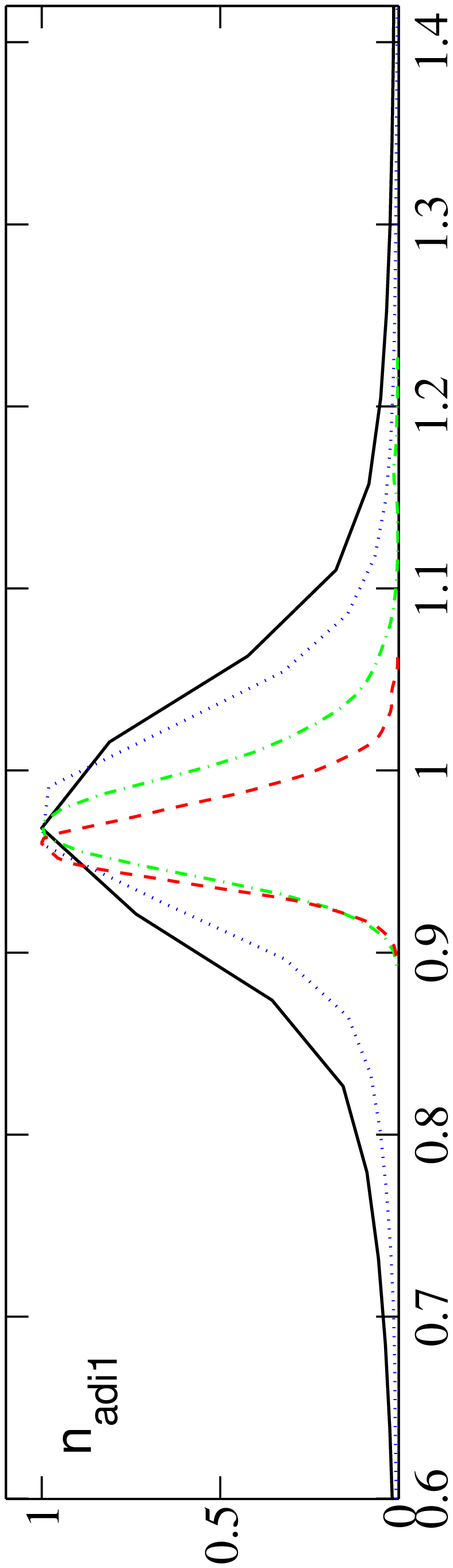}
  \includegraphics[angle=270,width=0.45\textwidth]{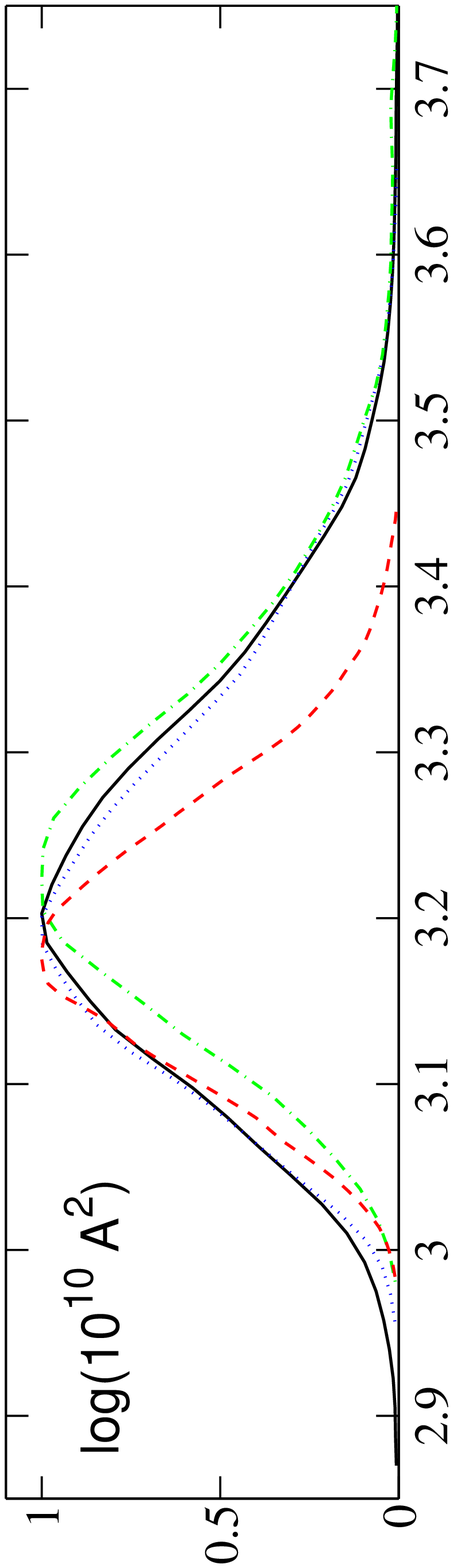}
  \includegraphics[angle=270,width=0.45\textwidth]{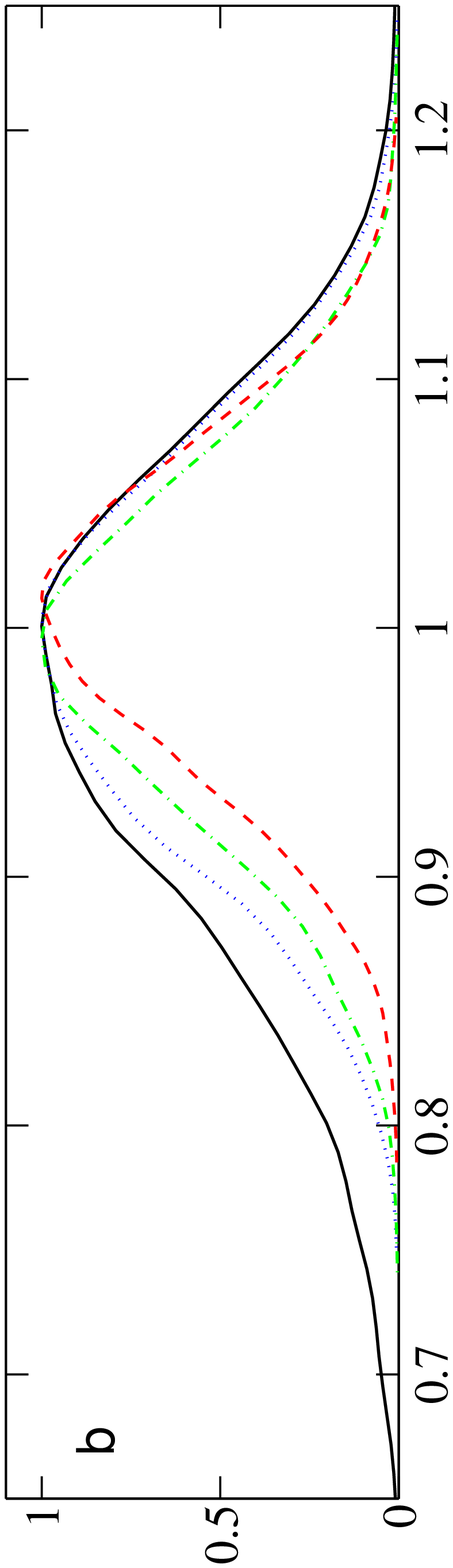}
  \caption{Marginalized likelihood functions for the standard cosmological
    parameters (i.e. those that exist in the adiabatic model).
    %We show all 7 independent parameters.
    The {\em solid} (black) line is the likelihood in
    our 11-parameter model, the {\em dashed} (red) line is for the adiabatic model.
    Other line types show the effects
    of additional priors discussed in the text: \emph{dotted} (blue) for
    Gaussian $\Omega_\Lambda = 0.70\pm0.04$, and \emph{dot-dashed} (green) for
    Gaussian $\gamma = 0.0\pm0.02$.}
  \label{fig:adiparam}
\end{figure}
\begin{figure}[h]
  \centering
  \includegraphics[angle=270,width=0.45\textwidth]{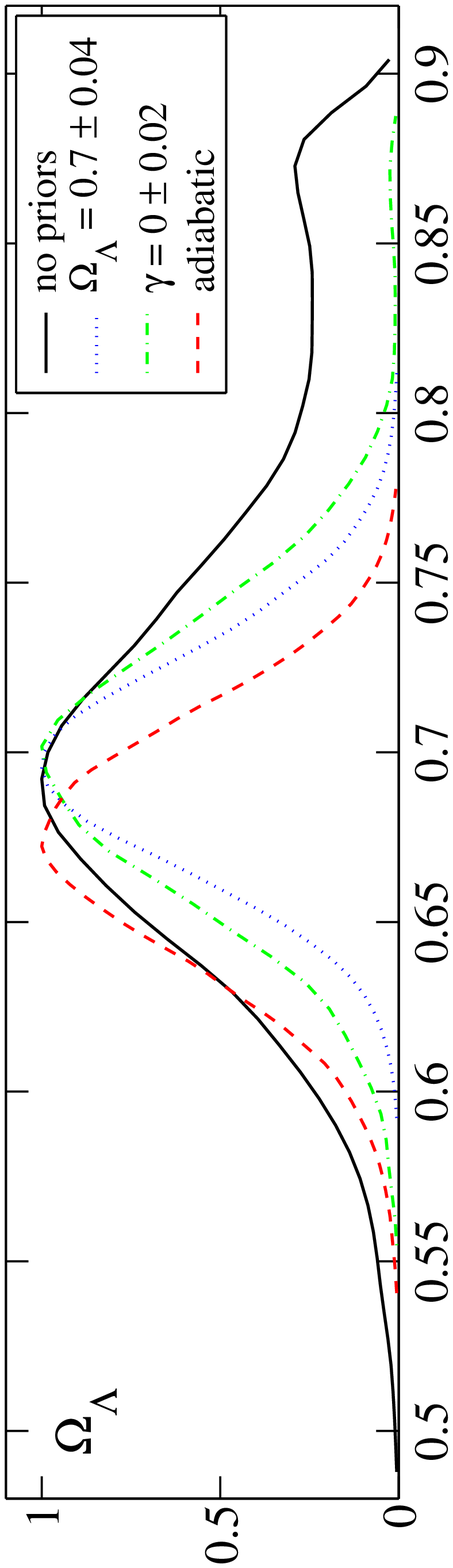}
  \includegraphics[angle=270,width=0.45\textwidth]{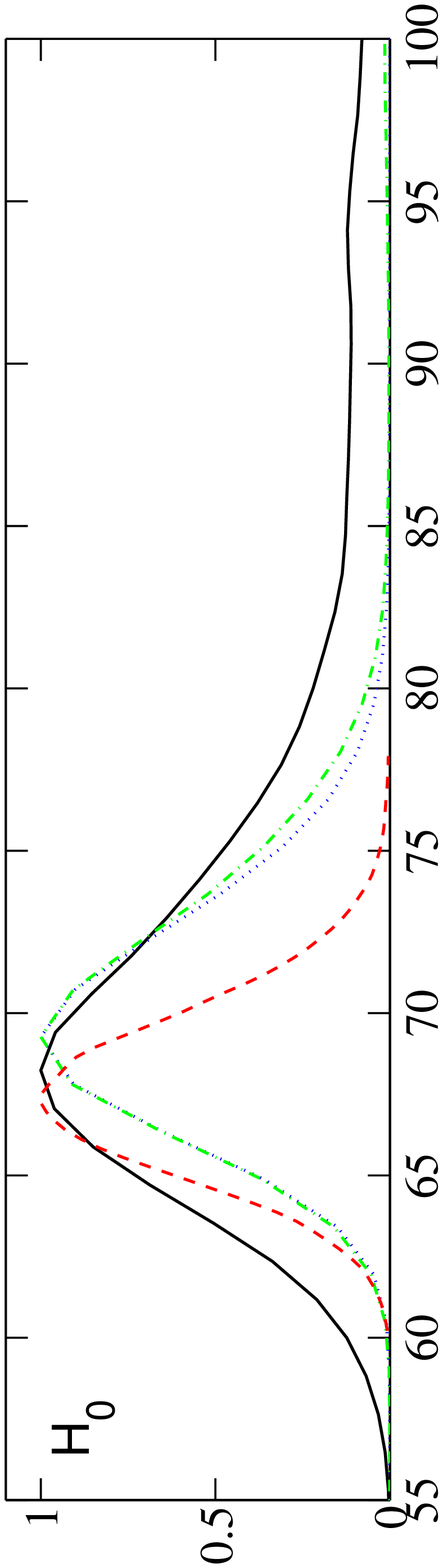}
  \caption{Marginalized likelihoods for two derived parameters,
    $\Omega_\Lambda$ and $H_0$.  The line styles
    have the same meaning as in Fig.~\ref{fig:adiparam}.}
  \label{fig:adiparamd}
\end{figure}
\begin{figure}[h]
  \centering
  \includegraphics[angle=270,width=0.45\textwidth]{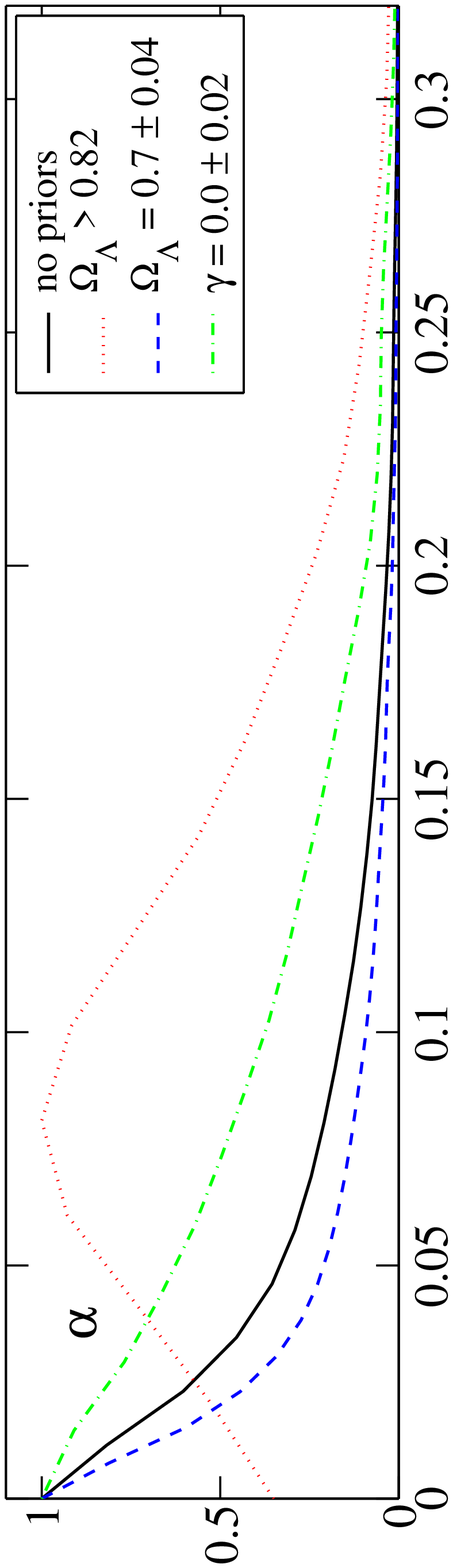}
  \includegraphics[angle=270,width=0.45\textwidth]{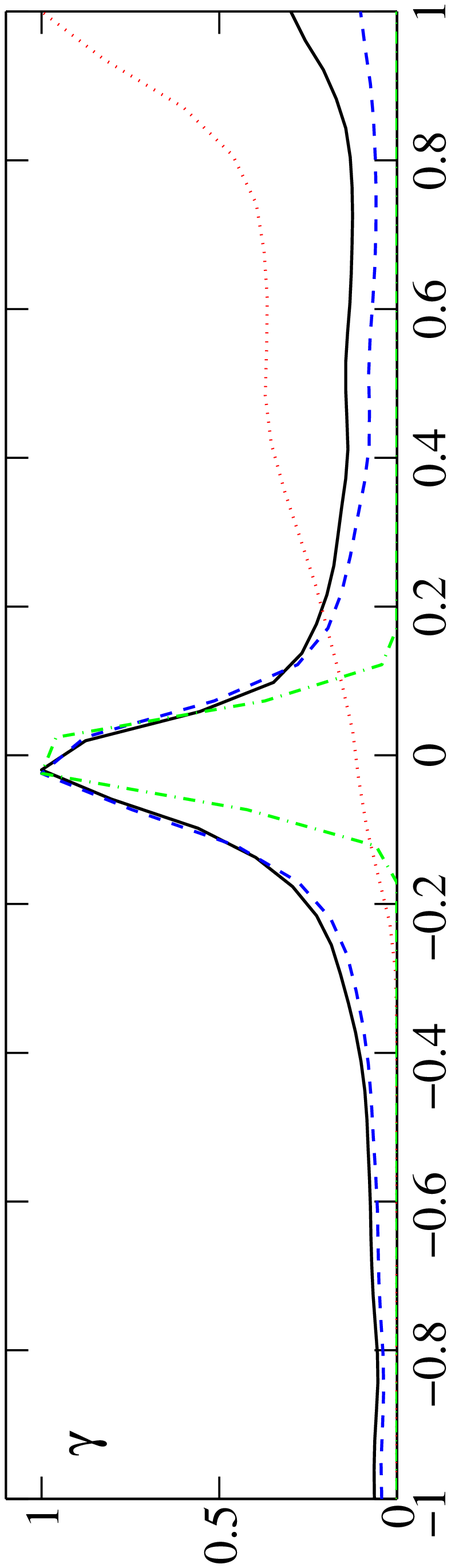}
  \includegraphics[angle=270,width=0.45\textwidth]{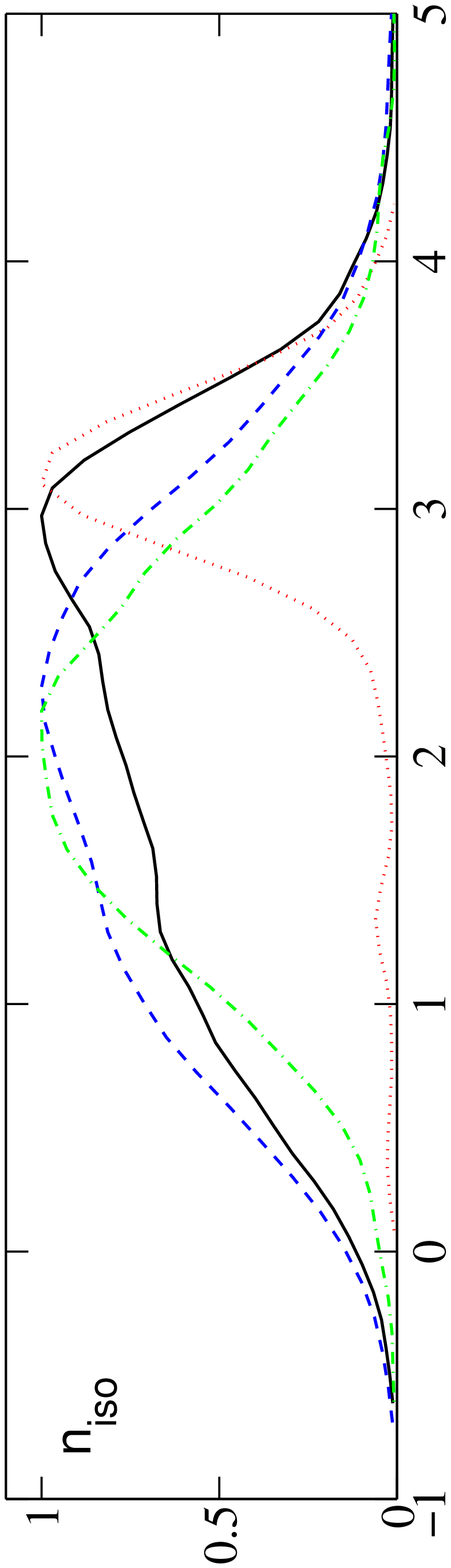}
    \includegraphics[angle=270,width=0.45\textwidth]{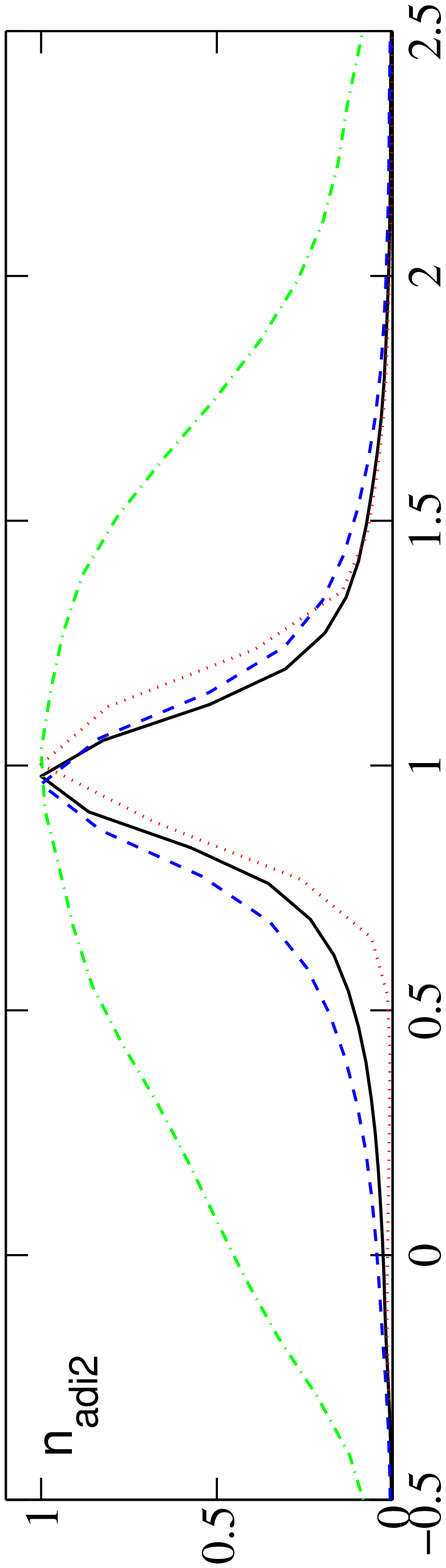}
  \caption{Marginalized likelihoods for the parameters related to the
    isocurvature mode and correlation.  We show the $4$ remaining independent
    parameters, $\alpha$, $\gamma$, $\niso$, $\nadiII$. The
   {\em solid} (black) line is the full likelihood,
    other line types show the effects
    of additional priors discussed in the text:
    {\em dotted} (red) for Gaussian $\Omega_\Lambda > 0.82$,
    {\em dashed} (blue) for  Gaussian $\Omega_\Lambda = 0.70\pm0.04$, and
    {\em dot-dashed} (green) for Gaussian $\gamma = 0.0\pm0.02$.}
  \label{fig:isoparam}
\end{figure}
\begin{figure}[h]
  \centering
  \includegraphics[angle=270,width=0.45\textwidth]{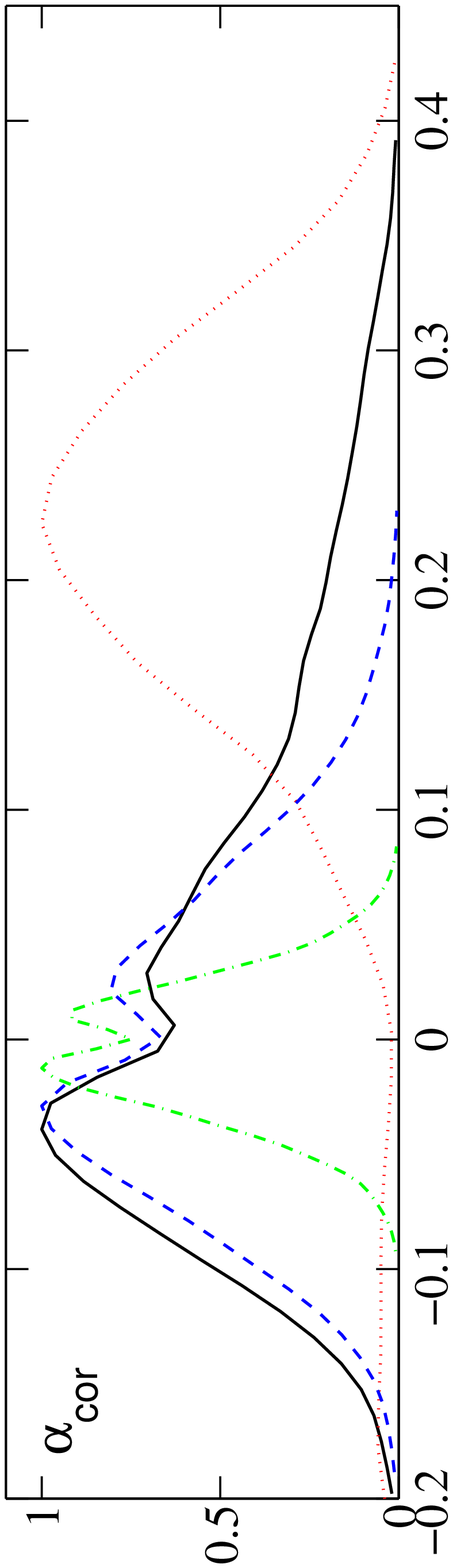}
  \includegraphics[angle=270,width=0.45\textwidth]{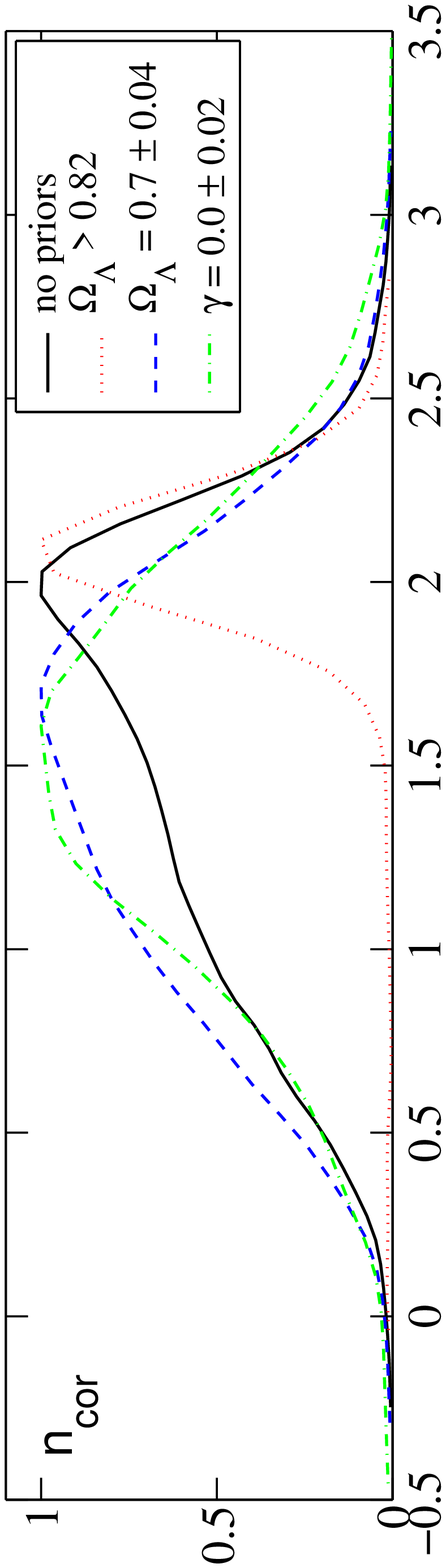}
  \caption{Marginalized likelihoods for the isocurvature-related
    derived parameters, $\alpha_{\mr{cor}}$ and $\ncor$. The line styles
    have the same meaning as in Fig.~\ref{fig:isoparam}.}
  \label{fig:isoparamd}
\end{figure}
\begin{figure}[h]
  \centering
  \includegraphics[angle=270,width=0.45\textwidth]{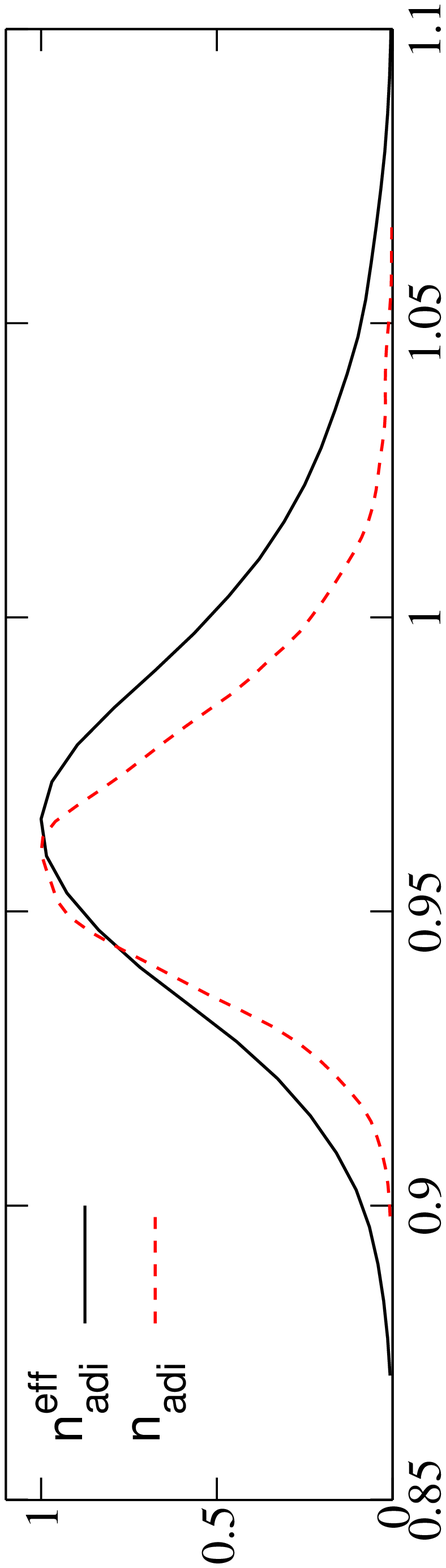}
  \includegraphics[angle=270,width=0.45\textwidth]{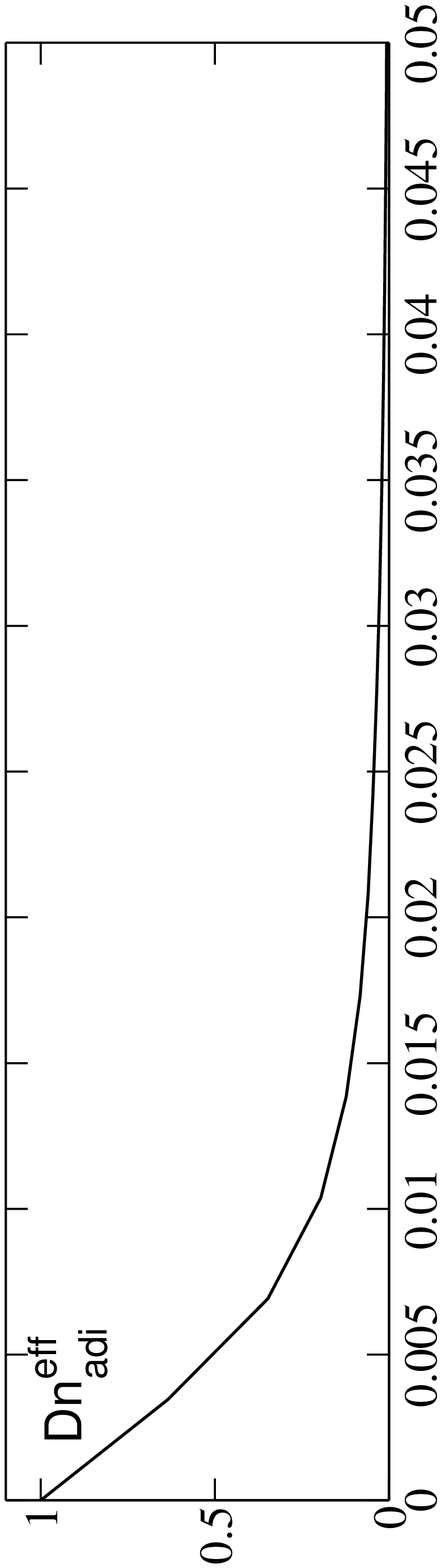}
  \caption{(a) Marginalized likelihoods for the effective adiabatic spectral
    index $\nadi^{\mr{eff}}$ (\emph{solid line}, black) compared to the spectral
    index of the pure adiabatic model $\nadi$ (\emph{dashed line}, red).
    (b)
    Marginalized likelihood for $d \nadi^{\mr{eff}} / d\ln \bar k$ at the pivot scale
    $k_0 = 0.01\mbox{ Mpc}^{-1}$.}
  \label{fig:neff}
\end{figure}

At our pivot scale we have $\bar k = 1$ and the above expressions simplify to
\begin{equation}
  \left. \nadi^{\mr{eff}}\right|_{k=k_0} =
  (\nadiI - 1) (1 - |\gamma|) + (\nadiII - 1) |\gamma| + 1\,,
  \label{eqn:nadieff}
\end{equation}
and
\begin{equation}
  \left. \frac{d \nadi^{\mr{eff}}(\bar k)}{d\ln \bar k}\right|_{k=k_0} =
  \Big(\nadiI - \nadiII\Big)^2 \Big(1-|\gamma|\Big)|\gamma| \,.
  \label{eqn:nadider}
\end{equation}
From Fig.~\ref{fig:adiparam} we observe that $\nadiI$ is much more loosely
constrained than the $\nadi$ of the adiabatic model. The distribution for
$\nadiII$ becomes even wider than the one for $\nadiI$, see
Fig.~\ref{fig:isoparam}. The reason is that the MCMC chains contain many models
with $|\gamma|$ close to zero allowing $\nadiII$ to take any value or
$|\gamma|$ close to 1 allowing $\nadiI$ to take any value. However, the
effective adiabatic spectral index (\ref{eqn:nadieff}) becomes nearly as
tightly constrained as the spectral index in pure adiabatic models. The 95\%
C.L. regions are $0.910 < \nadi^{\mr{eff}} < 1.050$ with median $0.968$ and
$0.923 < \nadi < 1.013$ with median $0.961$, see also Fig.~\ref{fig:neff}(a).
Moreover, the data disfavor (positive) running of the adiabatic spectral index.
For the 95\% C.L. upper limit we obtain $d \nadi^{\mr{eff}}/d\ln\bar k < 0.03$ at
$k_0 = 0.01$Mpc$^{-1}$, see Fig.~\ref{fig:neff}(b). The largest $k$ in the data
sets is about $k_\mr{max} \approx 0.15$Mpc$^{-1}$. So the maximum running from
$k_0$ to $k_{\mr{max}}$ is approximately $\Delta n = 0.03\times \ln
(k_{\mr{max}}/k_0) = 0.08$. The quadrupole ($l=2$) corresponds to $k_{\mr{min}}
\approx 1.4\times 10^{-4}$Mpc$^{-1}$ leading to $\Delta n = 0.03\times \ln
(k_{\mr{min}}/k_0) = -0.12$.

%%%%%%%%%%%%%%%%%%%%%%%%%%%%%%%%%%%%%%%%%%%%%%%%%%%%%%%%%%%%%%%%%%%%

\subsection{Small matter density models}
\label{sec:largelambda}

\begin{figure}[th]
  \centering
  \includegraphics[angle=270,width=0.45\textwidth]{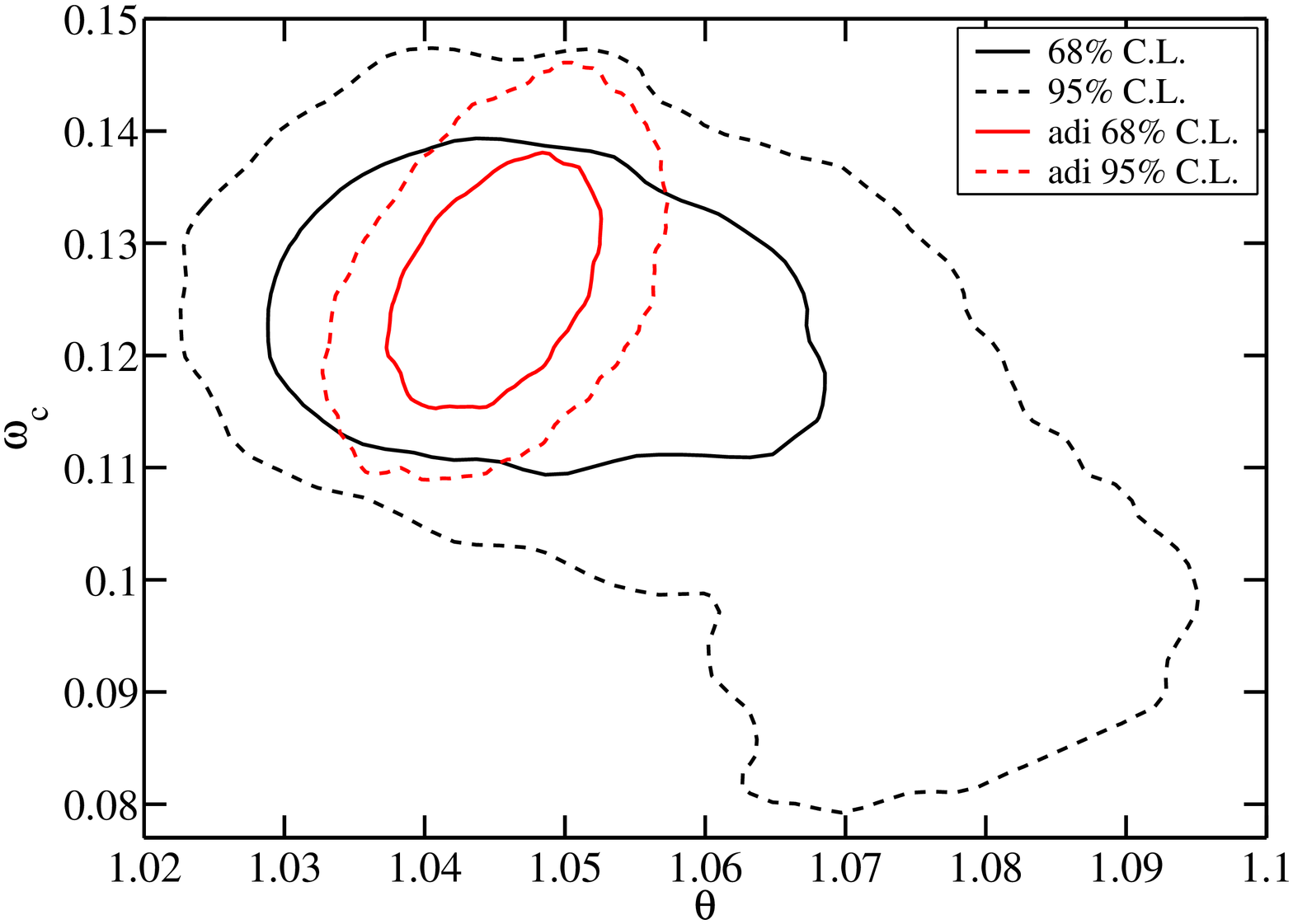}
  \includegraphics[angle=270,width=0.45\textwidth]{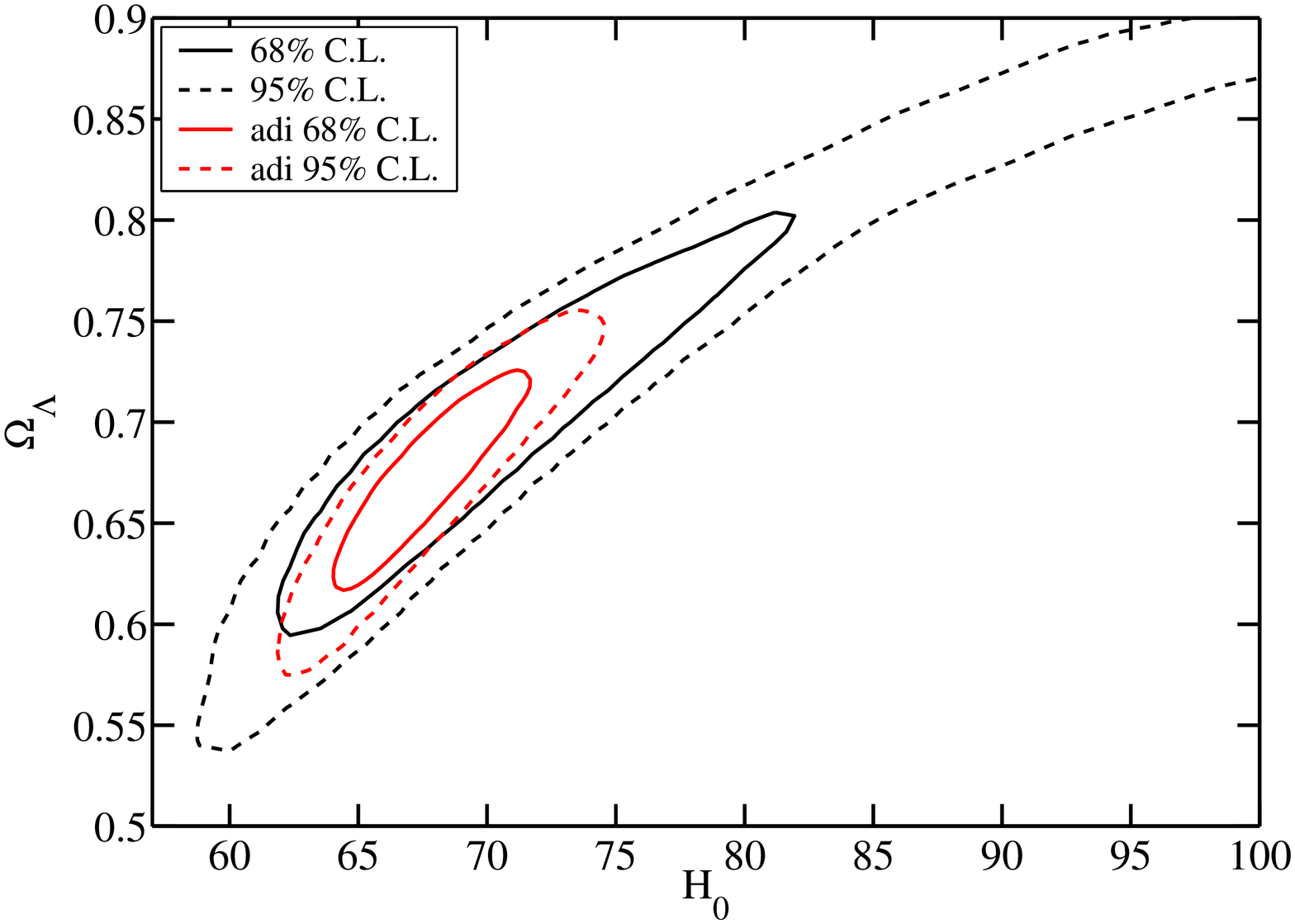}
  \caption{(a) 2-d marginalized likelihood for $\theta$ and $\omega_c$, showing
    that the large sound horizon angles $\theta$ are connected with low CDM
    densities $\omega_c$. We indicate
    the 68\% (\emph{solid}) and 95\% (\emph{dashed}) C.L. regions
    for our isocurvature model (\emph{black})
    and for the adiabatic model (\emph{red}).
    (b) 2-d likelihood for the two derived
    parameters $H_0$ and $\Omega_\Lambda$ closely related to the independent
    parameters $\theta$ and $\omega_c$.}
  \label{fig:bumpintwod}
\end{figure}

In Fig.~\ref{fig:adiparam} the most obvious difference from the adiabatic model
is the extension of the $\theta$ likelihood towards larger sound horizon angles
and the $\omega_c$ likelihood towards smaller densities.  These two features
are related as can be seen in Fig.~\ref{fig:bumpintwod}(a). The corresponding
effect is seen in the two derived parameters, $\Omega_\Lambda$, $H_0$, closely
related to $\theta$ and $\omega_c$, see Fig.~\ref{fig:bumpintwod}(b).
Compare to a simalar Fig. in \cite{Trotta:2002iz}.

The 1-d likelihood for the derived parameter $\Omega_\Lambda$ show
(Fig.~\ref{fig:adiparamd}) a second peak at $\Omega_\Lambda \sim 0.87$.  (This
feature is somewhat enhanced because the flat prior for our independent
parameters actually leads to an increasing prior for the derived parameter
$\Omega_\Lambda$, and larger values of $\Omega_\Lambda$ are cut off with our $h
\leq 1$ constraint.)

\begin{figure}[th]
  \centering
  \includegraphics[angle=270,width=0.45\textwidth]{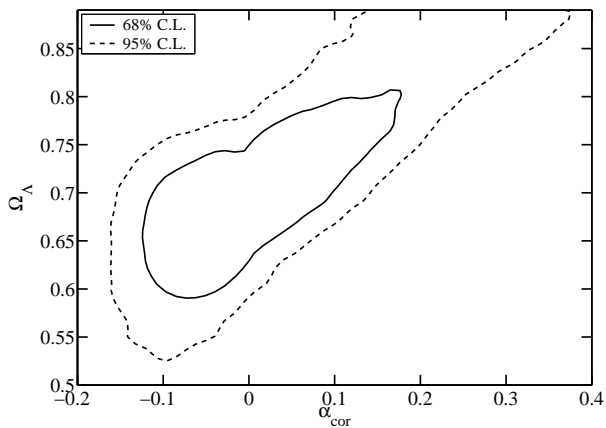}
  \caption{
    2-d marginalized likelihood for $\alpha_{\mr{cor}}$ and $\Omega_\Lambda$,
    showing that the large $\Omega_\Lambda$ are connected with a positively
    correlated isocurvature contribution.}
  \label{fig:corLambda}
\end{figure}

Thus the possibility of an isocurvature contribution leads to larger
$\Omega_\Lambda$ models becoming acceptable by the CMB and LSS data.
According to Fig.~\ref{fig:corLambda} these models have a
positive correlation between the adiabatic and isocurvature modes. Indeed, if
we cut to the subset of ``uncorrelated models'', $\gamma = 0.0\pm 0.02$, the
(large $\theta$, small $\omega_c$) feature disappears from the 1-d likelihoods.

In Figs.~\ref{fig:isoparam} and \ref{fig:isoparamd} we show the 1-d likelihoods
of the isocurvature-related parameters separately for the
large-$\Omega_\Lambda$ subset ({\em dotted red} lines) and with a $\gamma =
0.0\pm0.02$ prior that cuts the more correlated models off ({\em dot-dashed green}
lines).  We see clearly that the large-$\Omega_\Lambda$ models are associated
with a positive correlation between the isocurvature and adiabatic modes.

\begin{figure}[th]
  \centering
  \includegraphics[angle=270,width=0.45\textwidth]{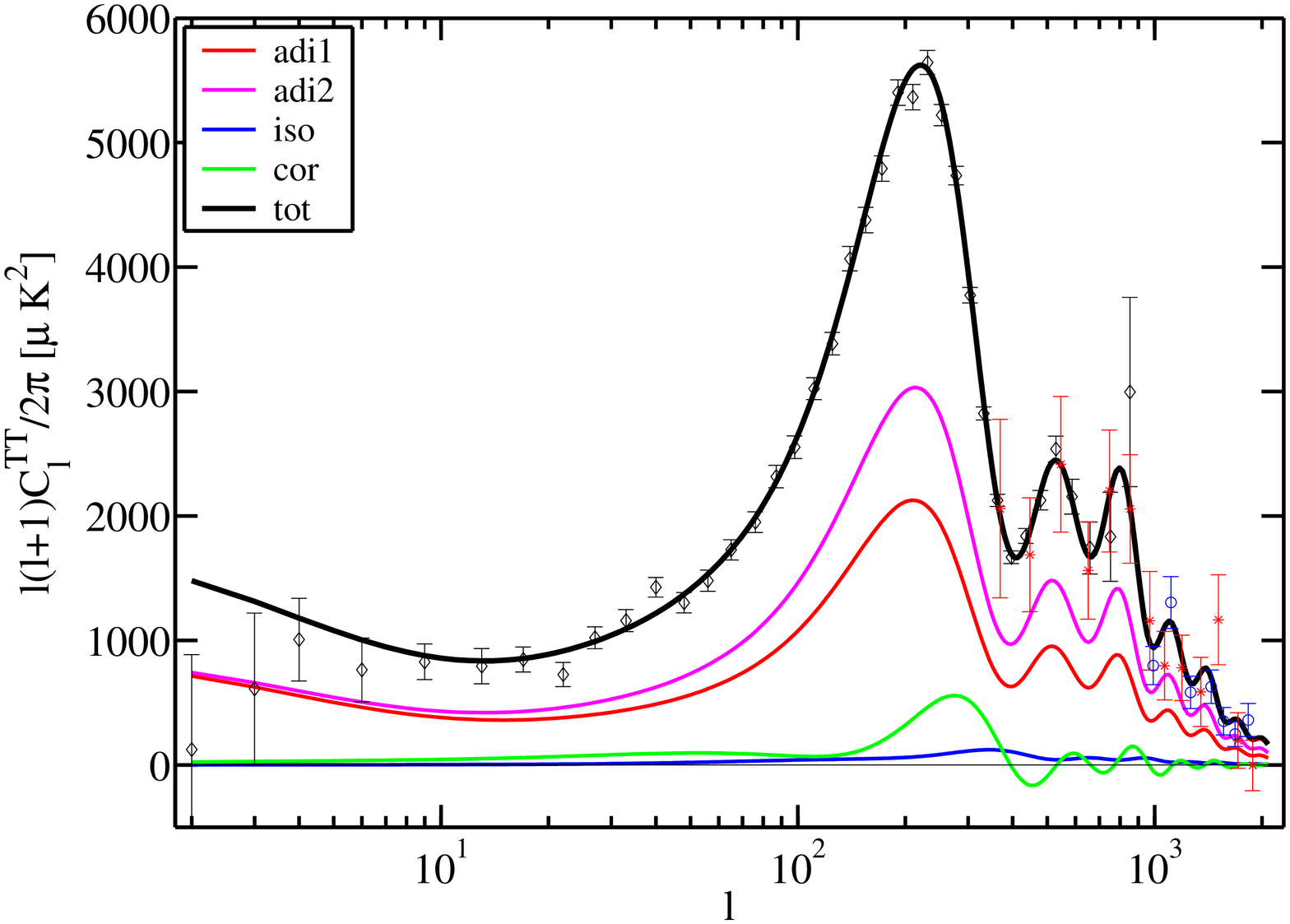}
  \includegraphics[angle=270,width=0.45\textwidth]{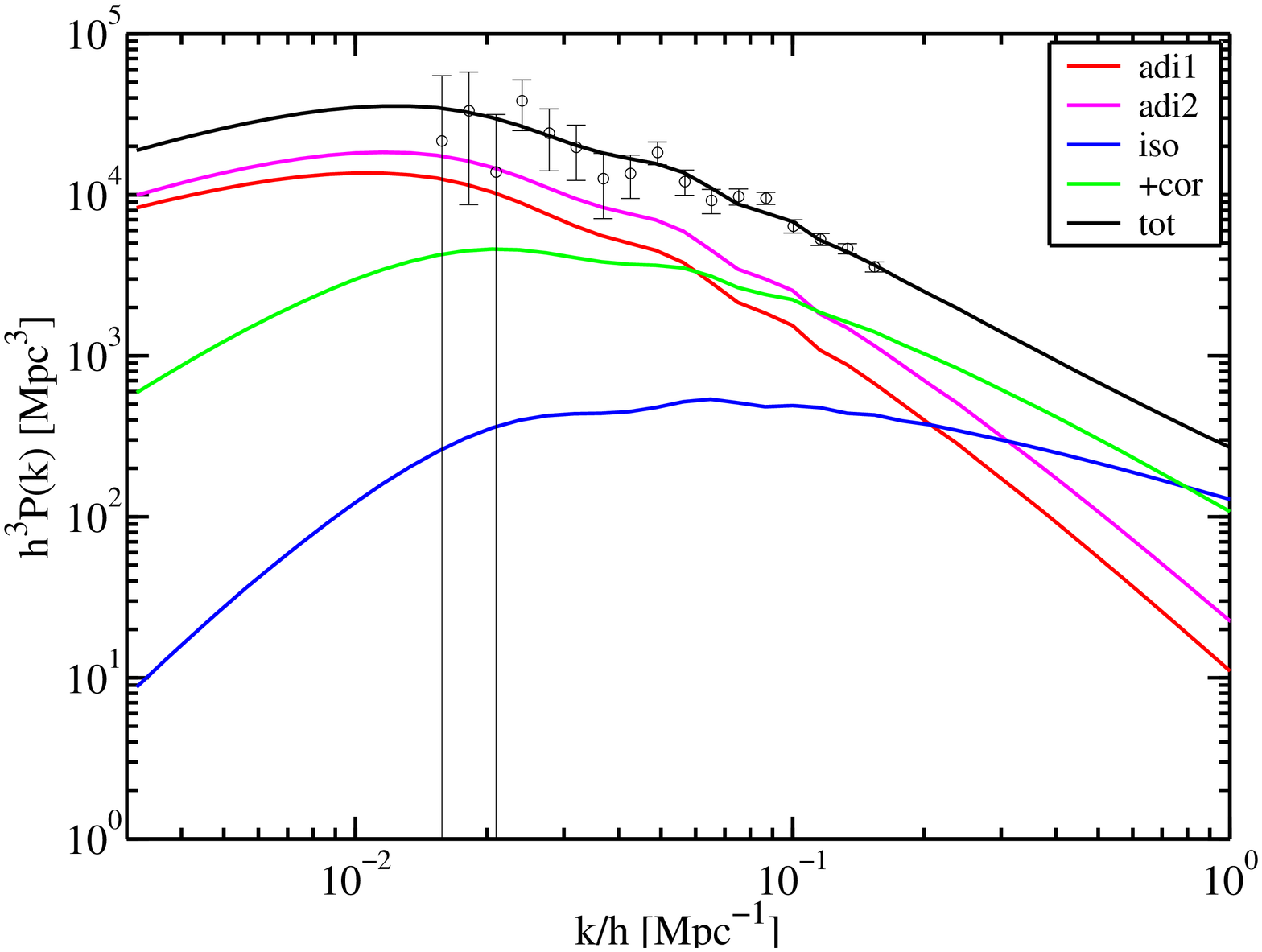}
  \caption{(a) The TT spectrum for the best-fit large
    $\Omega_\Lambda$ ($\Omega_\Lambda > 0.82)$ model, showing how the
    correlation part contributes towards shifting the acoustic peaks to the
    right.  This model has $\omega_b = 0.0221$, $\omega_c = 0.100$, $\theta = 1.082$,
    $\tau = 0.177$, $b = 0.796$, $\ln(10^{10} A^2) = 3.24$, $\nadiI = 0.949$,
    $\alpha = 0.108$, $\niso = 3.11$, $\gamma = 0.57$, $\nadiII = 1.043$.
    The correlated adiabatic component (``ad2'', \emph{magenta}) dominates
    over the uncorrelated adiabatic component (``ad1'', \emph{red}). The other
    curves are the correlation component (``cor'', \emph{green}) and isocurvature
    component (``iso'', \emph{blue}). The total angular power (``tot'', \emph{black})
    is a sum of these components.
    (b) The matter power spectrum, showing how the correlation contribution
    changes its shape. }
  \label{fig:bestLambda}
\end{figure}

The angular and matter power spectra of the best-fit
large-$\Omega_\Lambda$ model (from the subset $\Omega_\Lambda \geq 0.82$) are
shown in Fig~\ref{fig:bestLambda}. This model has $\chi^2 = 1461.86$.  Compared
to the best-fit adiabatic model, the somewhat worse fit, $\Delta \chi^2 =
2.21$, is due to 1) a worse fit to the SDSS data ($\Delta \chi^2 = 1.83$) and
2) a worse fit to the Sachs-Wolfe region ($2 \leq l \leq 21$) of the WMAP TT
data ($\Delta \chi^2 = 2.51$).  The latter is due to the increased late ISW
effect caused by the larger $\Omega_\Lambda$. This model fits the rest of the CMB
data better than the adiabatic model.

The reason the larger sound horizon angles (which shift the acoustic peaks
left, i.e., towards smaller $l$) are accepted is the correlation contribution
$C^{\mr{cor}}_{l}$, whose acoustic peaks are at somewhat larger $l$ than the
adiabatic ones, and thus adding it appears as a shifting of the peaks to the
right (i.e., towards larger $l$).  An uncorrelated isocurvature contribution
cannot do the same trick, since the isocurvature acoustic peaks are too much to
the right for adding them to appear as a shift in peak position. The
distribution of the isocurvature spectral index is concentrated at the upper
end of the allowed range for $\niso$ in these $\Omega_\Lambda
> 0.82$ models, see again Figs.~\ref{fig:isoparam} and \ref{fig:isoparamd}.
This is required for the correlation contribution to maintain roughly the same
relative power through the acoustic peak region.

Because of this large spectral index, especially the correlation contribution
also changes the shape of the matter power spectrum, see
Fig.~\ref{fig:bestLambda}(b). This allows for a smaller ``shape parameter''
$\Omega_m h$ to fit the SDSS data, than the SDSS result $\Omega_m h =
0.21\pm0.03$ for adiabatic models.  In the adiabatic model, large values of
$\Omega_\Lambda$ and $h$ would be allowed by either the CMB or the LSS data
alone, but not by the combined data sets, because either data set allows a
narrow region (the ``vanilla banana'' in Fig.~5 of \cite{Tegmark:2003ud}) in
the $(\Omega_\Lambda,h)$ plane, but these regions have somewhat different
orientations.  The correlation contribution makes both regions wider, in such a
way that their overlap is extended to higher $h$ and smaller $\Omega_m$ (larger
$\Omega_\Lambda$), or in terms of our independent parameters, towards smaller
$\omega_c$. In fact, even $h>1.0$, with $\Omega_m < 0.1$ (or $\Omega_\Lambda >
0.9$) would be allowed, but our $h\leq 1.0$ prior cuts them off.  These models
also favor smaller bias parameters $b$ and baryon densities $\omega_b$.

One might expect the above to work in the other direction too, negative correlation
allowing models with a smaller $\theta$, $\Omega_\Lambda$, $h$, and a larger
$\omega_c$, but apparently some other feature in the data prevents the larger
$\omega_c$ required.

The large values of $\Omega_\Lambda$ are ruled out by the high-$z$ Type Ia
Supernovae redshift-magnitude (SNIa) data\cite{Riess:2004nr}.  Therefore
the large-$\Omega_\Lambda$
feature was not seen in \cite{Beltran:2004uv}.  We did not use the SNIa data;
but to study the effect of a SNIa constraint we simulated it by importance
weighting our MCMC chains with a Gaussian $\Omega_\Lambda = 0.70\pm0.04$
distribution. We show the 1-d likelihoods both with and without this extra
prior in Figs.~\ref{fig:adiparam}--\ref{fig:isoparamd}.  The effect of this
SNIa constraint cutting the (large $\theta$, small $\omega_c$) models off is
clearly seen in them.

%%%%%%%%%%%%%%%%%%%%%%%%%%%%%%%%%%%%%%%%%%%%%%%%%%%%%%%%%%%%%%%%%%%%
\subsection{Baryon density and Hubble parameter}
\label{sec:obH}

In pure adiabatic models the baryon density is practically determined by the
heights of the first and second acoustic peaks (and the valley between them).
An isocurvature contribution modifies these heights and thus one expects looser
constraint for $\omega_b$ in mixed adiabatic and isocurvature models. However,
our constraints \footnote{%
%%%%
Note that allowing for $\tau>0.3$ could lead to higher upper bounds.
%%%%
} ($0.0220 < \omega_b < 0.0246$ at 68\% C.L., $0.0207 < \omega_b < 0.0263$ at
95\% C.L., median $0.0232$) are very close to the adiabatic model ($0.0221 <
\omega_b < 0.0240$ at 68\% C.L., $0.0213 < \omega_b < 0.0250$ at 95\% C.L.,
median $0.0230$). Moreover, we have checked that the isocurvature amplitude
($\alpha$) dependence of the constraints for $\omega_b$ is very weak within the
allowed range $\alpha<0.18$.

As can be seen in Fig.~\ref{fig:adiparam} the median of $\omega_b$ shifts only
marginally towards larger values regardless of the (extra) priors chosen. Our
result is consistent with \cite{Beltran:2004uv} where the 1-d likelihoods for
the adiabatic reference model and for the correlated CDM isocurvature model
were practically indistinguishable. (The neutrino isocurvature modes shifted
$\omega_b$ towards smaller values unlike in some other models to be discussed
below.)

Both our result and the likelihood for $\omega_b$ in \cite{Beltran:2004uv}
differ from
%other recent result achieved in
\cite{Moodley:2004nz} where the
median shifted significantly towards larger values ($0.023 < \omega_b < 0.029$
at 68\% C.L., median $0.026$). In their model the adiabatic, CDM isocurvature
and correlation spectral indices were kept equal, i.e. they had only one free
spectral index. Both \cite{Beltran:2004uv} and \cite{Moodley:2004nz} used CMB
and LSS data sets very similar to those used by us. On the other hand, in
\cite{Ferrer:2004nv} the curvaton decay calculation (see e.g.
\cite{Gupta:2003jc}) was extended to the case when the curvaton does not
necessarily behave like dust. The resulting correlated CDM isocurvature
perturbations from the mixed inflaton-curvaton decay (or e.g. from
double-inflation which produce primordial power spectrum of similar type) were
considered in the light of WMAP data alone. Then $\omega_b$ got even larger
values. The 68\% C.L. region obtained in \cite{Ferrer:2004nv} was $0.027 <
\omega_b < 0.042$ with median $0.032$. (The best-fit model had
$\omega_b=0.041$.) Although, the form of the primordial power spectrum in
\cite{Ferrer:2004nv,Ferrer:2005} looks quite similar to ours, the important
difference is that the adiabatic and isocurvature components have equal
spectral indices there. Note however, that as the curvature and entropy
perturbations are both a sum of two components the model has two independent
spectral indices insted of just one of \cite{Moodley:2004nz}. In
\cite{Bucher:2004an} a model with equal spectral indices for adiabatic, CDM
isocurvature and neutrino isocurvature modes yielded also very large
$\omega_b\sim 0.04$. The CMB data alone led to a bit larger $\omega_b$ than
CMB and LSS data together.

Three years ago Trotta, Riazuelo, and Durrer demonstrated in
\cite{Trotta:2001yw} that allowing for ``general isocurvature modes''
(adiabatic, CDM isocurvature and neutrino isocurvature with equal spectral
indices in their study) prevented one from obtaining an upper bound for
$\omega_b$ from that day's CMB data (COBE
\cite{Smoot:1992td,Bennett:1994,Tegmark:2000db} and Boomerang
\cite{Netterfield:2001yq}). In \cite{Trotta:2001yw} most of the new freedom for
$\omega_b$ was explained to come from the neutrino isocurvature density mode
which can adjust the height of the second acoustic peak more than other
isocurvature modes. Hence, one would expect more freedom for $\omega_b$ when
allowing also for a neutrino isocurvature density mode instead of just a CDM
isocurvature mode (or a neutrino isocurvature velocity mode). However, one can
not see this effect in \cite{Beltran:2004uv} where the median of $\omega_b$ was
shifted a bit towards smaller values (compared to other cases) and the width of
the 1-d distribution remained small. While part of the different effect of the
neutrino isocurvature density mode in \cite{Trotta:2001yw} and
\cite{Beltran:2004uv} could result from the different data sets used (the
former used CMB only, the latter used precision CMB and LSS), we think that the
fundamental explanation resides in spectral indices. The same applies to other
isocurvature modes.

To be explicit, our model and the model in \cite{Beltran:2004uv}
(both have three independent spectral indices) yield very little difference to
the adiabatic case whereas models studied in
\cite{Moodley:2004nz,Bucher:2004an,Ferrer:2004nv,Trotta:2001yw} (all have
``$\nadi = \niso$'') lead to larger medians and wider distributions for
$\omega_b$. While we have not studied in detail the reason for this difference,
we discuss one possibility. The CMB data forces the dominant adiabatic
spectrum close to scale invariance ($\nadi \sim 1$). When the spectral indices
are kept equal (``$\nadi = \niso$'') the isocurvature component also acquires
the same spectral index. The multipole dependence of the isocurvature
contribution to the CMB spectrum can now be seen easily from
Fig.~\ref{fig:clothers}(c). Isocurvature and correlation modify more the low
multipole end of the spectrum than high multipoles. There is a large difference
in the power coming from non-adiabatic contributions to the Sachs-Wolfe plateau
($l\lesssim 21$) and to the first or second acoustic peak. Hence, the acoustic
peak structure is distorted significantly leading to a need/possibility to
adjust it by $\omega_b$. However, if the isocurvature spectral index is a free
parameter it acquires a value that leads to a small isocurvature contribution
on all scales. This happens with $\niso \sim 3$ as will be demonstrated in
Fig.~\ref{fig:cltt_loglog2}. With this large $\niso$ the isocurvature
contribution to the $C_l$ and to the matter power can remain, e.g., at some
3.5\% level (for our median $\alpha=0.035$) on all scales. Then the different
peak structure of the isocurvature compared to the adiabatic one represents
only a marginal distortion from the pure adiabatic case. This explains why we
and \cite{Beltran:2004uv} end up with the ``adiabatic value'' for $\omega_b$.

\begin{figure}[th]
  \centering
  \includegraphics[angle=270,width=0.45\textwidth]{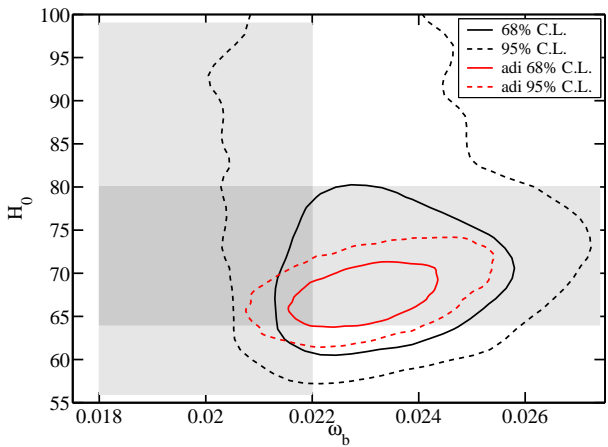}
  \caption{The 68\% (\emph{solid}) and 95\% (\emph{dashed}) C.L. regions
    in the $(\omega_b,H_0)$ plane for our isocurvature model (\emph{black})
    and for the adiabatic model (\emph{red}). The rectangular
    gray boxes represent a BBN constraint $\omega_b = 0.020\pm 0.002$
    \cite{Burles:2000zk}
    and the HST constraint $H_0= (72\pm 8)$ km/s/Mpc \cite{Freedman:2000cf}.}
  \label{fig:cor_omb_H0}
\end{figure}

The isocurvature effect on the determination of the Hubble parameter in our
model is more dramatic. Without extra priors we do not obtain an upper bound
for $h$ (within the analyzed range $0.4 < h < 1.0$), see
Fig.~\ref{fig:adiparamd} and the discussion of the previous subsection.
(However, with a different choice of the pivot scale we would miss the small
matter density models and hence obtain an upper bound for $h$. We will discuss
this in detail in Sec.~\ref{sec:pivotscale}.) Applying the SNIa result
$\Omega_\Lambda = 0.70\pm0.04$ to get rid of the large $h$ values we get a 95\%
C.L. region $0.64 < h < 0.77$.

In Fig.~\ref{fig:cor_omb_H0} we compare our model with the pure adiabatic model
by showing the C.L. contours in $(\omega_b,H_0)$ plane. We indicate also the
95\% C.L. result $\omega_b = 0.020\pm 0.002$ from a big bang nucleosynthesis
(BBN) calculation \cite{Burles:2000zk} and the Hubble space telescope key
project result $h=0.72\pm 0.08$ \cite{Freedman:2000cf}. The 95\% region (or
even the 68\% region) of our model certainly accommodates the HST result, but is
only marginally consistent with the BBN value of $\omega_b$ from
\cite{Burles:2000zk}. Actually, the same is true for the adiabatic model. On
the other hand, concordance is achieved with another BBN value
$\omega_b=0.022\pm0.002$ from \cite{Hagiwara:2002fs}.

For comparison, the similar contours opened up towards the upper right corner
of $(\omega_b,h)$ plane in \cite{Trotta:2001yw}. Moreover, the only ``excluded
region'' was ``an upper left corner'' of their $(\omega_b,h)$ plane where the
BBN and HST regions intersected  in their Fig. Again we stress that different
data sets were used in \cite{Trotta:2001yw}, also neutrino isocurvature modes
were allowed and there was only one spectral index. In any case, the
considerations in this subsection demonstrate that even within ``isocurvature
models'' the initial assumptions, e.g. the shape assumed for the primordial
spectrum, affect considerably the end results.

%%%%%%%%%%%%%%%%%%%%%%%%%%%%%%%%%%%%%%%%%%%%%%%%%%%%%%%%%%%%%

\subsection{Isocurvature parameters}
\label{sec:isoparam}

We now turn to the parameters related to isocurvature perturbations. The 1-d
likelihoods for the 4 independent parameters, the isocurvature fraction
$\alpha$, the isocurvature spectral index $\niso$, the adiabatic correlated
fraction $\gamma$, and the spectral index $\nadiII$, are
shown in Fig.~\ref{fig:isoparam}, and the two derived parameters, the
correlation fraction
 \begin{equation}
    \alpha_{\mr{cor}} \equiv \mr{sign}(\gamma)\sqrt{\alpha(1-\alpha)\abs{\gamma}}
 \end{equation}
and the correlation spectral index
 \begin{equation}
    \ncor \equiv \frac{\nadiII+\niso}{2} \,.
 \end{equation}
in Fig.~\ref{fig:isoparamd}.

We obtain an upper limit (95\% C.L.) to the isocurvature fraction
 \begin{equation}
    \alpha < 0.18 \,.
 \label{alphalimit}
 \end{equation}
One should be careful about the meaning of this.  First, $\alpha$ is defined as
the isocurvature fraction at our pivot scale $k_0 = 0.01\mbox{Mpc}^{-1}$.
Models with a small isocurvature fraction at this scale may have a large
isocurvature fraction at some other scale, depending on how the spectral
indices for the adiabatic and isocurvature fractions differ from each other.
Second, since $\alpha$ is defined in terms of the primordial curvature and
entropy perturbations, it does not give directly  the relative isocurvature
contribution to $C_l$, but that depends also on the shapes of the component
spectra $\hat{C}^{\mr{ad1}}_{l}$, $\hat{C}^{\mr{ad2}}_{l}$,
$\hat{C}^{\mr{cor}}_{l}$, and $\hat{C}^{\mr{iso}}_{l}$, (i.e., on the transfer
functions) which depend on the other cosmological parameters, and are typically
such that the isocurvature contribution to the total $C_l$ and $P(k)$ is
smaller than $\alpha$.  Thus the limit to an isocurvature signal in the data is
actually tighter than would appear from Eq.~(\ref{alphalimit}). (See
Sec.~\ref{sec:nonadicon}.) Third, because of the presence of poorly constrained
parameters, $\nadiII$ and $n_{\mr{iso}}$ in the case of small $\gamma$ or
$\alpha$, the likelihood functions, and thus upper limits, are sensitive to the
priors implied by the choice of parametrization.  We discuss this last point in
Sec.~\ref{sec:pivotscale}. Similar caveats apply to the other isocurvature
related parameters.

For the ``uncorrelated'' subset, $\gamma = 0.0\pm0.02$ the formal upper limit
is larger
 \begin{equation}
    \alpha < 0.22 \,.
 \end{equation}
The limit for correlated models is tighter, since the correlation contribution
to the data tends to be larger than the isocurvature contribution, due to the
transfer functions (see Figs.~\ref{fig:cltt} and \ref{fig:clothers}), and since
 $|\alpha_\mr{cor}| \equiv \sqrt{\alpha(1-\alpha)|\gamma|} > \alpha$ for
small $\alpha$ and moderate  $\gamma$.

\begin{figure}[th]
  \includegraphics[angle=270,width=0.40\textwidth]{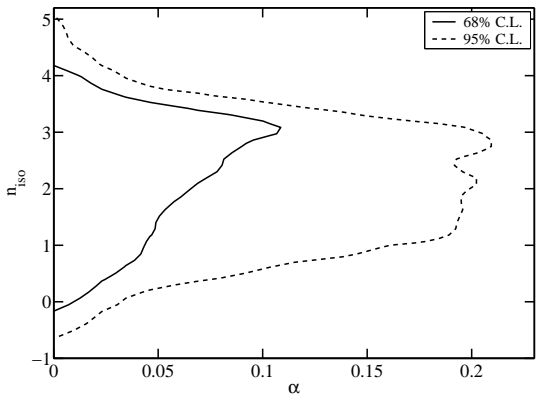}
  \caption{2-d marginalized likelihood for $\alpha$ and $\niso$.}
  \label{fig:alpha_niso}
\end{figure}

The isocurvature spectral index has a fairly wide distribution covering the
range $0 < \niso < 4$.  The median value is $\niso = 2.252$.  The
distribution is skew, so that the largest marginalized likelihood is at
somewhat larger values, $\niso \sim 3.0$.  This peak at $\niso$ comes
from the large-$\Omega_\Lambda$-models discussed in Sec.~\ref{sec:largelambda}.
Otherwise, values $1 < \niso < 3$ are preferred. Fig.~\ref{fig:alpha_niso}
shows the 2-d likelihood for $\alpha$ and $\niso$.

There are basically two reasons why the data selects this range for $\niso$.
Disregarding the peak structure in the $C_l$ spectrum, the overall distribution
of power in the data over different scales is such that for the adiabatic
models it favors a scale-independent $\nadi \sim 1$ primordial spectrum.  For
the isocurvature modes the $C_l$ transfer function falls more steeply with $k$
(see Figs.~\ref{fig:cltt} and \ref{fig:clothers}.) Thus, for the isocurvature
contribution not to disturb this overall distribution of power, it needs a
larger spectral index.  The other reason is in the more detailed shape of the
data. The CMB data clearly does not like the isocurvature contribution, since
it has the wrong peak structure. Too small (large) $\niso$ would cause it to
show up for small (large) $l$, even for small $\alpha$. With $k_0$ in the
middle of the data sets, this keeps $0<\niso<4$.

\begin{figure}[t]
  \centering
  \includegraphics[angle=270,width=0.45\textwidth]{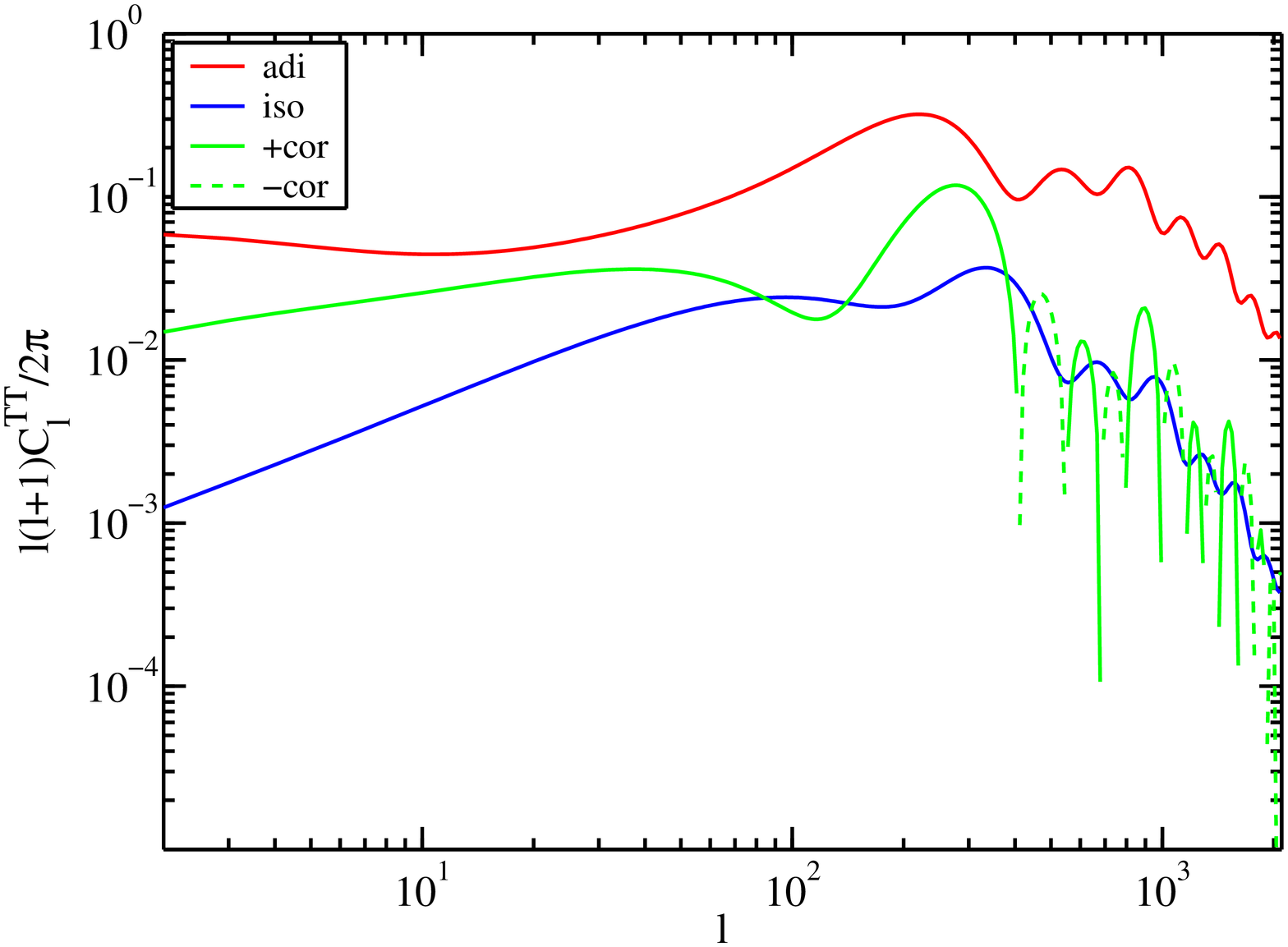}
  \caption{Like Fig.~\ref{fig:clothers}(c), but now with $\niso = 2.252$.}
  \label{fig:cltt_loglog2}
\end{figure}

In Fig.~\ref{fig:cltt_loglog2} we show the unit-amplitude component
$\hat{C}_l$ spectra, for $\nadi$ still at $1$, but $\niso$ at
the median value, $\niso = 2.252$.  Now the effective slope of the
adiabatic and isocurvature contributions is roughly the same, so that the
isocurvature contribution is kept low everywhere with moderately small
$\alpha$.

\subsection{Correlation Parameters}
\label{sec:corparam}

Zero correlation, $\gamma \approx 0$, is favored over any other particular
value for the correlation.  However, 61\% of the models have $|\gamma| \geq
0.1$ (Fig.~\ref{fig:isoparam}). Positive correlations are favored over negative
ones.

\begin{figure}[t]
  \centering
  \includegraphics[angle=270,width=0.40\textwidth]{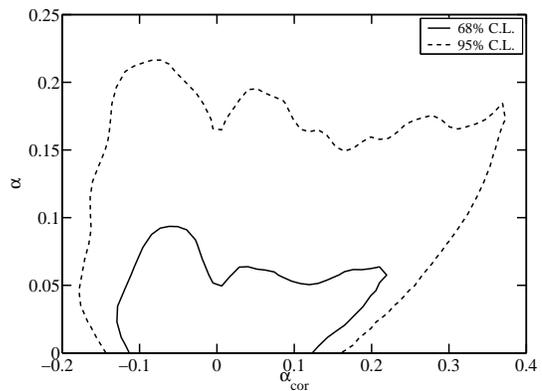}
  \caption{2-d marginalized likelihood for $\alpha_{\mr{cor}}$ and $\alpha$.}
  \label{fig:Acor_alpha}
\end{figure}

\begin{figure}[t]
  \centering
  \includegraphics[angle=270,width=0.40\textwidth]{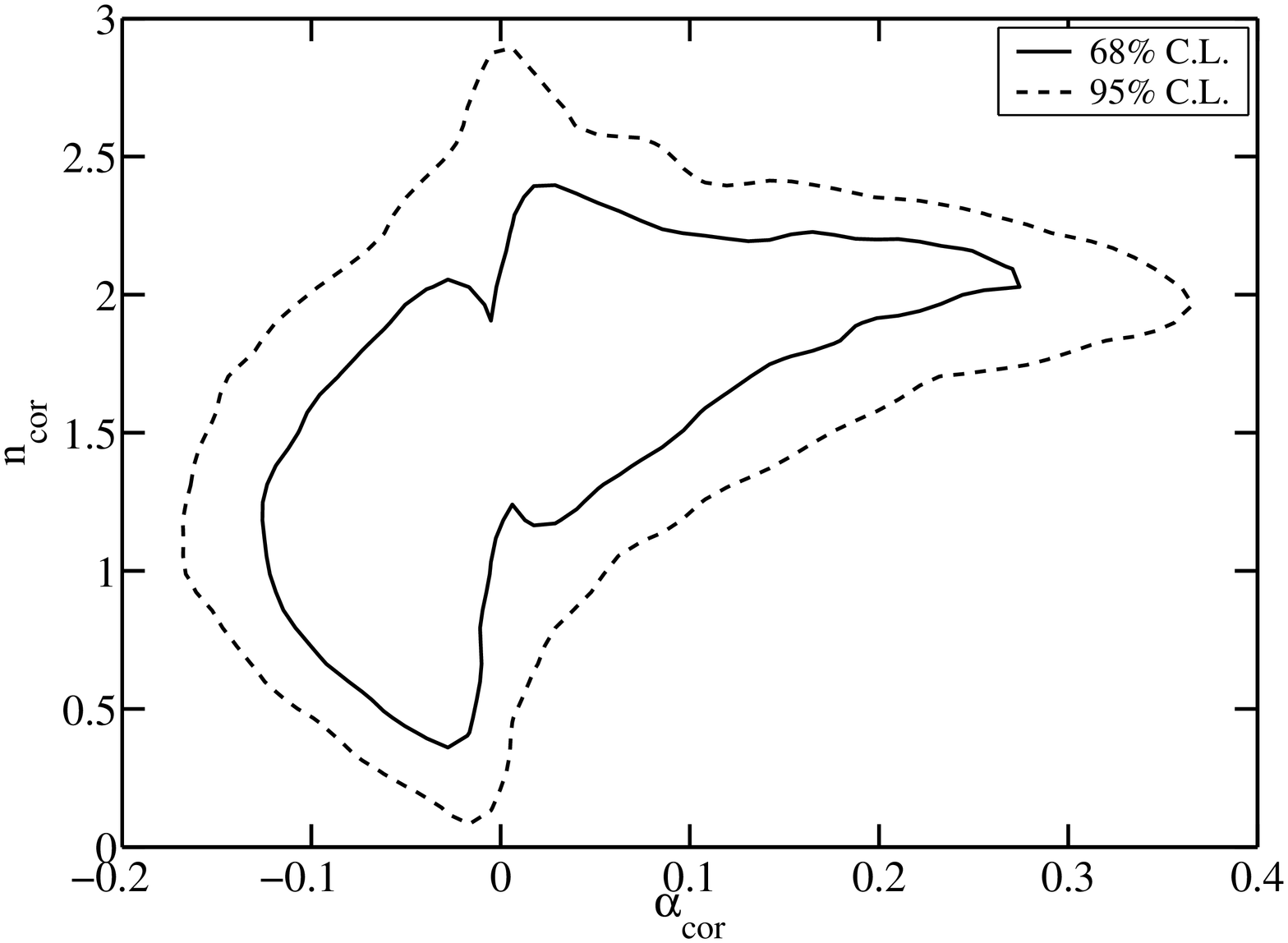}
  \caption{2-d marginalized likelihood for $\alpha_{\mr{cor}}$ and $\ncor$.}
  \label{fig:Acorncor}
\end{figure}

A strong correlation $\gamma$ between the adiabatic and isocurvature
perturbations however has little effect on the data, if the isocurvature
perturbations, with which the adiabatic perturbations are correlated, are
negligibly small.  The signature in the data is better measured by the derived
parameter $\alpha_{\mr{cor}}$ (which is restricted between
$\pm\sqrt{\alpha(1-\alpha)}$ by definition).  We see that the 1-d likelihood of
$\alpha_{\mr{cor}}$ is skew (Fig.~\ref{fig:isoparamd}); the preference for positive correlations that we
saw in $\gamma$ appears here as a long tail towards large $\alpha_{\mr{cor}}$.
If we add the Gaussian prior $\Omega_\Lambda = 0.70\pm0.04$ to represent SNIa
constraints, this tail goes away, and the 1d likelihood becomes rather
symmetric.  Thus the preference for positive correlations is due to the
large-$\Omega_\Lambda$ models discussed in Sec.~\ref{sec:largelambda}. The dip
at $\alpha_{\mr{cor}} = 0$ in Figs.~\ref{fig:isoparamd}, \ref{fig:Acor_alpha},
and \ref{fig:Acorncor} does not indicate that uncorrelated models would be
unfavored by the data; rather it comes because flat priors for $\alpha$ and
$\gamma$ lead to a prior for $\alpha_{\mr{cor}}$ which is small for small
$\alpha_{\mr{cor}}$. Fig.~\ref{fig:Acor_alpha} shows the 2-d likelihood of
$\alpha$ and $\alpha_{\mr{cor}}$.

In Fig.~\ref{fig:Acorncor} we show the 2-d likelihood of
$\alpha_{\mr{cor}}$ and $\ncor$.  It shows that positive correlations
are connected with larger spectral indices $\ncor$ than negative
correlations. This feature remains also after applying the $\Omega_\Lambda =
0.70\pm0.04$ prior, although the largest $\alpha_\mr{cor}$ values are cut off.
For large $|\gamma|$ we have $\nadiII \approx 1$, and thus
 \beq
   \ncor \equiv \frac{\nadiII+\niso}{2} \approx \frac{1+\niso}{2} \,.
 \eeq
%%%%%%%%%%%%%%%%%%
%
\begin{figure}[t]
  \centering
  \includegraphics[angle=270,width=0.40\textwidth]{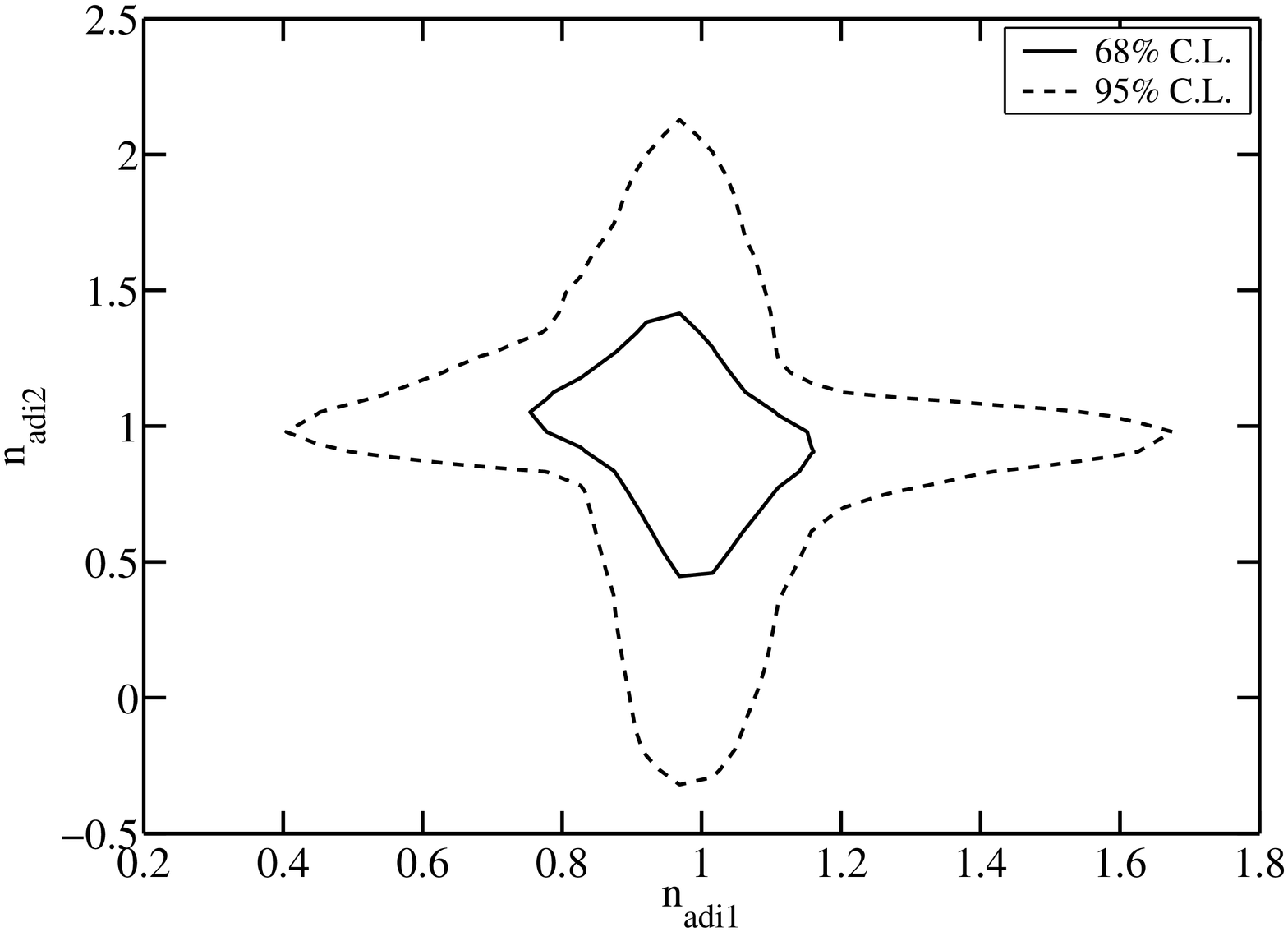}
  \caption{2-d marginalized likelihood for $\nadiI$ and $\nadiII$.}
  \label{fig:nadi1_nadi2}
\end{figure}
%
%%%%%%%%%%%%%%%%%
The reason negative correlations are favored with smaller $\ncor$ is that
then there is a significant isocurvature and correlation contribution to the
Sachs-Wolfe region of the TT spectrum, and the negative correlation now
subtracts from it, helping to fit the lowest $l$ WMAP data points which lie
below the adiabatic spectra (see Fig.~\ref{fig:2ndb}).

For larger $\ncor$ the correlation contribution is insignificant in the SW region, but
becomes important in the region of acoustic peaks and for the matter power
spectrum. Positive correlations are now favored for the reasons discussed in
Sec.~\ref{sec:largelambda}.  This effect remains after adding the Gaussian
prior $\Omega_\Lambda = 0.70\pm0.04$, in part since this SNIa result
favors somewhat larger $\Omega_\Lambda$ than the CMB+SDSS data applied to
adiabatic models.

Whenever one adiabatic component has negligible amplitude, the corresponding
spectral index (i.e., $\nadiII$ for $\gamma \approx 0$, and
$\nadiI$ for $|\gamma| \sim 1$) becomes unconstrained (see
Fig.~\ref{fig:nadi1_nadi2}), otherwise it is tightly constrained to be near
$1$. When both components are significant, there is a small anticorrelation
between $\nadiI$ and $\nadiII$ \cite{Valiviita:2003ty}, as a red
tilt in one of them can compensate for a blue tilt in the other one, making the
sum closer to scale-invariant (as preferred by the data).
This effect however introduces a positive $dn/d\ln k$ in the
combined spectrum (Sec.~\ref{sec:adiindex}), which the data does not like,
especially with the inclusion of LSS data, and therefore the anticorrelation
effect is now
more limited than in \cite{Valiviita:2003ty}.

In Fig.~\ref{fig:2ndb} we show the spectra for our best-fit model.  This is an
example of a low-$\niso$, negative-correlation model, where the
correlation contribution subtracts from the Sachs-Wolfe region in the $C_l$.
This model has
   $\omega_b = 0.0227$, $\omega_c = 0.129$, $\theta = 1.043$, $\tau = 0.144$,
   $b = 0.962$, $\ln(10^{10}A^2) = 3.30$, $\nadiI = 0.988$, $\alpha =
   0.00197$, $\niso = 0.388$, $\gamma = -0.67$, $\nadiII = 0.926$,
   and $\chi^2=1459.29$.

\begin{figure}[th]
  \centering
  \includegraphics[angle=270,width=0.45\textwidth]{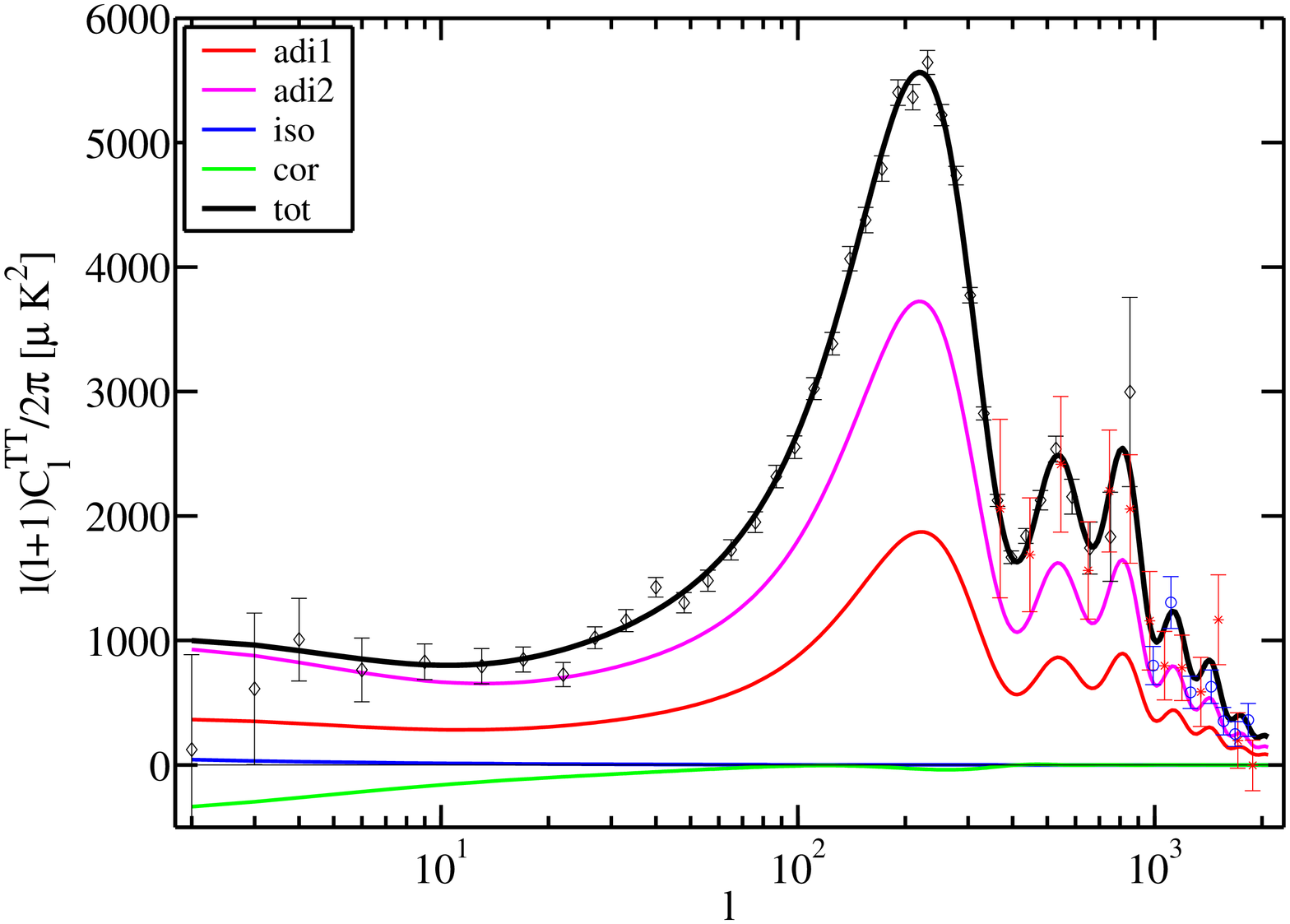}
  \includegraphics[angle=270,width=0.45\textwidth]{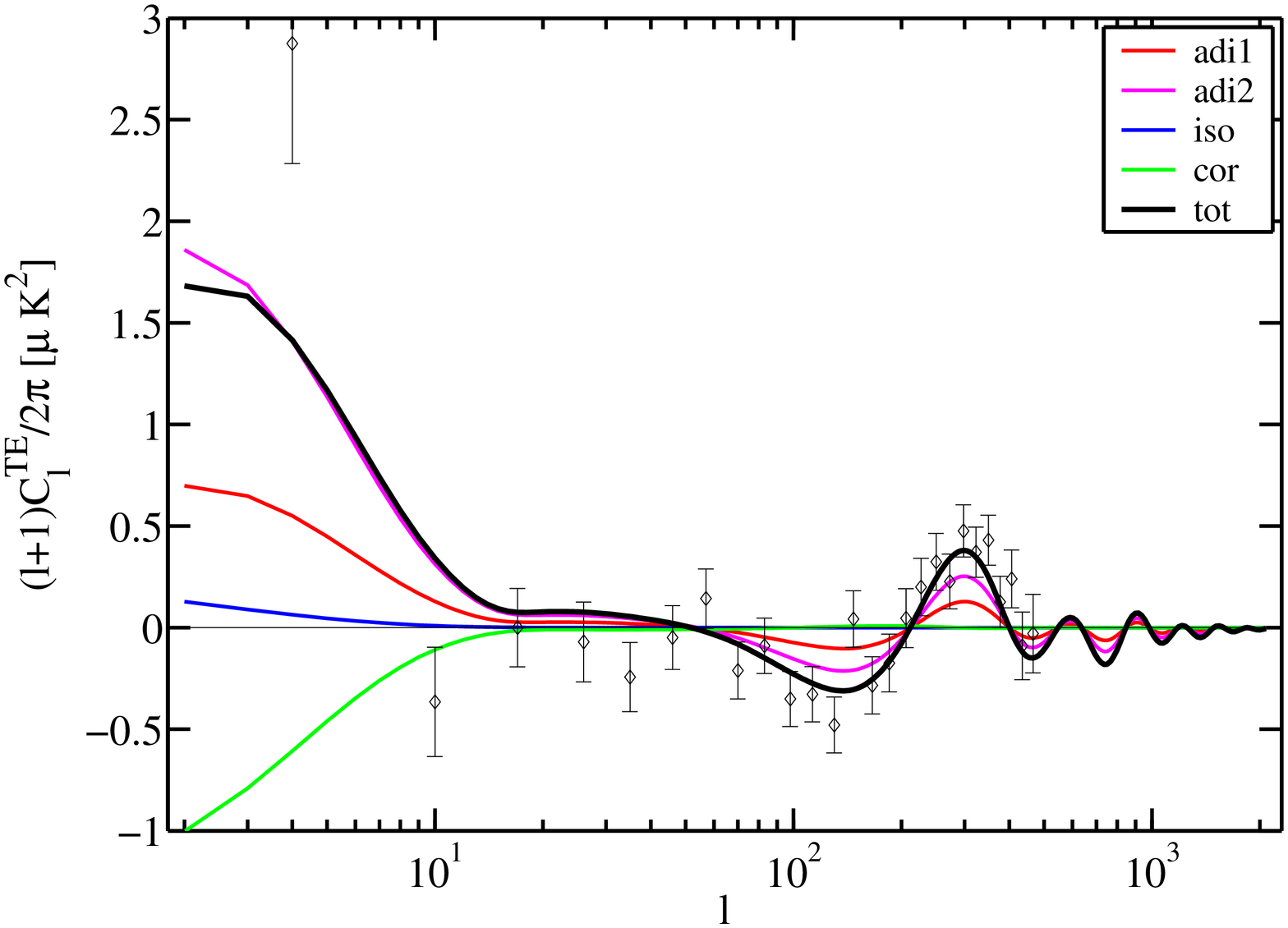}
  \includegraphics[angle=270,width=0.45\textwidth]{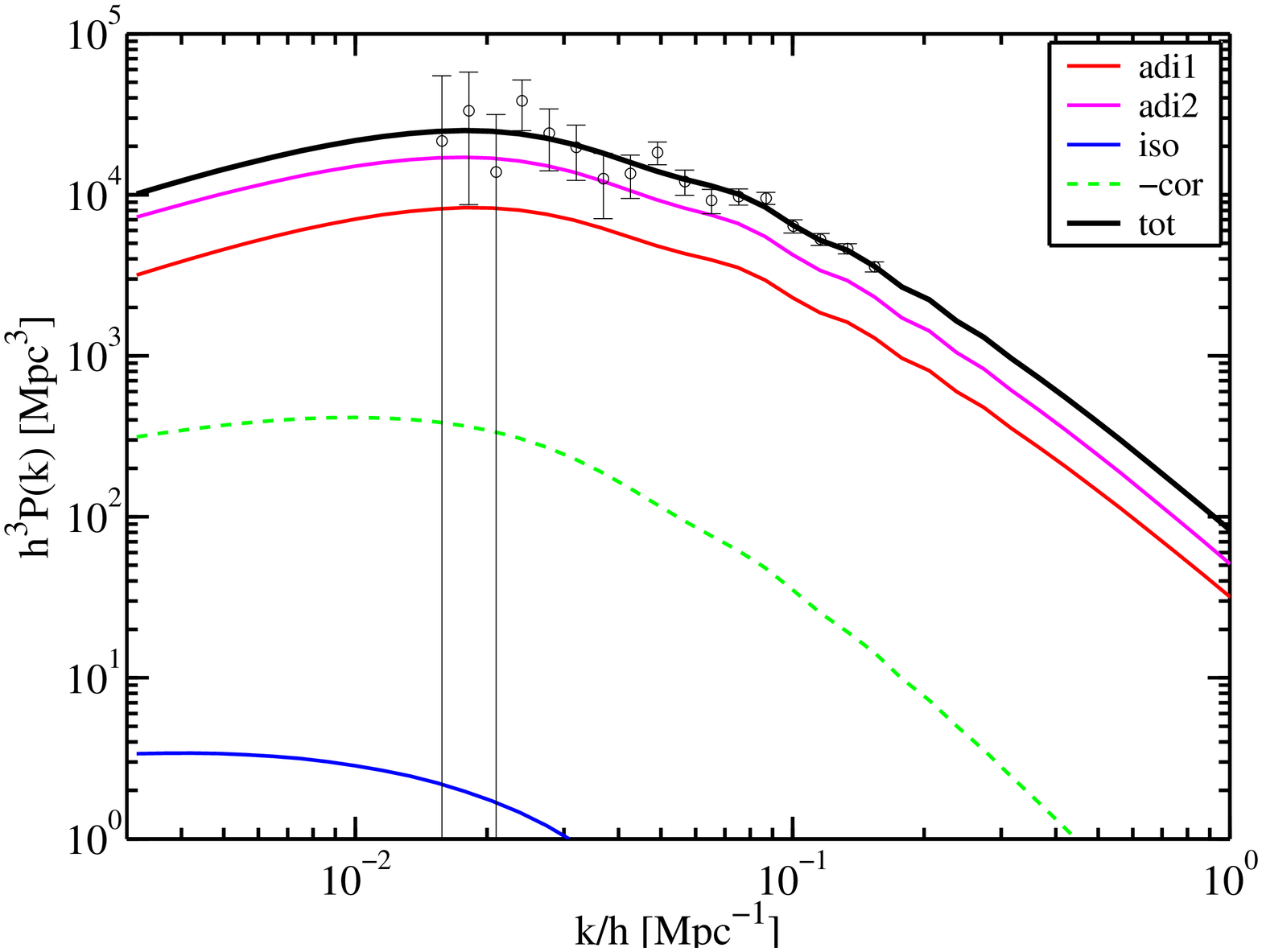}
  \caption{The CMB and matter power spectra for our best-fit model.
   The correlated adiabatic component (``ad2'', \emph{magenta}) dominates
   over the uncorrelated adiabatic component (``ad1'', \emph{red})
   and the negative correlation
   subtracts from the Sachs-Wolfe region of the CMB spectrum. 
 %This model has
 %  $\omega_b = 0.0227$, $\omega_c = 0.129$, $\theta = 1.043$, $\tau = 0.144$,
 %  $b = 0.962$, $\ln(10^{10}A^2) = 3.30$, $\nadiI = 0.988$, $\alpha =
 %  0.00197$, $\niso = 0.388$, $\gamma = -0.67$, $\nadiII = 0.926$.
  }
  \label{fig:2ndb}
\end{figure}

\subsection{Models with Very Large Isocurvature Spectral Index}
\label{sec:largeniso}

We set a very wide allowed range for the isocurvature spectral index.  While
most of the good models had $\niso$ in the range $0$ to $4$, one of our
MCMC chains found an apparently very good region where $\niso$ was
between $5$ and $6$.  In fact the highest likelihood ($\chi^2 = 1459.20$)  was
obtained in this region.  We did not have enough statistics to assess correctly
the relative importance of this disjoint good-likelihood region in the
parameter space. We discard this region for reasons explained below.  Thus we
have not included this chain in our full analysis.  (And therefore we do not
take our best-fit model from it.)  In fact, these models are obviously nonsense,
and we discuss them just as a warning.

\begin{figure}[t]
  \centering
  \includegraphics[angle=270,width=0.45\textwidth]{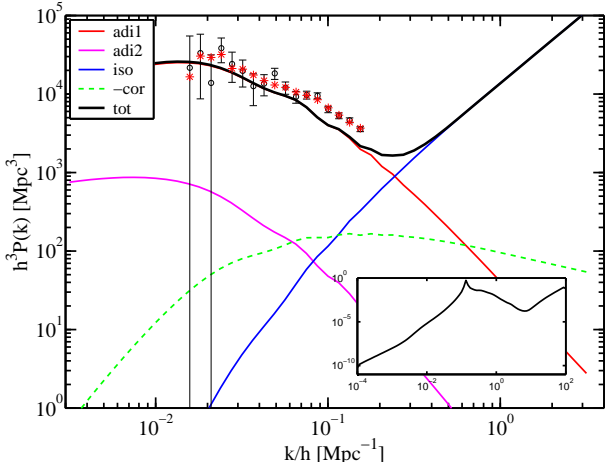}
  \caption{The matter power spectrum of a model with $\niso$=5.69, which has
   $\chi^2 = 1459.20$. The inset shows the SDSS window function for the 16th, i.e. second
    from right, data point. The red markers ($\ast$) indicate the values
    (theoretical matter power
    spectrum convolved with the window function) to be compared to the data
    points, showing that the fit ($\chi^2$) obtained this way is very good,
    although the power spectrum seems to lie below the data. The problem with
    this kind of matter power is that the window function picks most of the
    ``power'' from the non-linear regime. Moreover, we have calculated the matter
    power up to $k/h \sim 3$Mpc$^{-1}$ only. Had we calculated further,
    the rapidly rising power at large $k$ would have caused the red markers
    ($\ast$) to move up leading to worse $\chi^2$ for this model.
    }
  \label{fig:absbest_mat_win}
\end{figure}

These models necessarily have a very small $\alpha$; because of the large
$\niso$ the isocurvature contribution is steeply rising, and only
becomes noticeable at the smallest scales of our data set.  At scales smaller
than included in our data set, the isocurvature contribution then becomes
dominant, and $P(k)$ rises rapidly.

Thus for most of the data set, these models are essentially equal to the
adiabatic model. The improvement over the adiabatic model is then in the "other
CMB" and "SDSS" data which cover the smallest scales. The fit to the SDSS data
is however obtained in a rather unnatural way. Because the SDSS window
functions, that describe how the data points relate to the underlying power
spectrum, extend to much smaller scales (larger $k$) than the nominal $k$
values of the data points, for these models they pick up most of the
contribution at these very small scales (see Fig.~\ref{fig:absbest_mat_win}).
Since the perturbations are non-linear at these scales, our use of a linear
power spectrum does not give correct results. (We also suspect that the SDSS
window functions were not really meant to be used for this kind of spectra.)
Anyway, these models would be ruled out if some smaller scale constraints were
added.

Because our pivot scale is far enough to the left from the right (small-scale)
end of our data set, these models are forced to have a rather small $\alpha$,
which makes the measure of this region of parameter space rather small. If a
smaller pivot scale (larger $k_0$) is used, it becomes more likely for the MCMC
chains to end up in this questionable region (Sec.\ref{sec:pivotscale}).

%%%%%%%%%%%%%%%%%%%%%%%%%%%%%%%%%%%%%%%%%%%%%%%%%%%%%%%%%%%%%
\section{Non-adiabatic contribution to the observed spectra}
\label{sec:nonadicon}

So far we have constrained the non-adiabatic contribution to the primordial
spectrum in terms of $\alpha$ and $\alpha_{\mr{cor}}$ (or $\gamma$). Although
the isocurvature component can be as large as 18\% of the total primordial
power at our pivot scale $k_0$, its role in the observed $C_l$ (or matter
power) spectrum is less significant. This comes because of different behavior
of adiabatic and isocurvature transfer functions as discussed in
Secs.~\ref{sec:correlation} and \ref{sec:isoparam}. Moreover, the
non-scale-invariant spectral index complicates drawing conclusions for the
observed $C_l^{\mr{iso}}$ and $C_l^{\mr{cor}}$ from $\alpha$ and
$\alpha_{\mr{cor}}$, respectively. Thus we devote this section to finding
limits for non-adiabatic contributions to the \emph{observed spectra.}

We define a relative non-adiabatic contribution to $C_l^{\mr{TT}}$ by
\begin{equation}
\alpha_l = \frac{C_l^{\mr{TTiso}} + C_l^{\mr{TTcor}}}{C_l^{\mr{TT}}}\,,
\label{eqn:alphal}
\end{equation}
where $C_l^{\mr{TT}} = C_l^{\mr{TTad1}} + C_l^{\mr{TTad2}} + C_l^{\mr{TTiso}} +
C_l^{\mr{TTcor}}$.
%{\bf $C_l^{\mr tot}$ could be defined earlier...}
When creating MCMC chains we saved this quantity for $l = 2$, $140$, $200$, and
$700$ for each accepted step. By similar manner we define a non-adiabatic
contribution to the matter power at $k=k_i$
\begin{equation}
\alpha_{mi} = \frac{{P}^{\mr{iso}}(k_i) + {P}^{\mr{cor}}(k_i)}{{P}^(k_i)}\,.
\label{eqn:alphami}
\end{equation}
We saved this around the first SDSS data point $k_1 / h = 0.0154$~Mpc$^{-1}$ and
at the last data point $k_{17} / h =0.154$~Mpc$^{-1}$.

The range of possible values for $\alpha_l$ and $\alpha_{mi}$ is $[-\infty,
1]$. For example, $\alpha_l$ gets negative values whenever $C_l^{\mr{cor}} <
-C_l^{\mr{iso}}$. In the extreme case that $C_l^{\mr{iso}} = C_l^{\mr{ad2}}$
and $C_l^{\mr{cor}} = -2C_l^{\mr{iso}}$ the denominator approaches zero in the
absence of $C_l^{\mr{ad1}}$. On the other hand, the maximum value $1$ is
obtained with $C_l^{\mr{cor}} = C_l^{\mr{ad2}} = C_l^{\mr{ad1}} = 0$.

Recall that the $C_l^{\mr{TT}}$s are related to the variance of the CMB
temperature perturbation by
\begin{equation}
{\textstyle\left\langle\left( \frac{\delta T}{T} \right)^2 \right\rangle} =
\frac{1}{4\pi} \sum_{l = 0}^\infty (2l + 1) C^{TT}_l \,. \label{eqn:Cell}
\end{equation}

We have calculated the $C_l$ for $l = 2$--$2100$. In all well-fitting models
%%%%%%%%%%%%%%%%
%
%
\begin{figure}[t]
  \centering
  \includegraphics[angle=270,width=0.45\textwidth]{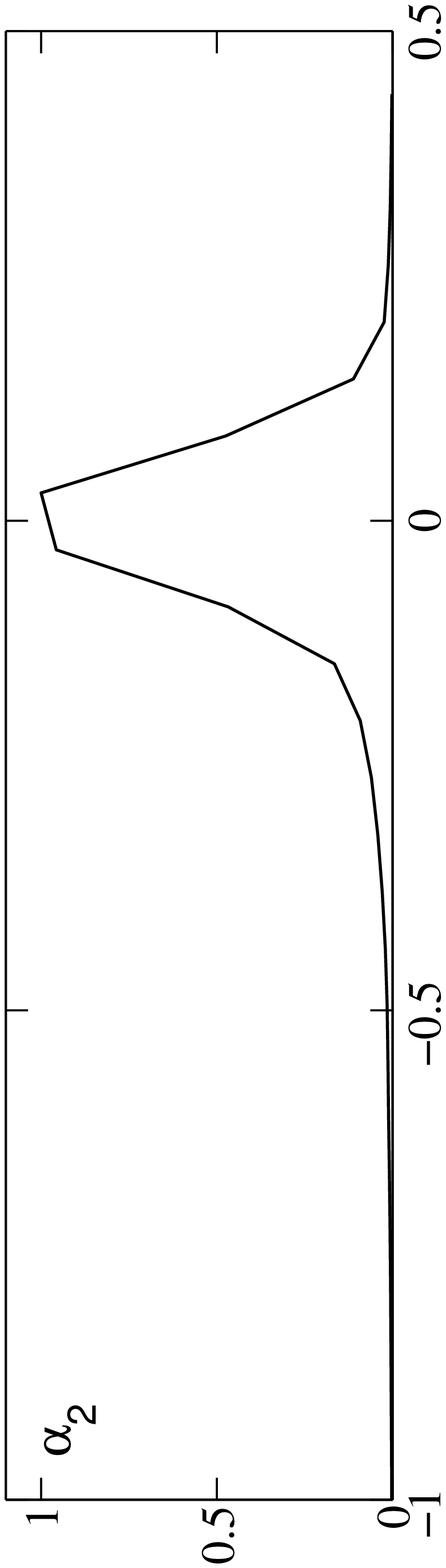}
  \includegraphics[angle=270,width=0.45\textwidth]{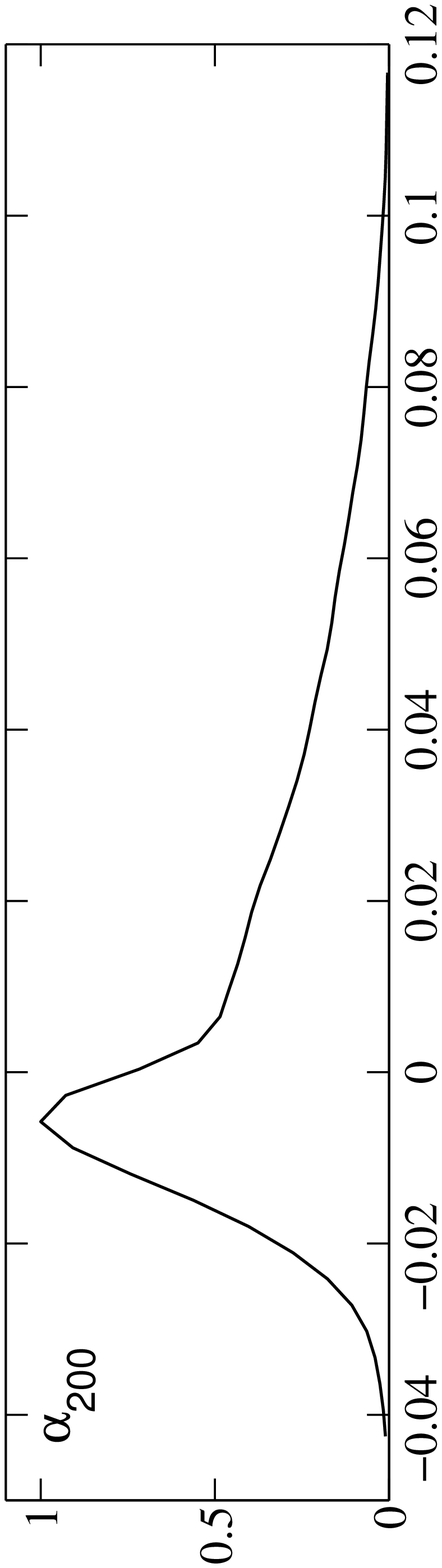}
  \includegraphics[angle=270,width=0.45\textwidth]{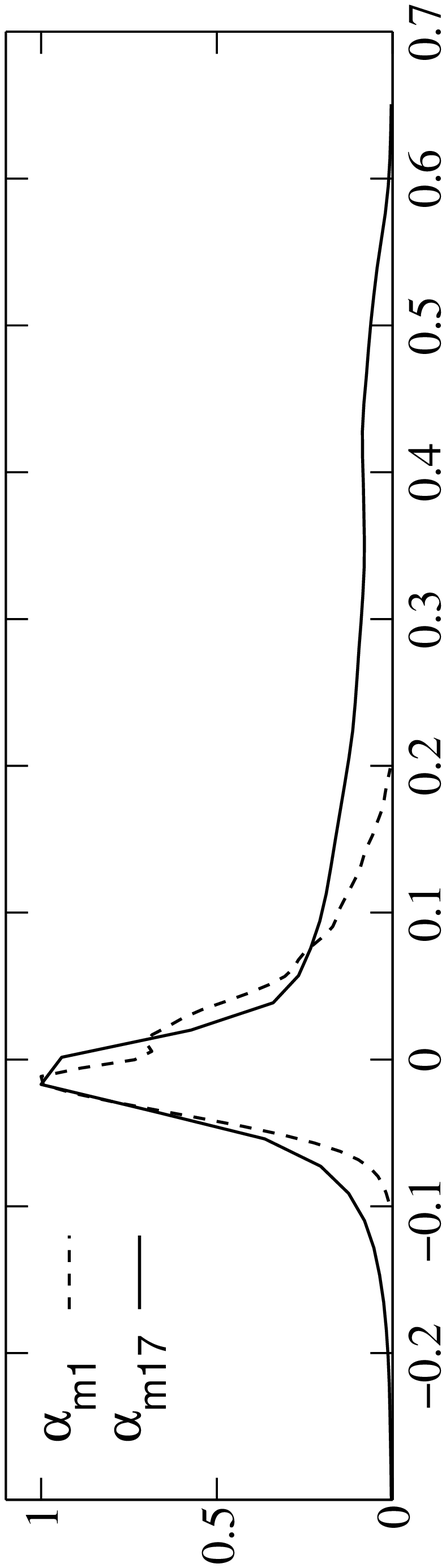}
  \includegraphics[angle=270,width=0.45\textwidth]{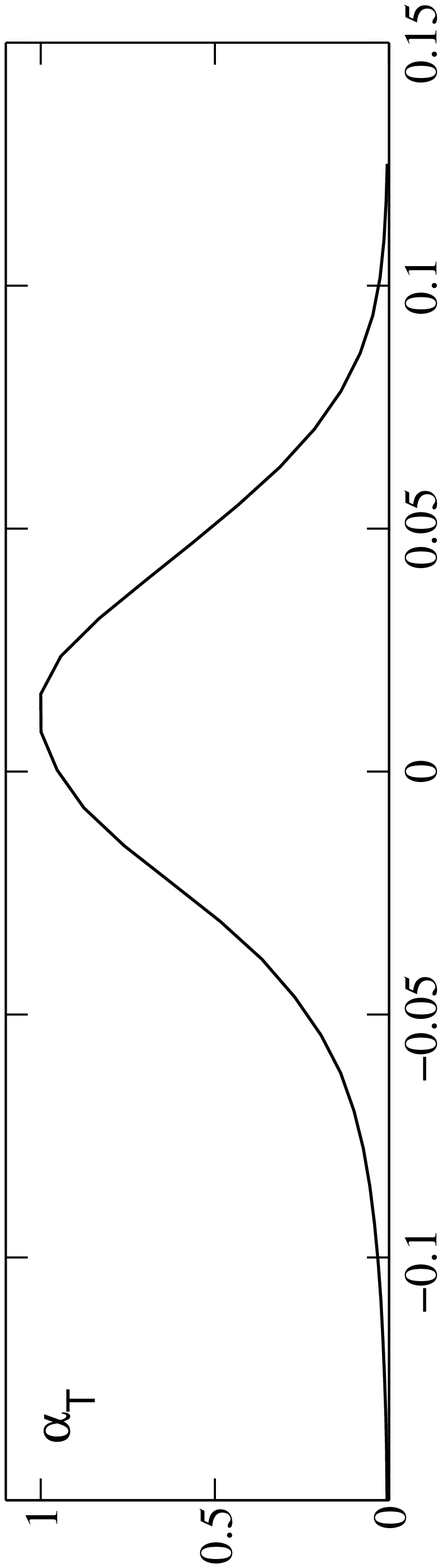}
  \caption{Marginalized likelihoods of non-adiabatic contributions
           to the observed spectra in our 11 parameter model.}
  \label{fig:non-adi_contr}
  \vspace{19mm}
\end{figure}
%
%%%%%%%%%%%%%%%%
the power at $l=2100$ is negligible due to diffusion damping. These
considerations lead us to one more measure of the non-adiabatic contribution
\begin{eqnarray}
\alpha_T & = &
\frac{\langle (\delta T^{\mr{non-ad}})^2 \rangle}{
\langle (\delta T^{\mr{total}})^2 \rangle}  \nonumber \\
& = &
\frac{\sum_{l = 2}^{2100} \left(2l + 1\right)
 \left(C_l^{\mr{TTiso}} + C_l^{\mr{TTcor}}\right)}{\sum_{l = 2}^{2100} \left(2l + 1\right)
 C_l^{\mr{TT}}}\,.
\label{eqn:alphaT}
\end{eqnarray}

\emph{Correlated models.} In Fig.~\ref{fig:non-adi_contr} we plot the
marginalized 1-d likelihoods for $\alpha_2$, $\alpha_{200}$, $\alpha_{m1}$,
$\alpha_{m17}$, and $\alpha_T$. At the quadrupole ($l=2$) a long tail of
$\alpha_2$ towards negative non-adiabatic contribution appears, since the
measured quadrupole is rather low compared to typical pure adiabatic models.
The 95\% C.L. region spans an interval $-0.46 < \alpha_2 < 0.10$. Around the
first acoustic peak the non-adiabatic contribution is much more constrained,
$-0.024 < \alpha_{200} < 0.079$, and  at $l = 700$ the allowed contribution
becomes even smaller, $-0.011 < \alpha_{700} < 0.026$.
In the matter power, the limits are $-0.06 < \alpha_{m1} <
0.14$ and $-0.11 < \alpha_{m17} < 0.51$. The latter, quite large values come because
we do not use any data from larger $k$. So the spectrum is practically
unconstrained after $k_{17}$. (Recall also our warning example in Figure
\ref{fig:absbest_mat_win}.) The likelihood for the total non-adiabatic
temperature perturbation is quite symmetric with the median at $\alpha_T =
0.009$ and a 95\% C.L. interval $-0.075 < \alpha_T < 0.075$. Hence, we conclude
that the non-adiabatic contribution to the observed temperature perturbation is
less than 7.5\%.

\emph{Uncorrelated models.} Four years ago we \cite{Enqvist:2000hp} found upper
limits for an uncorrelated CDM isocurvature contribution using the first data
releases of Boomerang \cite{deBernardis:2000gy} and Maxima \cite{Balbi:2000tg}
together with COBE data \cite{Tegmark:2000db}. The 95\% C.L. limits were
$\alpha_2 < 0.56$ (called $\alpha$ in \cite{Enqvist:2000hp}) and $\alpha_{200}
< 0.13$. Let us update these numbers to reflect the dramatically increased
accuracy of the data. We approximate uncorrelated models by applying a Gaussian
prior $\gamma = 0.00 \pm 0.02$ when analyzing the chains.
%%%%%%%%%%%%%%%%%%
%
\begin{figure}[t]
  \centering
  \includegraphics[angle=270,width=0.45\textwidth]{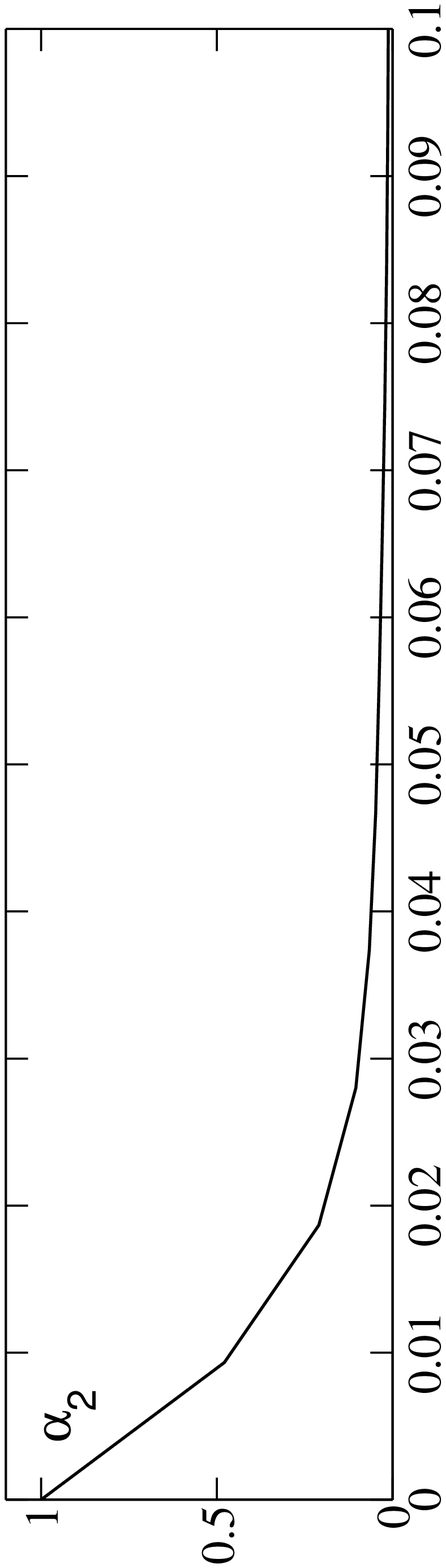}
  \includegraphics[angle=270,width=0.45\textwidth]{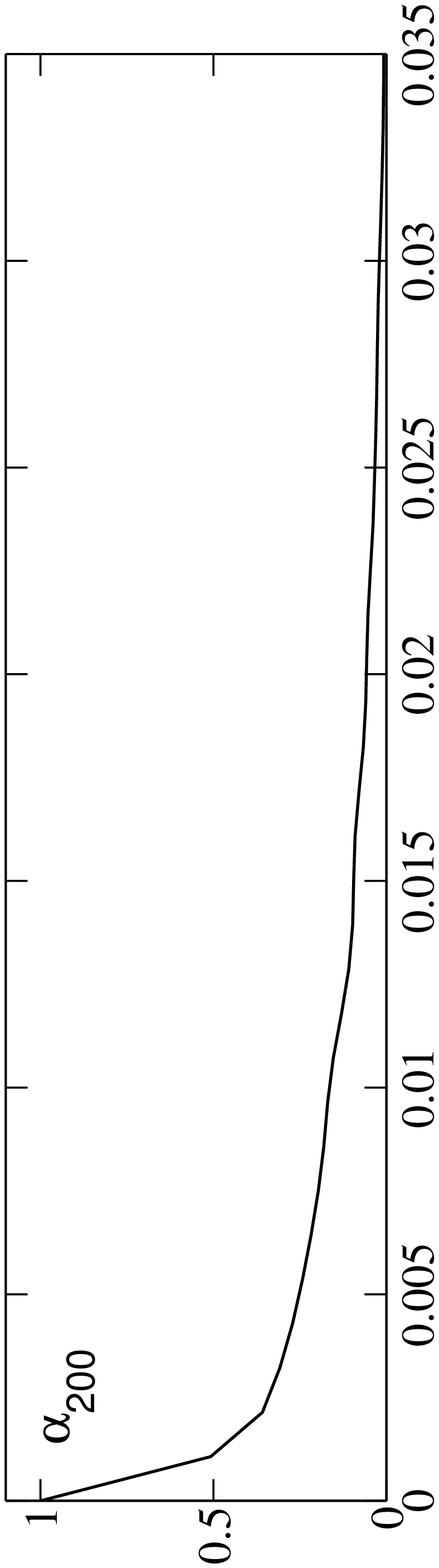}
  \includegraphics[angle=270,width=0.45\textwidth]{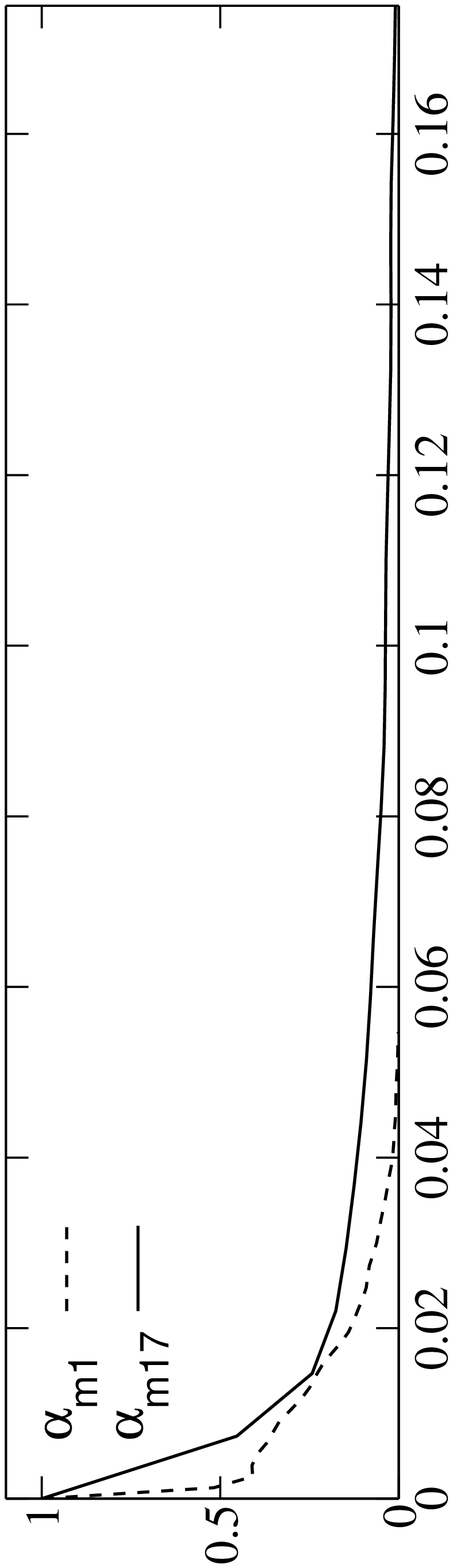}
  \includegraphics[angle=270,width=0.45\textwidth]{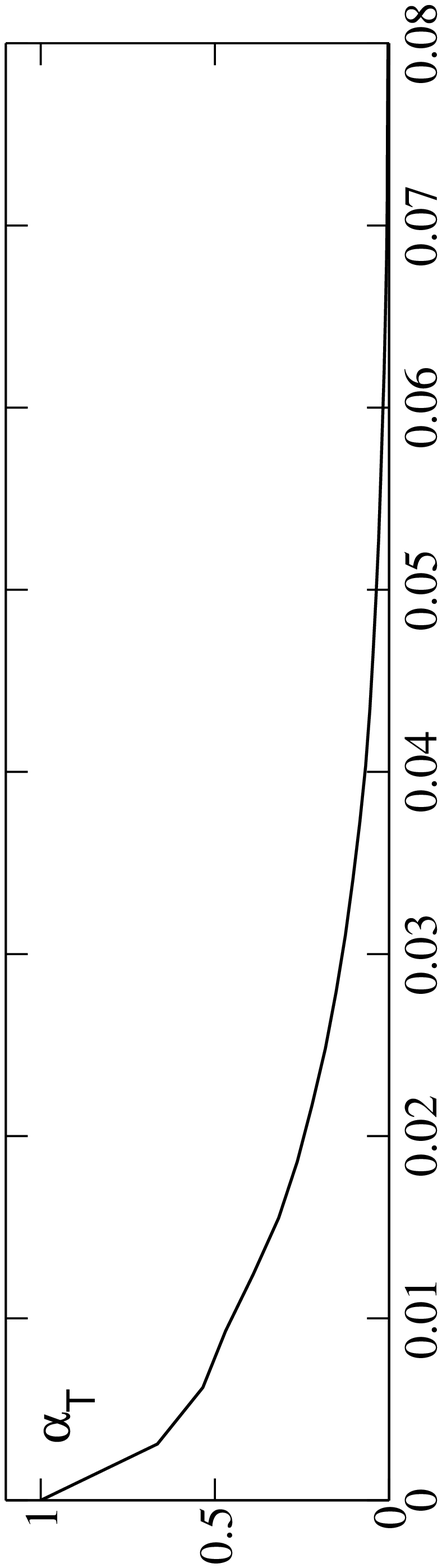}
  \caption{Marginalized likelihoods of non-adiabatic contributions
           to the observed spectra  in ``uncorrelated models''
           ($\gamma = 0.0 \pm 0.02$).}
  \label{fig:non-adi_contr_sg}
\vspace{13mm}
\end{figure}
%
%%%%%%%%%%%%%%%%%%%%
Since the data does
not favor correlation (see Figure \ref{fig:isoparam}), the sampling of models
with small $|\gamma|$ is very good. For uncorrelated models the correlation
component is missing from definitions (\ref{eqn:alphal}), (\ref{eqn:alphami}),
and (\ref{eqn:alphaT}). Then the range for $\alpha_l$, $\alpha_{mi}$ and
$\alpha_T$ is $[0,1]$.
1-d likelihoods are given in
Fig.~\ref{fig:non-adi_contr_sg}. The 95\% C.L. limits are $\alpha_2 < 0.085$
and $\alpha_{200} < 0.023$. So, the allowed isocurvature contribution in the
uncorrelated case has dropped to about one sixth part of the limits obtained
four years ago. Finally,  the allowed total non-adiabatic contribution
($\alpha_T$) to the observed temperature perturbation signal becomes less than
4.3\%.

\section{Effect of Choice of Pivot Scale}
\label{sec:pivotscale}

When the modes have different spectral indices, the relative amplitude
parameters $\alpha$ and $\gamma$ become dependent on the choice of pivot scale
$k_0$.  In the literature, different pivot scales have been used, e.g., $k_0 =
0.002\mbox{ Mpc}^{-1}$ and $k_0 = 0.05\mbox{ Mpc}^{-1}$,
whereas we have chosen an intermediate value $k_0 = 0.01\mbox{ Mpc}^{-1}$.

One can convert the results obtained using one pivot scale $k_0$ to what one
would get with another pivot scale $\tilde{k}_0$, by using the parameter
transformation
\begin{widetext}
  \begin{equation}
    \tilde{\alpha} =
%    \frac{\tilde{B}^{2}}{\tilde{A}^{2}} =
    \frac{\alpha\hat{k}^{\niso-1}}
    {(1-\alpha)(1-\abs{\gamma})\hat{k}^{\nadiI-1} +
      (1-\alpha)\abs{\gamma}\hat{k}^{\nadiII-1} + \alpha \hat{k}^{\niso-1}}
    \label{eq:newalpha}
  \end{equation}
  \begin{equation}
    \tilde{\gamma}
%    = \mr{sign}(\tilde{A}_{\mr{s}}\tilde{B})
%    \frac{\tilde{A}^{2}_{\mr{s}}}{\tilde{A}^{2}_{\mr{r}}\tilde{A}^{2}_{\mr{s}}}
    = \frac{\gamma\hat{k}^{\nadiII-1}}
    {(1-\abs{\gamma})\hat{k}^{\nadiI-1} + \abs{\gamma}\hat{k}^{\nadiII-1}}
    \label{eq:newgamma}
  \end{equation}
  \begin{equation}
    \tilde{A}^{2} = A^{2}\left[(1-\alpha)(1-\abs{\gamma})\hat{k}^{\nadiI-1} +
      (1-\alpha)\abs{\gamma}\hat{k}^{\nadiII-1} +
      \alpha\hat{k}^{\niso-1}\right] \,,
    \label{eq:newA}
  \end{equation}
where $\hat{k} \equiv \tilde{k}_{0}/k_{0}$, and weighting the likelihoods with
the Jacobian determinant of this parameter transformation,
  \begin{equation}
    \label{eq:Jacobian}
    J = \frac{\hat{k}^{\nadiI-1}\hat{k}^{\nadiII-1}
      \hat{k}^{\niso-1}}
    {\left[(1-\abs{\gamma})\hat{k}^{\nadiI-1} +
        \abs{\gamma}\hat{k}^{\nadiII-1}\right]
      \left[(1-\alpha)(1-\abs{\gamma})\hat{k}^{\nadiI-1} +
      (1-\alpha)\abs{\gamma}\hat{k}^{\nadiII-1} +
      \alpha\hat{k}^{\niso-1}\right]^{2}} \,.
  \end{equation}
\end{widetext}
This weighting gives the effect of changing from flat priors for $\alpha$,
$\gamma$, and $\ln A$ to flat priors for $\tilde{\alpha}$, $\tilde{\gamma}$
and $\ln\tilde{A}$.

Typically we have $\nadiI \approx \nadiII \approx 1$, so that
 \bea
   \tilde{\alpha} & \approx & \frac{\alpha\hat{k}^{\niso-1}}
   {1-\alpha + \alpha\hat{k}^{\niso-1}}
   \sim \alpha\hat{k}^{\niso-1}  \label{alphaboost} \\
   \tilde{\gamma} & \approx & \gamma \\
   \tilde{A}^2 & \approx & A^2\left(1-\alpha+\alpha\hat{k}^{\niso-1}\right) \sim
   A^2
 \eea
and
 \beq
    J \approx
    \frac{\hat{k}^{\niso-1}}{\left(1-\alpha+\alpha\hat{k}^{\niso-1}\right)^2}
    \sim \hat{k}^{\niso-1} \,,
 \eeq
where the ``$\sim$'' are for small $\alpha$.  Thus, if $\tilde{k}_0 > k_0$, the
likelihood of models with large $\niso$ is increased and that of small $\niso$
is decreased and the opposite holds if $\tilde{k}_0 > k_0$.

To study the effect of varying $k_0$, we both (1) applied the above
transformation to our results from our main run and (2) did shorter MCMC runs
(8 chains) using $k_0 = 0.002\mbox{ Mpc}^{-1}$ and $k_0 = 0.05\mbox{
Mpc}^{-1}$. Both methods should give the same result if the MCMC runs have
sufficient statistics.  In practice, the results were close to each other for
$k_0 = 0.002\mbox{ Mpc}^{-1}$, but for $k_0 = 0.05\mbox{ Mpc}^{-1}$, our
original run had insufficient sampling at large $\niso$, for the
reparameterization to give meaningful results.  We show in Fig.~{\ref{fig:pivotscale}
the resulting marginalized likelihoods for the (primary) parameters most affected.
For $k_0 = 0.002\mbox{ Mpc}^{-1}$ the result shown
is by method (1), but for $k_0 = 0.05\mbox{ Mpc}^{-1}$ by method (2), since it
had better statistics.  However, these results should only be
taken as indicative, especially for $k_0 = 0.05\mbox{ Mpc}^{-1}$
as the statistics was not nearly as good as in our
main case, $k_0 = 0.01\mbox{ Mpc}^{-1}$.
\begin{figure}[th]
  \centering
  \includegraphics[angle=270,width=0.40\textwidth]{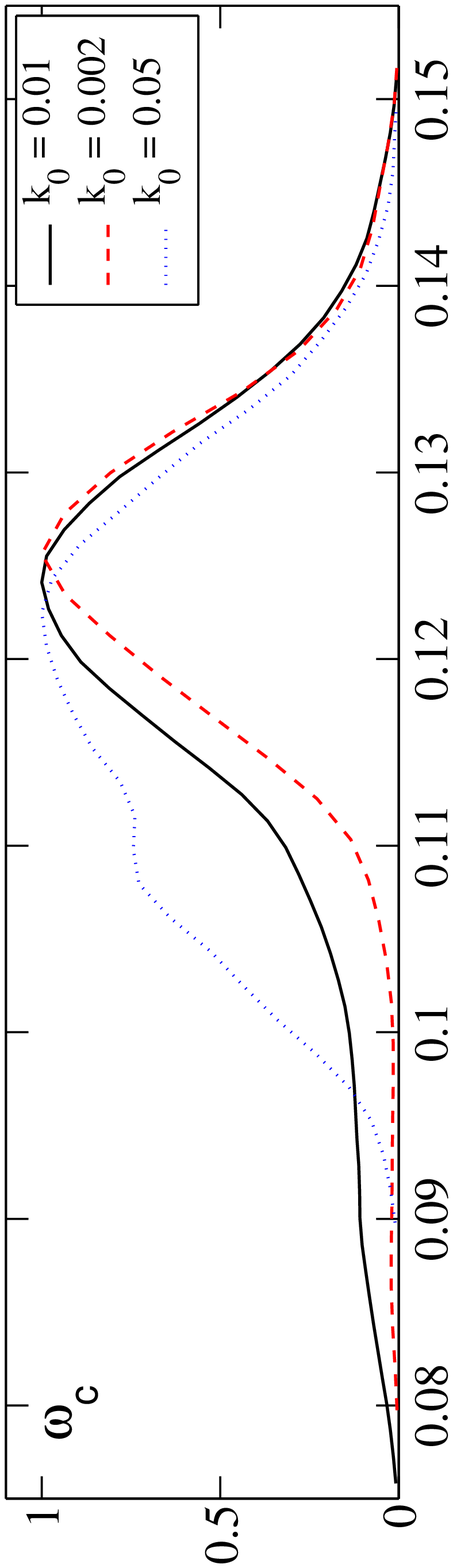}
  \includegraphics[angle=270,width=0.40\textwidth]{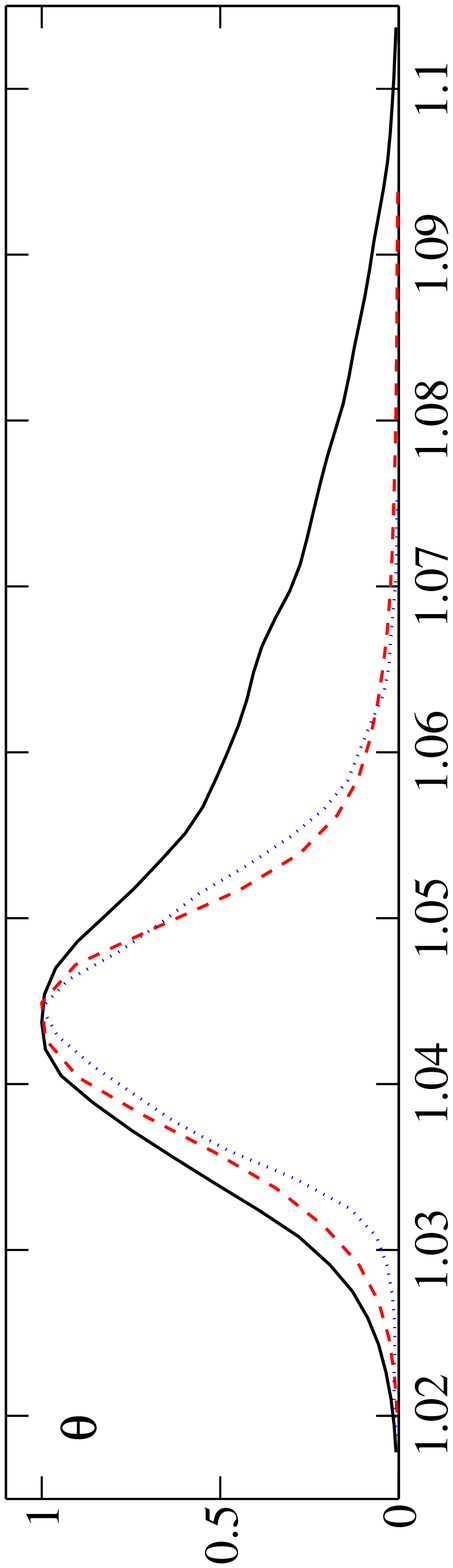}
  \includegraphics[angle=270,width=0.40\textwidth]{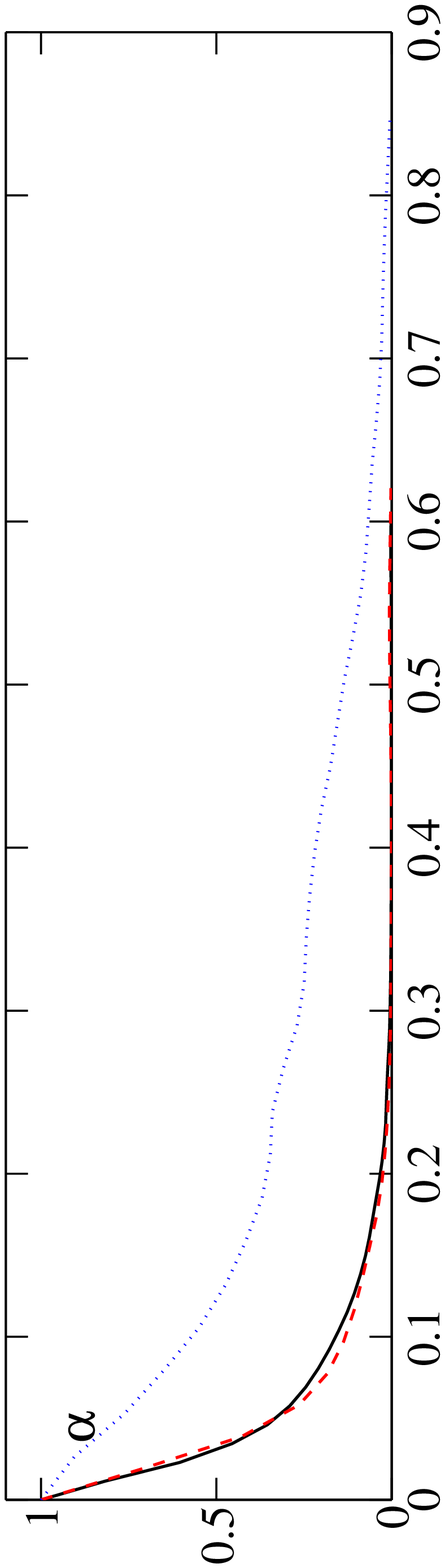}
  \includegraphics[angle=270,width=0.40\textwidth]{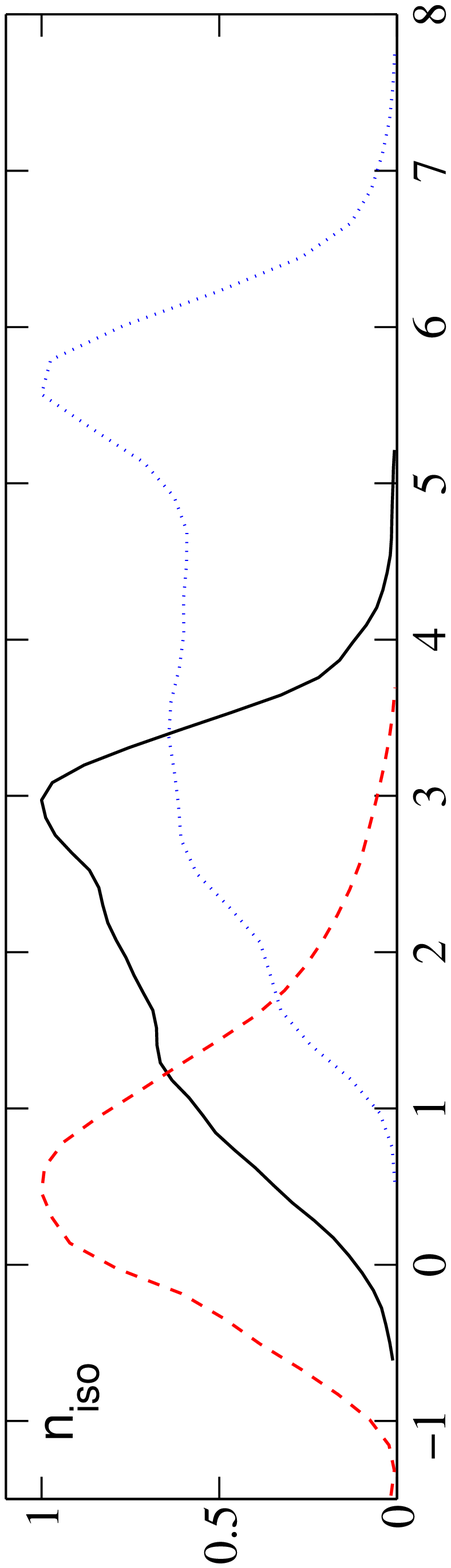}
  \includegraphics[angle=270,width=0.40\textwidth]{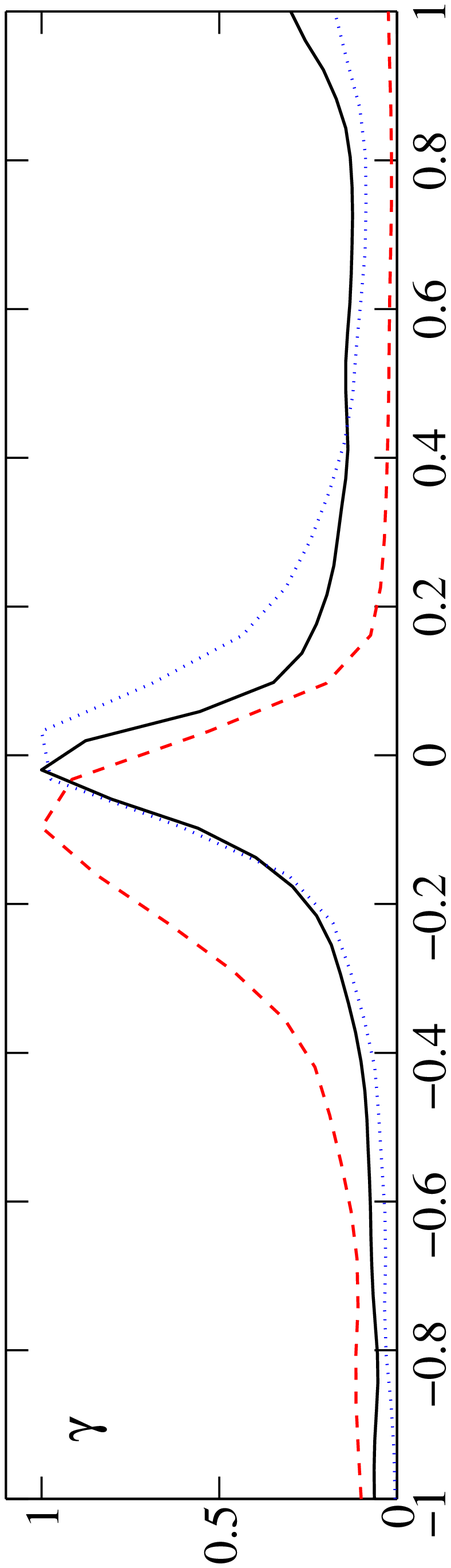}
  \includegraphics[angle=270,width=0.40\textwidth]{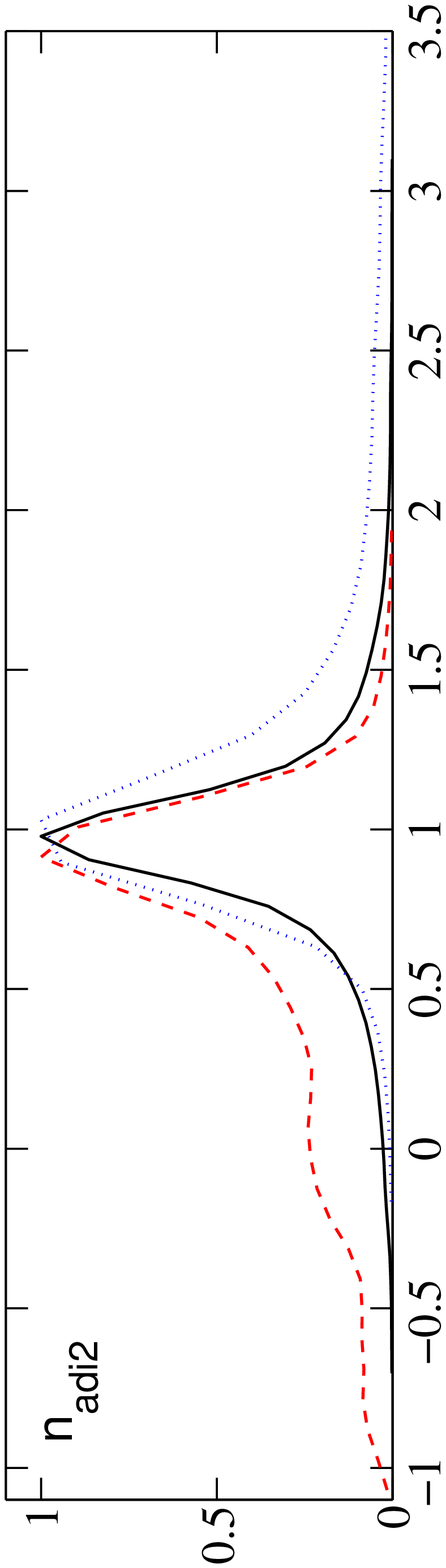}
  \caption{1-dimensional likelihoods for $\omega_c$, $\theta$,
    $\alpha$, $\niso$, $\gamma$, and $\nadiII$, using three different
    pivot scales, $k_0 = 0.002\mbox{Mpc}^{-1}$ ({\em dashed}, red), $k_0 =
    0.01\mbox{Mpc}^{-1}$ ({\em solid}, black), and $k_0 = 0.05\mbox{Mpc}^{-1}$ ({\em
      dotted}, blue).}
  \label{fig:pivotscale}
\end{figure}

The 1-d likelihoods of $\omega_b$, $\tau$, $b$, and $\nadiI$ did not change
significantly.  Thus these parameters are not sensitive to the choice of pivot
scale.  The parameter affected the most is $\niso$, where we see very clearly the
shift to smaller (larger) $\niso$ as the pivot scale is increased (decreased).

Consider first the change to a large pivot scale $k_0 = 0.002\mbox{ Mpc}^{-1}$,
corresponding to $l_0 \sim 28$.  Now a lot of weight is given to models with a
``red'' isocurvature spectrum, $\niso<1$.  For these models the isocurvature
contribution is significant in the SW region of the CMB spectrum, and
negligible elsewhere.  Accordingly, negative correlation $\gamma$ is favored,
since it subtracts power in the SW region where the data is below the adiabatic
model prediction.  A red correlated adiabatic index $\nadiII$ is favored, as
the low-$l$ boost in the negative $C^\mr{cor}_l$ tends to win over that in the
positive $C^\mr{ad2}_l$.  Because of the negative correlation contribution,
somewhat larger amplitudes $A$ are favored (not shown in Fig.~\ref{fig:pivotscale}).
With very little weight
at $\niso \sim 3$, the large-$\Omega_\Lambda$ models are eliminated, so the
tails in the $\omega_c$ and $\theta$ distributions disappear.

The 1-d likelihood for the isocurvature fraction $\alpha$ is surprisingly close
to the $k_0 = 0.01\mbox{Mpc}^{-1}$ case.  Thus our upper limit $\alpha < 0.18$
seems to be more robust than one might have thought, and applies over a fairly
large range of scales.

Consider then the change to a small pivot scale $k_0 = 0.05\mbox{ Mpc}^{-1}$,
corresponding to $l_0 \sim 700$.  This has the effect that the problematic
``high likelihood'' region around $\niso \sim 5$--$6$, discussed in
Sec.~\ref{sec:largeniso}, acquires a much larger measure in the parameter
space, increasing the marginalized likelihood of these $\niso$ values. These
models have now a large weight in the 1-d likelihoods of all parameters.  While
they had a very small $\alpha(k_0 = 0.01\mbox{Mpc}^{-1})$, they have a rather
large $\alpha(k_0 = 0.05\mbox{Mpc}^{-1})$ (see Eq.~(\ref{alphaboost})), and
thus the $\alpha$ distribution now extends to large values.
At the 95\% C.L. we obtain $\alpha(k_0 = 0.05\mbox{Mpc}^{-1}) < 0.56$.
The ``bump'' in
the $\omega_c$ distribution around $\sim 0.11$ is also due to these models. The
way the SDSS window functions collect power from smaller scales
(Sec.~\ref{sec:largeniso}) of $P^\mr{iso}(k)$ allows a smaller ``shape
parameter'' $\Omega_m h$ for $P^\mr{ad}(k)$, which leads to the smaller
$\omega_c$.  As explained in Sec.~\ref{sec:largeniso}, we do not take these
models with $\niso \sim 5$--$6$ seriously.  Thus the $k_0 = 0.05\mbox{ Mpc}^{-1}$
case here should just
be taken as a warning for what may happen with extreme values of spectral
indices in this kind of studies.

In general, the pivot scale should be chosen to be in the middle of the data
set.  If $k_0$ is near either end of the range of scales covered by the data,
the spectral indices of components which are subdominant in the main data
region become unconstrained in the direction which causes this component to
blow up outside, or near the edge, of the data set.

\section{Comparison to Beltran et al}
\label{sec:beltran}

Due to the many differences in approach discussed in Sec.~\ref{sec:correlation}
the comparison of our results to those of \cite{Beltran:2004uv} is not
straightforward. If we include an $\Omega_\Lambda = 0.70\pm0.04$ prior to mimic
their use of SNIa data, the (large $\theta$, small $\omega_c$) models we found
but they did not, disappear from our results. The remaining minor differences
in the determination of standard (adiabatic) cosmological parameters are mainly
due to a different choice of pivot scale $k_0$.  When we shift to the same
pivot scale, $k_0 = 0.05\mbox{ Mpc}^{-1}$ they used, our results approach
theirs. Our results for this pivot scale are however contaminated by
problematic very large $\niso\sim5$--$6$ models, whereas they have imposed an
upper limit $\niso<3$, so the results are not directly comparable even in this
case.

The parameters related to isocurvature perturbations are defined with respect
to the pivot scale.  Our upper limit to the isocurvature fraction $\alpha$ at
pivot scales $k_0=0.002\mbox{ Mpc}^{-1}$ and $k_0=0.01\mbox{ Mpc}^{-1}$ is much
tighter than their upper limit of about 60\% at $k_0=0.05\mbox{ Mpc}^{-1}$. Our
unreliable $k_0=0.05\mbox{ Mpc}^{-1}$ upper limit $\alpha \leq 0.56$ agrees
with that limit.

Because of different choice of correlation parameters, the results for
correlation are best compared in terms of our $\alpha_\mr{cor}$, which equals
their $-\beta\sqrt{\alpha(1-\alpha)}$ plotted in Fig.~3 of
\cite{Beltran:2004uv}. Our discovery of a preference for positive correlations
at large $\niso$ is in agreement with their result (with the opposite sign
convention).

\section{Discussion}
\label{sec:discussion}

We have used CMB and large-scale structure data to constrain models where the
primordial perturbations have both an isocurvature and an adiabatic component,
allowing for different spectral indices for these components, and a possible
correlation between them.  We restricted these models to a spatially flat
($\Omega = \Omega_\Lambda + \Omega_m = 1$) background universe.

The basic conclusion is that the data clearly disfavors the presence of
isocurvature perturbations. This makes a likelihood study of such models
problematic, since once the isocurvature contribution is small, the related
spectral indices become unconstrained. When some of the independent parameters
are unconstrained, the likelihood function becomes sensitive to the implied
prior due to the parametrization used.  We demonstrated this by changing the
pivot scale used to define our isocurvature and correlation fraction
parameters.

The problem with spectral indices does not occur when a model has only one
independent spectral index. It would also not occur if the data would clearly
favor a nonzero fraction for any component whose spectral index we have as an
independent parameter.

Perhaps a better parametrization of isocurvature models would be to use the
amplitudes at two different scales (e.g. at $k_\mr{min}$ and $k_\mr{max}$) as
the independent parameters for the likelihood analysis, instead of an amplitude
at one scale and a spectral index. The spectral index would then become a
derived parameter. We suggest that one tries this approach in future studies,
since it might: 1) Lead to a much faster convergence of the MCMC chains because
the unconstrained spectral indices would be missing. 2) Remove a possible bias
towards zero isocurvature amplitude models, which was a result of blowing up
the parameter space volume upon marginalization caused by unconstrained $\niso$
in case of small $\alpha$. 3) Prevent the feature that with too large $k_0$ the
integration measure (weight) of models with extremely large $\niso$ becomes
arbitrary large.

For models with the largest isocurvature fractions at the pivot scale $k_0 =
0.01\mbox{Mpc}^{-1}$, which is roughly in the middle of the data set used, the
isocurvature spectral index is constrained to be in the range $0.5 \lesssim
\niso \lesssim 3.5$ which prevents the isocurvature contribution from rising
too high either in the small- or large-scale ends of the data used.  If one
moves the pivot scale to smaller (larger) scales the upper (lower) limit to
$n_{\mr{iso}}$ is relaxed, or removed, as the rising part of the isocurvature
spectrum moves outside the data range.

Of the standard (adiabatic model) cosmological parameters, the determination of
the baryon density $\omega_b$, the primordial perturbation amplitude $A$, the
adiabatic spectral index $\nadi$, the optical depth due to reionization
$\tau$, or the bias parameter $b$, is not significantly affected by a possible
isocurvature contribution.  On the other hand, models with a smaller CDM
density $\omega_c$ and a larger sound horizon angle $\theta$ become acceptable.
This means that we cannot even rule out models with $H_0 > 100 \mbox{
km/s/Mpc}$ and $\Omega_m < 0.1$  (at 95\% C.L.) using CMB and LSS data alone.

We obtained an upper limit $\alpha < 0.18$ (95\% C.L.) for the CDM isocurvature
fraction for models where correlation is allowed between the isocurvature and
adiabatic contributions.  This limit is somewhat tighter than the corresponding
limit for uncorrelated models, since correlation causes a stronger signature in
the data than an uncorrelated isocurvature perturbation.

Here $\alpha$ is defined as the ratio
$\mc{P}_\mc{S}/(\mc{P}_\mc{R}+\mc{P}_\mc{S})$ of the primordial entropy and
curvature perturbation power spectra, at a pivot scale $k_0$, and our upper
limit applies for both $k_0 = 0.002 \mbox{ Mpc}^{-1}$ and $k_0 = 0.01 \mbox{
Mpc}^{-1}$, and presumably also for the range in between. The value $\alpha =
0.18$ corresponds to $f_\mr{iso} \equiv (\mc{P}_\mc{S}/\mc{P}_\mc{R})^{1/2} =
0.47$. For smaller scales (larger $k$) our results are less conclusive, since
there the constraint on $\alpha$ relies more on the large-scale structure
(SDSS) data, whose use is problematic for a steeply rising (large $\niso$)
isocurvature contribution. However, our results for $k_0 = 0.05 \mbox{
Mpc}^{-1}$ are not in disagreement with the upper limit of 60 \% for $\alpha$
obtained in \cite{Beltran:2004uv} using this pivot scale.

In the observed temperature anisotropy signal the amount of non-adiabatic
contribution is $|\alpha_T| < 0.075$ at 95\% C.L. in our correlated
isocurvature model. The upper limit becomes tighter in the uncorrelated case,
$\alpha_T < 0.043$ at 95\% C.L.

In models with a large isocurvature spectral index, $\niso \sim 2\mbox{--}4$, a
positive correlation between the adiabatic and isocurvature perturbations is
favored.  The correlation contribution appears then in the acoustic peak
region, where the effect of a positive correlation is to shift the acoustic
peaks towards larger multipoles $l$, which then favors a larger sound horizon
angle $\theta$ to push the peaks back to where the data has them.  To satisfy
also the large scale structure data, smaller CDM densities $\omega_c$ are then
favored.  These effects translate into a larger $H_0$ and a smaller $\Omega_m$
(larger $\Omega_\Lambda$).

In models with a small isocurvature spectral index, $\niso \sim 0\mbox{--}2$, a
negative correlation is favored.  Here the correlation contribution appears in
the Sachs-Wolfe region, where this negative correlation brings the $C_l$ down
to better agree with the small large-scale CMB anisotropy seen by WMAP.

\begin{acknowledgements} We thank the CSC - Scientific Computing Ltd. (Finland)
for computational resources.  HKS would like to thank Sarah Bridle for
introducing him to CosmoMC.  VM was supported by the Magnus Ehrnrooth
Foundation and the Graduate School in Astronomy and Space Physics.
JV was supported by the Magnus Ehrnrooth foundation
and by the Research Foundation of the University of Helsinki
(Grant for Young and Talented Researchers).

\end{acknowledgements}

%\begin{thebibliography}{99}
\bibliography{hxreferences2}
%\end{thebibliography}

\end{document}